\documentclass[3p]{elsarticle}
\usepackage[dvipsnames]{xcolor}
\usepackage{soul}
\usepackage{lineno}
\usepackage{booktabs}
\usepackage{multirow}
\usepackage{algorithmic,algorithm}
\usepackage{graphicx}
\usepackage{array}
\usepackage{todonotes}
\usepackage{gensymb}
\usepackage{amsmath}
\usepackage{amsfonts}
\usepackage{amssymb}
\usepackage{comment}
\usepackage{pgfplots}
\usepackage{svg}
\usepackage[export]{adjustbox}
\usepackage[caption=false,font=footnotesize]{subfig}

\usepackage{url}%

\biboptions{numbers,sort&compress}
\pgfplotsset{compat=1.16} 

\newcommand{\tablescale}{0.75}
\definecolor{blue}{HTML}{1F77B4}
\newcolumntype{R}[1]{>{\raggedleft\let\newline\\\arraybackslash\hspace{0pt}}m{#1}}

\makeatletter
\def\ps@pprintTitle{%
 \let\@oddhead\@empty
 \let\@evenhead\@empty
 \def\@oddfoot{}%
 \let\@evenfoot\@oddfoot}
\makeatother

\begin{document}

\begin{frontmatter}
\title{Simulation study on the fleet performance of shared autonomous bicycles}


\author[mit]{Naroa Coretti Sanchez \corref{cor1}%
\fnref{fn1}}
\ead{naroa@mit.edu}

\author[unav]{Iñigo Martinez \fnref{fn1}}
\ead{imartinezl@alumni.tecnun.es}

\author[mit]{Luis Alonso Pastor}
\ead{alonsolp@mit.edu}

\author[mit]{Kent Larson}
\ead{kll@mit.edu}

\address[mit]{MIT Media Lab, Cambridge, MA, USA}
\address[unav]{University of Navarra, San Sebastian, Spain}

\cortext[cor1]{Corresponding author}
\fntext[fn1]{These authors contributed equally to this work.}


\begin{abstract}
Rethinking cities is now more imperative than ever, as society faces global challenges such as population growth and climate change. The design of cities can not be abstracted from the design of its mobility system, and, therefore, efficient solutions must be found to transport people and goods throughout the city in an ecological way. An autonomous bicycle-sharing system would combine the most relevant benefits of vehicle sharing, electrification, autonomy, and micro-mobility, increasing the efficiency and convenience of bicycle-sharing systems and incentivizing more people to bike and enjoy their cities in an environmentally friendly way. Due to the uniqueness and radical novelty of introducing autonomous driving technology into bicycle-sharing systems and the inherent complexity of these systems, there is a need to quantify the potential impact of autonomy on fleet performance and user experience. This paper presents an ad-hoc agent-based simulator that provides an in-depth understanding of the fleet behavior of autonomous bicycle-sharing systems in realistic scenarios, including a rebalancing system based on demand prediction. In addition, this work describes the impact of different parameters on system efficiency and service quality and quantifies the extent to which an autonomous system would outperform current bicycle-sharing schemes. The obtained results show that with a fleet size three and a half times smaller than a station-based system and eight times smaller than a dockless system, an autonomous system can provide overall improved performance and user experience even with no rebalancing. These findings indicate that the remarkable efficiency of an autonomous bicycle-sharing system could compensate for the additional cost of autonomous bicycles.
\end{abstract}


\end{frontmatter}


\section{Introduction}

The global population is projected to grow to 9.7 billion by 2050 \cite{UNreport2019}. Combined with the effect of urbanization, this process could add 2.5 billion people to urban areas \cite{UNreport2018}. In 2018, 55\% of the world's population lived in cities, and this number is expected to grow to 68\% by 2050 \cite{UNreport2018}. Population growth poses a challenge for sustainable development \cite{UNreport2019}, making the need for rethinking cities each time more imperative. Challenges such as climate change, along with other issues such as social inequality or the recent COVID-19 pandemic, are constantly reminding us of the urgency of designing cities to be more sustainable, equitable, and resilient. 

The design of cities cannot be abstracted from the design of their transport systems \cite{ALESSANDRINI2015145}; therefore, when considering the future of cities, the transportation of people and goods throughout the city becomes a key challenge. At the MIT Media Lab City Science group, we envision future cities to have a human scale, composed of dense and diverse districts where all the resources and amenities needed in daily life can be found within a walkable distance \cite{alonso2018cityscope}. In these districts, most of the trips would be one-person, short distance, and low speed, reducing car dependency and favoring walking and sustainable micro-mobility systems \cite{alonso2018cityscope, grignard2018cityscope, yao2019idk}.  In this scenario, micro-mobility systems could be a solution for these short-distance trips and provide the first- and last-mile connection to mass transit. Furthermore, micro-mobility solutions could also be used for package or food delivery when not in use by riders \cite{yao2019idk}. 

With this image of a future city in mind, the MIT Media Lab City Science research group has developed several mobility solutions over the last decades. These solutions include vehicles such as the CityCar lightweight electric folding automobile, or the PEV, an autonomous three-wheeler designed for shared use \cite{PEV, CityCar}. One of the most recent projects is the MIT Autonomous Bicycle Project: a novel system that aims to bring autonomy into bicycle-sharing systems (BSS) \cite{sanchez2020autonomous}. Autonomous driving capabilities would transform BSS into an on-demand mobility system, radically increasing the convenience of bike-share \cite{sanchez2020autonomous}.  In an autonomous BSS, a user would be able to request a trip through a mobile app, and an autonomous bicycle would drive to the user's location. Then, the user would ride the autonomous bicycle just like a regular bicycle. Upon arrival to the destination, the bicycle would drive autonomously to pick up another user, to a charging station, or towards wherever the demand is predicted to occur. 

We envision this bike to be part of the ecosystem of shared and autonomous micro-mobility in future walkable cities, as well as a new approach to current bicycle-sharing systems. Governments all over the world have been promoting public BSS, intending to improve mobility and public health and introduce a sustainable mode of transportation  \cite{shaheen2010bikesharing}. Especially in the last 15 years, BSS have quickly proliferated, increasing from  17 station-based systems worldwide in 2005 to over 2000 in 2019 \cite{sanchez2020autonomous}.

However, current BSS still face several challenges, among which it is worth highlighting the rebalancing problem caused by uneven travel patterns and the oversupply of bicycles that often happens in dockless systems \cite{ScientometricReview, curran2008, shaheenguzman2011, gu2019, shi2018}. Autonomous bicycles could be the solution to these challenges: For the system operators, it would solve the rebalancing problem and make the sharing system more efficient, with higher vehicle utilization rates and smaller fleet sizes. Moreover, having bicycles with autonomous driving capabilities would bring the convenience of mobility-on-demand into the current BSS, improving user experience. This article aims to describe and quantify to which extent an autonomous system would outperform the current BSS.

Most of the previous research around the performance of shared autonomous vehicles has been focused on cars or taxis. Narayanan et al. \cite{narayanan2020shared} provide an overview of the relevant studies in the field of shared autonomous vehicle (SAV) services published from 1950 to 2019. This overview covers the most relevant work done in terms of simulations SAV \cite{fagnant2014travel, fagnant2018dynamic, martinez2015urban, moreno2018shared, martinez2017assessing, lokhandwala2018dynamic, jager2017agent, boesch2016autonomous, loeb2018shared} and provides a comparison of the main results. While this overview explicitly warns that bike-sharing and scooter-sharing systems are not considered in the study, there are several similarities between simulations of other shared autonomous vehicles, such as cars, taxis, or buses, and shared autonomous bikes. Therefore, we recommend visiting the overview mentioned above for further details. With regards to simulations related to autonomous micro-mobility, Grignard et al. \cite{grignard2018impact}  proposed a generic simulation framework that allows for modeling different future mobility modes and understanding their impact on traffic and congestion. Recently, Kondor et al. \cite{kondor2021estimating} presented a simulation study of the impact of a fleet of autonomous scooters.

Due to the uniqueness and radical novelty of introducing autonomous driving technology into BSS and the inherent complexity of these systems, there is the need to quantify the potential impact that autonomy might have on the fleet performance and user experience. Consequently, this paper presents an ad-hoc simulator that provides an in-depth understanding of the fleet behavior of an autonomous BSS, how different parameters affect the performance, and a quantification of the extent to which it outperforms the current BSS. These simulations were conducted considering the most realistic possible scenarios, including a rebalancing system based on demand prediction.

The remainder of the paper is organized as follows. Section \ref{vision} describes some central aspects for the mobility modes of future walkable cities. Section \ref{bikes} analyzes how autonomous bicycles can pose a new and more efficient approach to BSS. Then, Section \ref{methodology} describes the architecture of the proposed agent-based, discrete event simulation framework along with its main features. The three BSS under study are also defined and formalized in this section. Section \ref{results} gathers the results of the simulation for the autonomous system and current station-based and dockless systems, comparing their performance under different scenarios. This section also quantifies the performance of a rebalancing system based on demand prediction and analyzes how the different configuration parameters impact the performance metrics. Finally, Section \ref{conclusions} concludes by highlighting the main conclusions of this study and providing an outline for future work.

\section{Mobility in future walkable cities: the power of combining transformative ideas} \label{vision}

Cities are in need of a change in the approach to urban design. The vision of the City Science group is that humans, and not cars, should be at the center of how we design cities, promoting sustainable transportation modes such as walking and the use of micro-mobility systems. These human-scale cities would be composed of a network of walkable districts that have mixed land uses and offer places to live and work to a diverse group of people. In addition, the amenities needed to cover people's needs in their daily lives  (e.g., cafeterias, schools, hospitals, grocery stores) will also be included in these districts, drastically reducing the need for commuting.

In this scenario, most of the trips would be one person, short distance, and low speed. As a consequence, walking or micro-mobility systems could be a solution for these short-distance trips, as well as providing the first- and last-mile connection to mass transit when traveling between districts \cite{alonso2018cityscope, grignard2018cityscope, yao2019idk}. These micro-mobility systems should be adequately designed to create an ecological, efficient, and equitable urban transportation network that solves current urban challenges. We believe that to achieve these goals, vehicle sharing, electrification, autonomy, and micro-mobility have to be wisely combined in a way that increases their potential.

Vehicle sharing combines the efficiency of public transportation with the convenience of privately owned vehicles\cite{he2017service} and, according to various studies, can lead to a very significant reduction in the number of vehicles \cite{millard2005car, martin2010impact, martinez2015urban}. Moreover, as the cost of ownership is shared, it can promote the use of electric vehicles, which have a high purchasing cost but low operating costs \cite{he2017service}. 

Electrification has been at the center of the innovation efforts of automotive companies for many years already \cite{gruenig2011overview}. Some studies highlight that electric vehicles show lower life-cycle greenhouse gas emissions than conventional internal-combustion-engine vehicles, both for the vehicle and fuel \cite{miotti2016personal}, which is very relevant considering that, for example, in 2018, the transportation sector accounted for the 28.3\% of the greenhouse gas emissions in the US \cite{us-ghg}. It must be noted that other authors warn that there are some other indicators, such as human toxicity and terrestrial acidification, that are worsened and should be taken into consideration in the vehicle electrification process \cite{lombardi2017comparative}. However, as the overall benefit of electrification depends on the energy mix of the country, there is a clear potential for greater benefits if we think of a future where most of the energy will be generated by renewable energy \cite{lombardi2017comparative}.  Another remarkable benefit of electric vehicles is the possibility of using the batteries of the electric vehicles to improve the performance of the electric grid. The energy storage capacity in vehicles could be used to compensate for the irregular supply of some renewable energy sources such as photo-voltaic panels in cloudy weather, as well as to balance the supply and demand of the electric grid \cite{mitchell2010reinventing}.

With regards to autonomous driving technology, there is some evidence that privately-owned autonomous vehicles could increase the number of vehicles in the streets due to the reduction in the value of travel time losses \cite{van2016autonomous} and the increase in the segment of the population that can drive \cite{duarte2018impact}. Autonomous vehicles could also introduce extra empty trips to, for example, pick up a delivery. Nevertheless, facts change when considering autonomy in shared fleets. According to Cervero \cite{cervero2017mobility}, the combination of autonomy and car-sharing could be the next paradigm shift for mobility in the US.  While there might still be a slight increase in vehicle kilometers traveled (v.k.t.) due to rebalancing and induced demand, increased efficiency would lead to reduced emissions, increased vehicle utilization rates, and consequently smaller fleet sizes \cite{narayanan2020shared}. The convenience of vehicle sharing systems could also be increased by the introduction of autonomous vehicles thanks to the vehicles being able to drive to wherever the demand is emerging, instead of the user having to find the vehicle \cite{boesch2016autonomous}. Finally, autonomous vehicles could serve areas that are not well connected by public transportation, working as a first or last-mile solution and improving access to mass transit \cite{huang2021use}.
 
While the term 'autonomous vehicles' generally refers to cars, working with smaller and more lightweight vehicles can present numerous advantages. In the US, the number of micro-mobility trips increased from 35 million in 2017 to 84 million in 2018 and 136 million in 2019 \cite{nacto2019}. Lighter vehicles consume less energy \cite{buekers2014health}, at lower speeds, the computational demand is decreased, and collisions at lower speeds and lighter vehicles are less deadly  \cite{ballesteros2004pedestrian, tefft2013impact}. Hence, from the point of view of regulation, it is likely that autonomous micro-mobility solutions may be deployed at a large scale in cities sooner than autonomous cars.

Vehicle sharing, electrification, autonomy, and micro-mobility sharing can have the greatest impact if combined. According to Greenblatt et al. \cite{greenblatt2015autonomous}, by 2030, each autonomous taxi implemented could reduce the GHG emission per mile traveled in the US 87-94\% below internal combustion engine vehicles, as a result of sustainable power generation, fleet size reduction, and more cost-effective high-performance electric vehicles. We believe that translating these characteristics into micro-mobility systems could lead to the ideal mobility solution for future walkable cities. 

In light of the above, it can be concluded that, while some of the above concepts have previously been explored, if combined in a new way, they can lead to solutions that transform urban mobility, making cities more livable, equitable, and sustainable.

\section{Autonomous bicycles as a new generation of bicycle-sharing systems} \label{bikes}

\subsection{Insight into current bicycle-sharing systems}\label{background}

\begin{figure}[!htb]
    \centering
    \includegraphics[width=0.5\linewidth]{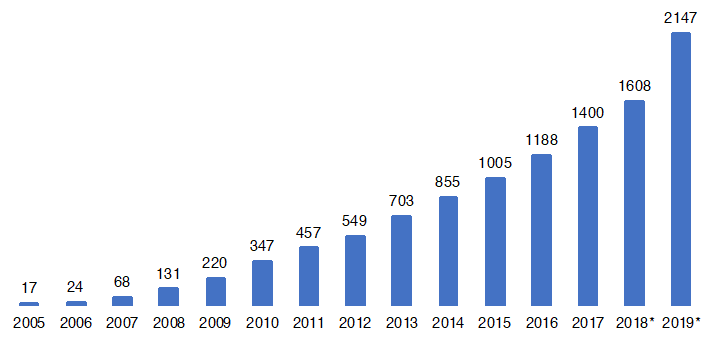}
    \caption{ Evolution of the number of station-based bicycle-sharing schemes worldwide \cite{sanchez2020autonomous}.} 
    \label{fig:Evolution of the number of station-based bicycle-sharing schemes}
\end{figure}

In recent years, BSS have become a mainstream solution for inner-city transportation in countries such as the US \cite{nacto2019}  with the goal of introducing a sustainable mobility mode that can help to alleviate urban congestion problems \cite{shaheen2010bikesharing}. As a result, the number of station-based BSS grew from 17 systems in 2005 to 2147 in 2019 worldwide (Figure \ref{fig:Evolution of the number of station-based bicycle-sharing schemes}). 

The first BSS are dated as early as 1965, but it took several generations of introducing new technologies, mainly to avoid theft and vandalism before BSS became successful \cite{demaio2009}. BSS typically provide a short-term rental of bicycles that can be picked up and returned in different locations. Currently, there are two main types of BSS: station-based systems and dockless systems. Station-based systems have a number of stations with several docking racks where users can either pick up or return bicycles \cite{luo2019comparative}. In this system, the bicycles can only be picked up from or returned at stations, limiting the use of BSS to the area where stations are available. More recently, GPS enabled a new way of bike-sharing which is called dockless bike-sharing \cite{ScientometricReview}. In this case, users can find available bicycles via a mobile app, and bikes can be dropped off at any location within the service region \cite{luo2019comparative}.

However, despite the improvements made throughout the various generations of BSS and its current popularity, some challenges must be faced in terms of efficiency, sustainability, and user experience. The remainder of this section describes two main challenges that are system rebalancing and bicycle oversupply. 

System rebalancing has been stated to be one of the biggest challenges of BSS by numerous studies: H Si et al. \cite{ScientometricReview} claim that the imbalance of the systems is the most urgent problem since the implementation of third-generation bicycle shares. Curran \cite{curran2008} refers to successful bicycle redistribution as one of the critical success factors in bicycle sharing schemes. Additionally, other sources such as Shaheen and Guzman \cite{shaheen2011worldwide} also highlight that the problem of finding available bikes and docks is one of the main complaints from users. 

The content and evolution of the research in recent years also showcase the importance of the rebalancing problem. In a scientometric literature review,  Hongyun Si et al. \cite{ScientometricReview} found that seven out of the ten most co-cited articles on bike-sharing systems between 2010 and 2018 referred to large-scale systems, demand prediction, and rebalancing problems.

The rebalancing problem occurs as a consequence of the spatial and temporal imbalance in commuting patterns and topographic landmarks like hills because users travel more often downhill than uphill \cite{lin2012geo}. In station-based systems, each station has a limited capacity, and the rebalancing problem leads to stations frequently lacking bicycles or docks \cite{Fricker2016}. In dockless systems, even if there is not a capacity restriction related to stations, finding available bicycles remains a problem due to trip flow imbalance \cite{wang2019}.

These experiences increase travel times and are detrimental to the user experience. Some studies found that not being able to find available bicycles and docks was the most frequent complaint among users \cite{Kaltenbrunner2010}, which led to frustration and loss of reliability in the system \cite{nair2013, lin2012geo}. This loss of reliability is particularly important since unreliability results in a negative effect in mode choice for commute trips, even for those who have flexible work schedules \cite{bhat2006impact}. This issue generates distrust in the system and could eventually lead to shifting to another mode of transportation.

Operators use vans and trucks to transport bicycles from saturated areas and mitigate the impact of the rebalancing problem. However, this process has a very high economic and ecological cost; redistribution is both the main contributor to the operational cost of bicycle-sharing schemes \cite{OBIS2011} and also the primary source of GHG emissions in the life-cycle of both station-based and dockless systems \cite{luo2019comparative}.

Bicycle oversupply in dockless systems has been another source of great concern. Dockless systems quickly proliferated around 2016 in China, the country with the largest dockless bike-sharing market \cite{DocklessBS}. Dockless bike-sharing was introduced in China in late 2015 \cite{shen2018understanding}, and from 2016 to 2017, the fleet of dockless systems grew from 2 million to 23 million \cite{DocklessBS}. 

Not being restricted by the number of docks, dockless systems can be up to 10 to 60 times larger than the existing station-based systems \cite{DocklessBS}. These fleet sizes vastly exceed the infrastructure capacities of cities, provoking urban problems such as the presence of dockless bicycles parked illegally clogging sidewalks \cite{zhao2020bike}. The number of bicycles also exceeds user demand, causing low levels of use of the bicycles and system inefficiency \cite{ScientometricReview}. 

As a consequence of these problems, dockless systems are currently facing a backlash from cities. Some cities have started to limit the number of bikes, while others have forced operators to cease their programs \cite{DocklessBS, shi2018}.  In China, from 2017 to 2019, the growth rate of dockless bike-sharing decreased from 631.2\% to 10.3\% \cite{zhao2020bike}. The saturation of the market has caused many companies to go bankrupt, and the amount of discarded bicycles has generated an enormous recycling crisis \cite{zhao2020bike}.

Overall, the severity of the problems faced with current BSS shows a need for disruptive innovation for bicycle-sharing to be the efficient, convenient, economical that our cities need.

\subsection{The MIT Autonomous Bicycle Project}\label{mit_bicycle}

\begin{figure}[!htb]
    \centering
    \includegraphics[width=0.45\linewidth]{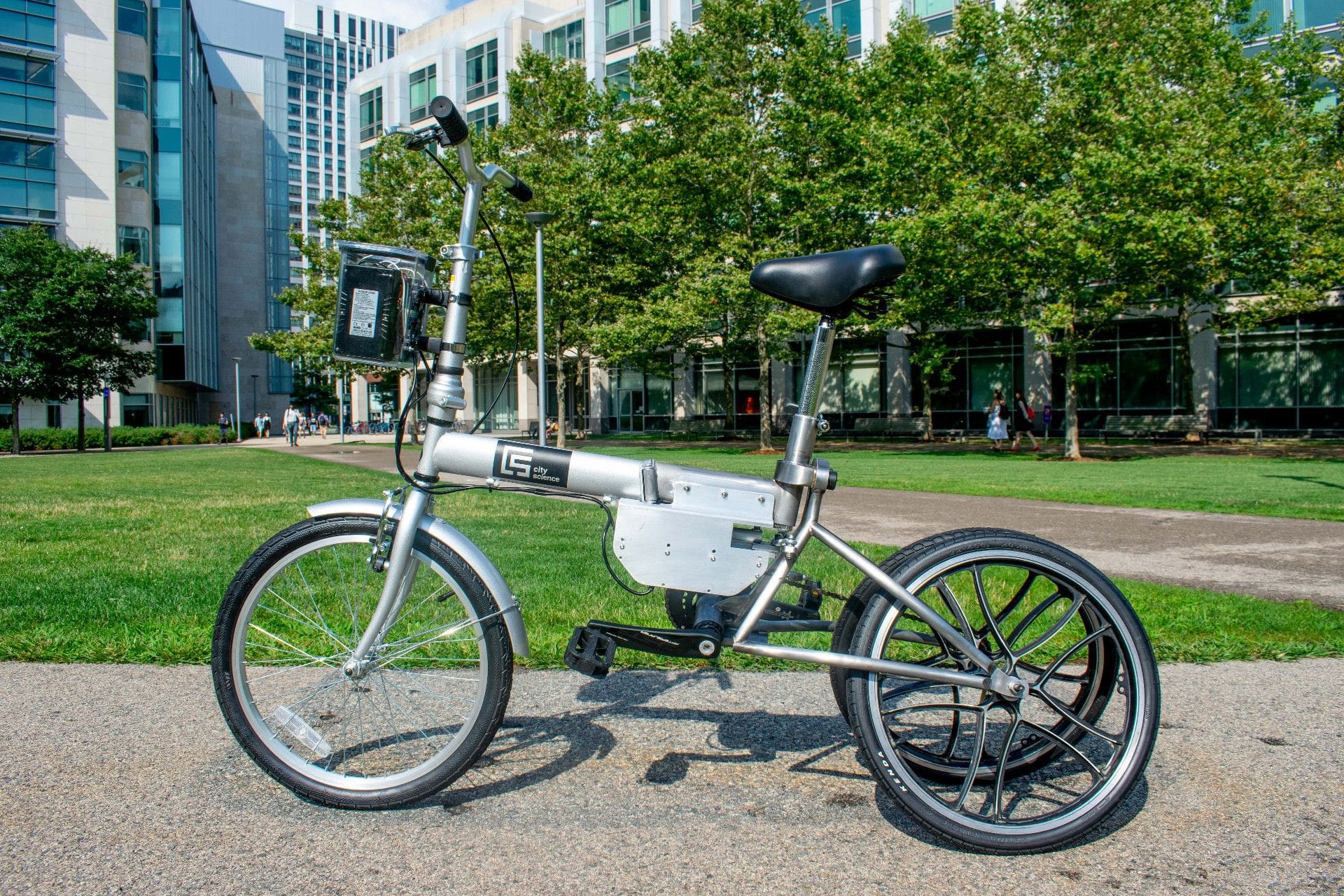}
    \caption{Prototype being tested at the MIT campus (Aug. 2019).}
    \label{fig:Prototype at the MIT campus}
\end{figure}

The MIT Media Lab City Science Group proposes  MIT Autonomous Bicycle as a mobility solution for future walkable cities (Figure \ref{fig:Prototype at the MIT campus}). As an alternative for short-distance trips and providing an efficient connection to mass transit, they could encourage more people to bike and make trips in environmentally friendly ways. At the same time, autonomous bicycles could serve as an alternative to the current BSS, providing a convenience that has not been achieved with other systems. 

Using this new system, whenever a user requested a ride, a bicycle would drive autonomously to pick up the user. Then, the user would ride it just as a regular bicycle and, upon arrival to the destination, the bicycle would go back to autonomous mode to drive for the next user or find a charging station.

Compared to other autonomous vehicles, bicycles have the additional challenge of lack of lateral stability. As bicycles have just two wheels, bicycles tend to tip over to the side without additional support.  We have tackled this issue by designing a novel system that allows the bicycle to transform into a tricycle for autonomous driving. 

\begin{figure*}[!htb]
    \centering
    \subfloat[Bicycle configuration]{\includegraphics[width=0.225\linewidth, valign=b]{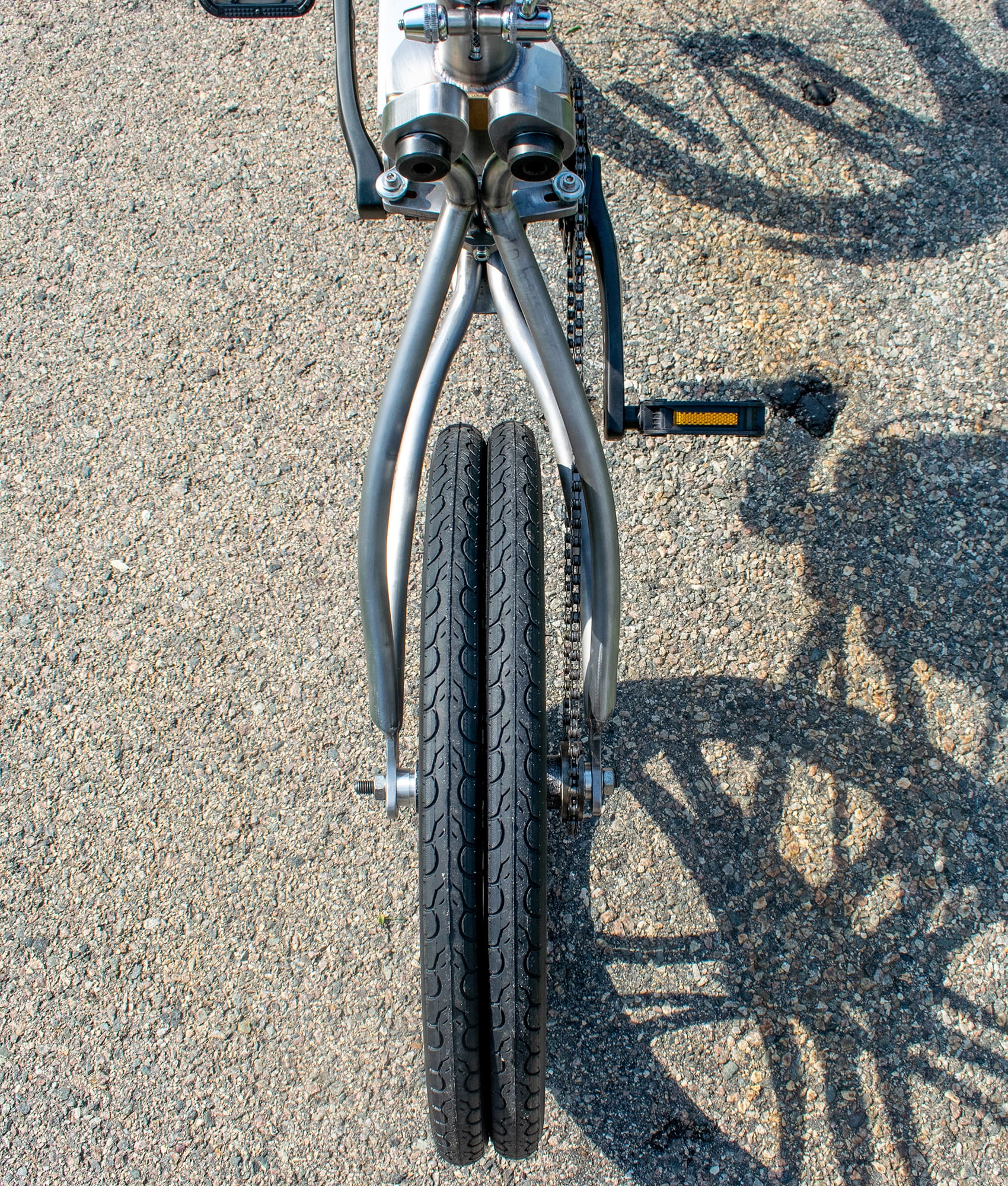} \label{fig:top-view-closed} }
    \subfloat[Tricycle configuration]{\includegraphics[width=0.225\linewidth,
    valign=b]{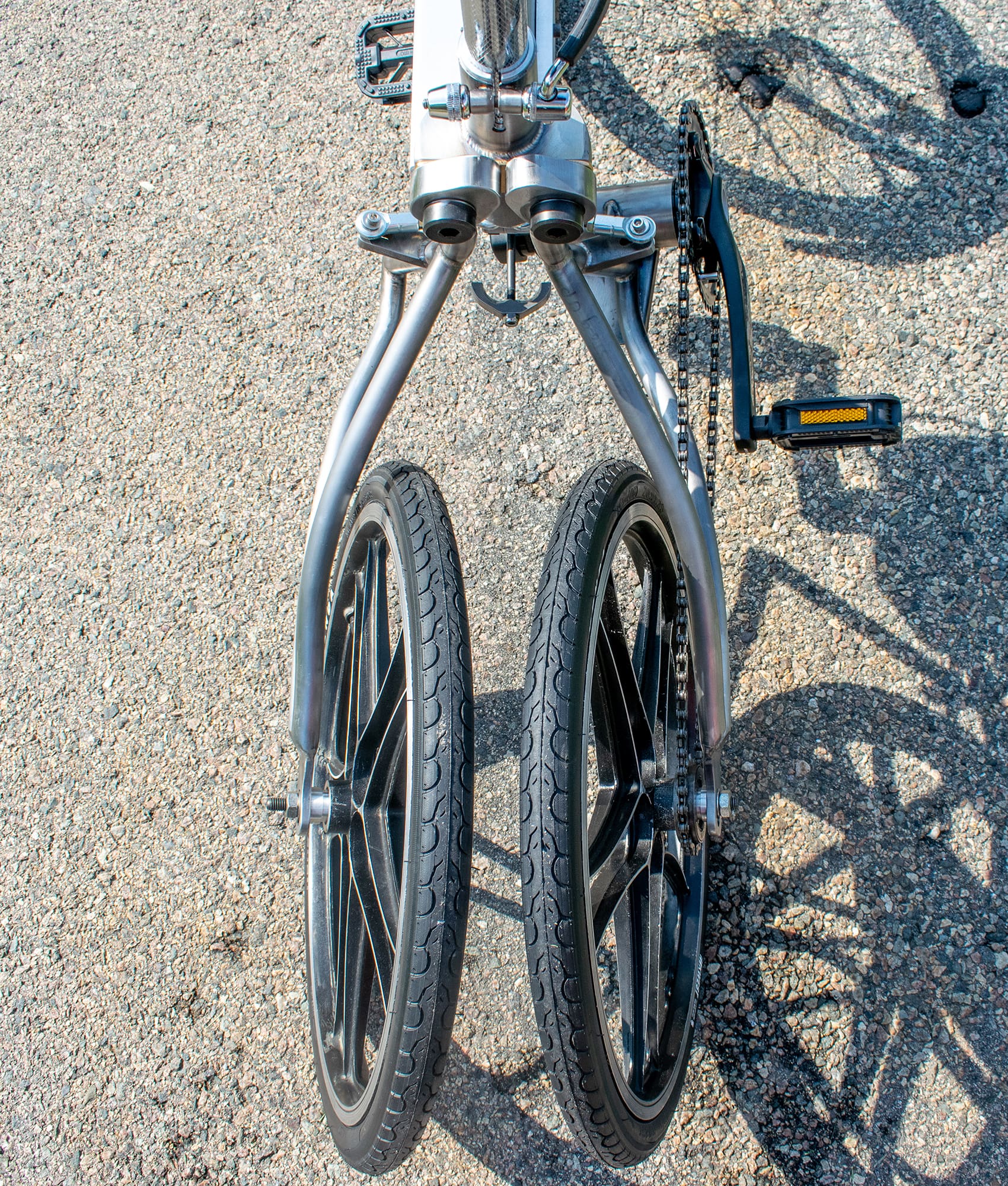}\label{fig:top-view-open} }
    \caption{Top view of the rear wheels: a) mechanism in bicycle configuration, with the two wheels acting as one b) mechanism in tricycle configuration, providing the necessary stability to drive autonomously.} 
    \label{fig:top_view}
\end{figure*}

When a user is riding it, the system is in bicycle mode. In this configuration, the two rear wheels act as a single wheel, and the riding experience remains unchanged from riding a regular bike (Figure \ref{fig:top-view-closed}).  In contrast, when the bicycle is driving autonomously, the bicycle is in tricycle mode. In this case, the wheels separate and provide the necessary lateral stability for autonomous drive (Figure \ref{fig:top-view-open}). We have designed, built, and tested a full-scale functional prototype of this mechanism, proving the viability and robustness of this novel design. 

\subsection{Potential benefits}\label{autonomous_benefits}

Here we specify some of the potential benefits of a BSS with autonomous bicycles over current systems from a conceptual point of view; the quantification of the benefits can be found in Section \ref{results}. 

First and foremost, with autonomous bicycles, there would be no need for rebalancing because bicycles would relocate by themselves. For users, it would eliminate the problem of finding available bikes or docks, and users would be able to use their time for other purposes while they wait for the autonomous bicycle to arrive. Furthermore, bikes that are not in use could be directed towards where the demand is predicted to occur, reducing wait times and increasing levels of use of the system. Increasing system reliability and improving user satisfaction could incentivize more people to choose biking as the transportation mode for commuting and enjoy the city by traveling in an environmentally friendly way. 

Moreover, not needing rebalancing could lead to cost savings for the operators \cite{sanchez2020autonomous}. Autonomous rebalancing and demand prediction could make the use of bicycles more efficient by increasing the number of trips per bike and day and decreasing the needed fleet size required for a given demand. As bicycles could drive autonomously to charging stations, the management of charging would be easier than with dockless e-bikes. Furthermore, data generated from its sensors could support the decision-making process to design more efficient, sustainable, and equitable systems \cite{Rico20}.

Lastly, stations are currently located in active socioeconomic areas, which creates an access inequality based on race and income levels \cite{ursaki2015quantifying}. Autonomous rebalancing could improve accessibility by reducing the impact of the location of the stations \cite{fuller2013}.

\section{Methodology} \label{methodology}

In this article, an activity-based multi-agent simulation framework is proposed to quantify the aforementioned potential impact of autonomous bicycles on the fleet performance of BSS. This section summarizes the key features of the proposed framework and formalizes the activity processes of the three modes of transportation under study: station-based, dockless, and autonomous.

\subsection{Agent-based simulator}\label{agent_simulator}

Agent-based simulators can be implemented in two ways: a) continuous simulation and b) event-based simulation. Continuous simulations have a fixed time-step, and the system state is updated in every step. For these simulations, it is critical to select an appropriate period parameter, which indicates how much time elapses between state updates. Furthermore, these simulations can be highly inefficient, as there may not be any changes from one step to the next. Conversely, in discrete event-based simulations, the system is only updated when a new event occurs. The simulator processes new events in sequential order as they are fired or triggered by the simulated entities or agents. An event-based approach is followed for this work.

\subsubsection{Related work}\label{related_simulator}

There is a plethora of open-source software solutions available with regard to BSS agent-based simulators.
GAMA \cite{taillandier2019building} is a modeling and continuous simulation development environment for building spatially explicit agent-based simulations. Several works have been developed to model BSS using GAMA \cite{lu2019considering, kaziyeva2021simulating, veldhuis2018applications}.
MATSim \cite{horni2016multi} is an activity-based, extendable, multi-agent simulation framework implemented in Java that performs an integral microscopic simulation of resulting traffic flows and the congestion they cause. MATSim employs a microscopic description of demand by tracing the daily schedule and travelers’ decisions. 
MATSim does not natively support BSS, but other authors have extended it and incorporated BSS plugins. Hebenstreit et al. \cite{hebenstreit2018dynamic} introduced two new MATSim modules that allow to simulate bicycle traffic and bike-sharing in combination with other modes. This work provided a high level of detail for gaining very precise information of BSS, i.e., usage, load factor, the catchment area of bike-sharing, but at the cost of increased computation effort. 
In the same fashion, Hörl \cite{horl2017agent} presented an agent-based simulation approach to capture the dynamic interplay between a supply of autonomous vehicle fleets and a population of artificial people based on MATSim. Their work shows how agents react to the new travel options. 
Similarly, Ruch et al. \cite{ruch2018amodeus} developed an open-source MATSim software package for the accurate and quantitative analysis of autonomous mobility-on-demand systems. They used an agent-based transportation simulation framework to simulate arbitrarily configured autonomous systems with static or dynamic demand.

Fernandez et al. \cite{fernandez2020bike3s} built a microscopic agent-based simulator for a station-based BSS that allows evaluating and testing different management decisions and strategies (station capacities, station distributions, and balancing strategies). This work was based on a simulator designed specifically to evaluate incentive-based rebalancing strategies \cite{fernandez2018bike}.
Lin et al. \cite{jin2019simulation} presented a simulation framework developed in Java to optimize BSS rebalancing and maintenance activities while satisfying customers’ needs over a service area. 
Soriguera et al. \cite{soriguera2018simulation} developed an open-source agent-based simulator for BSS, written in MATLAB.
Casado et al. \cite{casado2016simulation} created an agent-based simulation model developed also in MATLAB to emulate a BSS that allows the optimization of the main strategic and tactical system variables. 
Romero et al. \cite{romero2015simulation} presented a methodology for modeling an urban transportation system, integrating public bicycles into a multi-modal network. Developed with Python, it proposes a bike cost function that takes into account the effect of slopes on cycling speeds and the effect of traffic levels on the attractiveness of cycling routes. 
Saltzman et al. \cite{saltzman2016simulating} created an animated discrete-event simulation model of a BSS developed with Arena 14 software package that can be used to investigate critical issues faced by BSS operators. 
Also built in ARENA, Lin et al. \cite{lin2017simulation} presented a simulation model for solving bicycle repositioning problems in BSS. Their system decides the optimal number of bicycle for repositioning that minimizes customers' waiting time.


In summary, existing open-source BSS simulators have been designed with specific goals in mind. Some are more general than others, but none of them adequately address the questions presented in this research. For instance, traffic computations are unnecessary for our research questions. When comparing the performance of various modes of transportation, the impact of traffic on the results can be neglected. 

Overall, these simulation frameworks lack essential characteristics required for this research, in which we seek a framework that is able to:
a) create a high-resolution spatial virtual environment;
b) scale linearly to the number of bikes and the number of trips;
c) generate demand (static or dynamic) with any spatial distribution across the urban area;
d) be transferred and adapted to different cities and geospatial data;
e) compute fast routing solutions;
f) configure user behavior and charging/rebalancing strategies;
and finally, very importantly, g) to include different bike-sharing transportation modes.



\subsubsection{Architecture}\label{architecture}

\begin{figure}[!htb]
    \centering
    \includegraphics[width=0.65\linewidth]{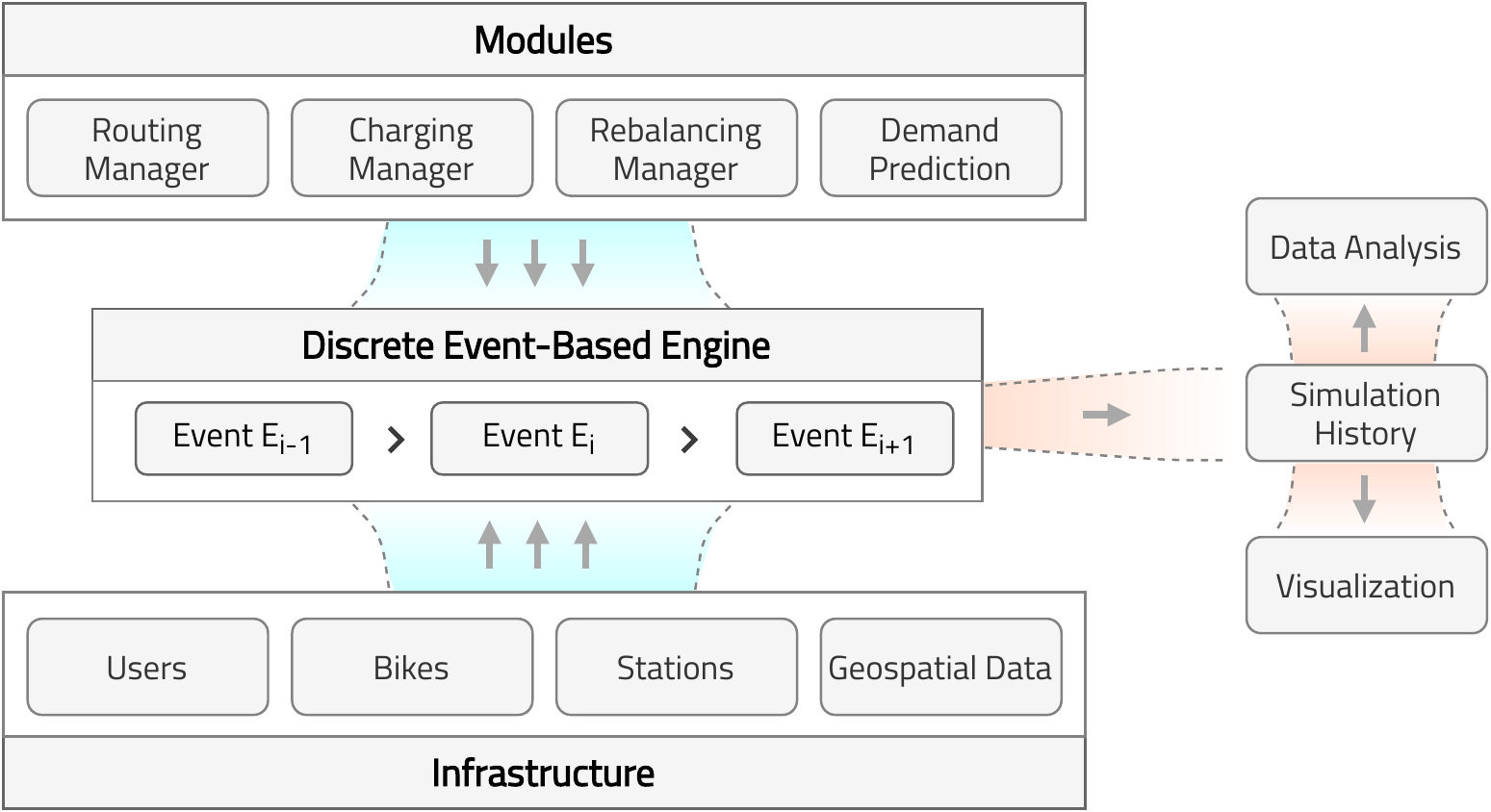}
    \caption{Simulator architecture}
    \label{fig:simulator-architecture}
\end{figure}

In this section, the architecture of the proposed agent-based simulator is described. 
The core of the simulator is composed of a discrete event-based engine that manages every activity carried out by the agents during their life-cycle using a priority event queue, ordered by time. The main building blocks of the simulator are depicted in Figure \ref{fig:simulator-architecture}. The event-based engine is at the core of the simulator and is fed by events produced by the agents. Any entity that can interact with or produce events is considered to be an agent. 
During the simulation execution, events are handled sequentially, in chronological order. Whenever any agent does an action or takes a decision, it generates and inserts new events into the priority queue.

The infrastructure block encloses the set of fundamental physical facilities and systems that support the functionality of the bike-sharing system. More specifically, it contains the required data to generate users, bikes, stations, and the road network. The block containing the modules is composed of four sub-blocks: routing manager, charging manager, rebalancing manager, and demand prediction module. Finally, as actions and activities occur, each event is registered on the simulation history to be further exploited for visualization and data analysis purposes. Let us now go over each of the simulator's components in detail. 

\paragraph{Engine}\label{sim_engine}
The simulator was implemented from scratch using Python 3.7.9 \cite{python}. An object-oriented paradigm was adopted, where each agent is a class instance.  The engine was developed on top of the SimPy 4.0.1  \cite{matloff2008introduction} library, a process-based discrete-event simulation framework. Under this paradigm, \textit{processes} are used to model the behavior of active components, such as bicycles and users. Processes live in an \textit{environment} and interact with the environment and with each other via \textit{events}. The most important event type for our application is the \textit{timeout}, which allows a process to sleep for the given time, determining the duration of the activity. Events of this type are triggered after a certain amount of simulated time has passed. 

\paragraph{Infrastructure}\label{sim_infrastructure}
The workflow of the simulator is described as follows: initially, Simpy's environment is created, and the provided infrastructure data is used to generate users, bikes, stations, and the road network. Users are introduced into the environment at a given location and departure time, and their task is to move to the specified target location using the defined mode of transportation. The location data is given in terms of longitude and latitude. The characteristics of the bike-sharing system are defined at the \textit{bikes} module. If required, station data is also included in its corresponding module. Parameters for users, bikes, and stations are specified in their respective modules. 

\paragraph{Geospatial data}\label{sim_geospatial}
The geospatial information is provided by the \textit{geospatial data} module, which contains a) the road graph, b) buildings data, and c) geospatial indexing system. Among these, only the road graph is strictly necessary to perform the simulation. Buildings data is optional, as it is used to generate users' origin and destination locations inside the buildings. This process yields realistic locations and avoids geographical obstacles such as highways or rivers. The geospatial indexing system is also optional, as it is used for the demand prediction and rebalancing manager, explained more in detail in Sections \ref{demandprediction} and \ref{rebalancing}. 

Geospatial data was obtained using OpenStreetMap \cite{OpenStreetMap} services. Given the bounding box of the city under study, OpenStreetMap \textit{highway} tag is queried, downloaded, and converted into a directed and weighted graph, denoted as road network from this point onward. A highway in OpenStreetMap ``is any road, route, way, or thoroughfare on land which connects one location to another and has been paved or otherwise improved to allow travel by some conveyance, including motorized vehicles, cyclists, pedestrians, horse riders, and other'' \cite{OpenStreetMap}. 

\paragraph{Routing Manager}\label{sim_routing}
The routing manager is in charge of choosing the most appropriate route (usually the shortest path) to transport people and vehicles around the urban space. This is a critical service and needs to be computed fast and with high resolution to yield results as close to reality. For the task of routing in road networks, an optimized fork of Pandana \cite{foti2012generalized} Python library was implemented, as it uses contraction hierarchies (CH) to calculate super-fast travel accessibility metrics and shortest paths. The numerical code is in C++. 

The contraction hierarchies algorithm is a speed-up technique for finding the shortest path in a graph, and it consists of two phases: preprocessing and query. To achieve its speed-up, CH relies on the fact that road networks do not change frequently. Given a directed, weighted graph $G(V,E,C)$ with vertex set $V$, edge set $E$ and cost function $C: E \rightarrow \mathbb{R}^+$, the goal is to preprocess $G$ in such a way that the subsequent shortest path queries specified by a source node $s$ and a target node $t$ can be answered very quickly. In the preprocessing phase, $G$ is augmented by additional edges $E'$, which are shortcuts that represent the shortest paths in the original graph $G$. In addition, a natural number is assigned to each node $v \in V$, called $level(v)$ \cite{geisberger2008contraction}. In the query phase, a bidirectional Dijkstra algorithm is applied on the augmented graph $G^\star$. Amongst all nodes settled from the Dijkstra, the one where the added distances from $s$ and to $t$ are minimal determines the shortest path from $s$ to $t$. The query is highly efficient because the modified Dijkstra can discard the majority of the nodes and edges of $G^\star$ while visiting only a small portion of the graph $G^\star$ \cite{geisberger2008contraction}.

\paragraph{Additional modules}\label{sim_modules}
The charging manager, the rebalancing manager, and the demand prediction modules will be discussed in greater depth under each mode of transportation. 

\paragraph{Simulator benefits}\label{sim_benefits}
The proposed simulation framework is highly configurable and flexible, with many parameters at different levels: geospatial data, user behavior, bike features, charging, and rebalancing strategies, among others. At the moment, it includes three modes of transportation (station-based, dockless, and autonomous BSS), but other mobility modes can be integrated with little effort. It can also combine multiple modes of transportation simultaneously, allowing researchers to test the effects of various mode choice models. Furthermore, it can be easily transferred to other cities by just changing the OpenStreepMap query. In this sense, the simulator works with geospatial data of high resolution and precision. The simulation framework is able to perform and scale linearly with respect to the number of trips and to the number of bikes. On the contrary to the use of MATSim, the simulation time is not restricted to one day.

\paragraph{Performance}\label{sim_performance}
In terms of performance, the simulator takes around 80 seconds to compute a station-based BSS run with 70000 bike trips (that elapse seven days of real data). In the case of the dockless system, this time increases up to 205 seconds due to the larger fleet size and the creation of KD-trees for closest bike queries every time a trip is requested. Regarding the autonomous system, it takes around 60 seconds to compute a run, given its reduced fleet size. These experiments were run with Python 3.7.4 in a Ubuntu 18.04 Linux system with an Intel i5-9400F processor with 6 threads running at 2.90 GHz using 64 GB of RAM. Even though these time-benchmarks are very satisfactory to run multiple simulations, perform sensitivity analysis and study the parameters' influence, we believe it could be further improved by integrating the event-based engine with a low-level programming interface. 




\subsection{Definition of bike-sharing systems}

In this section, the three bike-sharing systems under study (station-based, dockless, and autonomous) are defined and formalized.

\subsubsection{Station-based system}

A station-based system consists of bicycles that can be borrowed from a docking station and must be returned to another station belonging to the same system. The docking stations are special bike racks that lock the bike and only release it by computer control. Figure \ref{fig:process_user_station} depicts the steps a user takes when using station-based BSS mode of transportation. 

Users are initialized at a given location and departure time. Typically, a user checks the availability of bicycles via a smartphone application that displays the number of bikes and docks available in each station in real-time. In case there are no stations with available bikes, a bike request is made with probability $\beta$. This rebalancing action relocates a bike that is available at the nearest station, provided that the number of bikes does not fall below a predetermined minimum. If there are no stations within a walkable distance, the user might opt to use another transportation mode. In that case, the process is finalized, and the user does not complete the trip. Similarly, if there are stations but no available bicycles, the user might also decide to change the mode of transportation. 

If there are bikes within a walkable distance, the user will select the closest one to their location and will walk to that station. Sometimes, the station gets empty while the user is walking and finds no bicycles at arrival. In such cases, the user will check again for bikes in nearby stations and will repeat the process. 

Once the user has a bike, they will choose a station with available docks as close as possible to the destination. In this case, too, it might also happen that the station is full for when the user arrives, and the user will have to look for another station with available docks and bikes. Once the user has found a dock for the bike, they will lock the bicycle and walk to the final destination.

Bikes are initialized at a specified station, given that the dock capacity of that station is not exceeded. Bikes are then made available to any user. If requested, a user can unlock one bike from the station dock, becoming an unavailable bike. After being transported to another station, the bike is locked by the user and becomes available once again. This process was illustrated on Figure \ref{fig:process_bike_station}.

The configuration parameters for station-based systems are: 
\begin{itemize}
\itemsep 0em
\item \textit{Fleet size}: number of bikes in the system
\item \textit{Maximum walking radius}: the maximum distance a user is expected to walk to or from a station
\item \textit{Average riding speed}: average speed at which users ride the bikes including stopping times
\item \textit{Average walking speed}: average speed at which users walk
\item \textit{Rebalancing parameter}: probability of requesting a rebalanced bike from/to a nearby station
\item \textit{Minimum number of bikes/docks per station}: minimum bikes or docks to hold during rebalancing trips
\end{itemize}

\def\figformat{1}

\if\figformat1
\begin{figure*}[!htb]
    \centering
    \subfloat[Station-based user]{\includegraphics[width=0.4817\linewidth, valign=b]{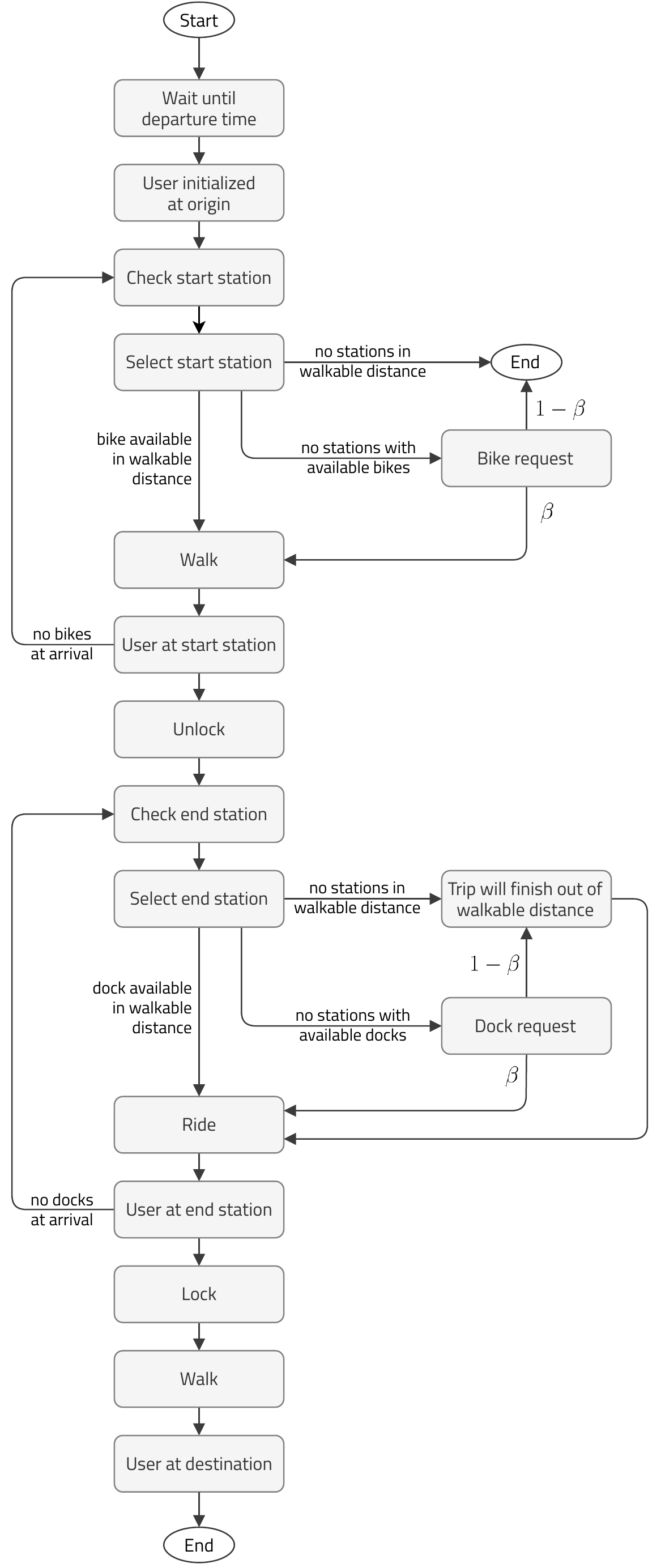} \label{fig:process_user_station} }
    \subfloat[Dockless user]{\includegraphics[width=0.2955\linewidth,
    valign=b]{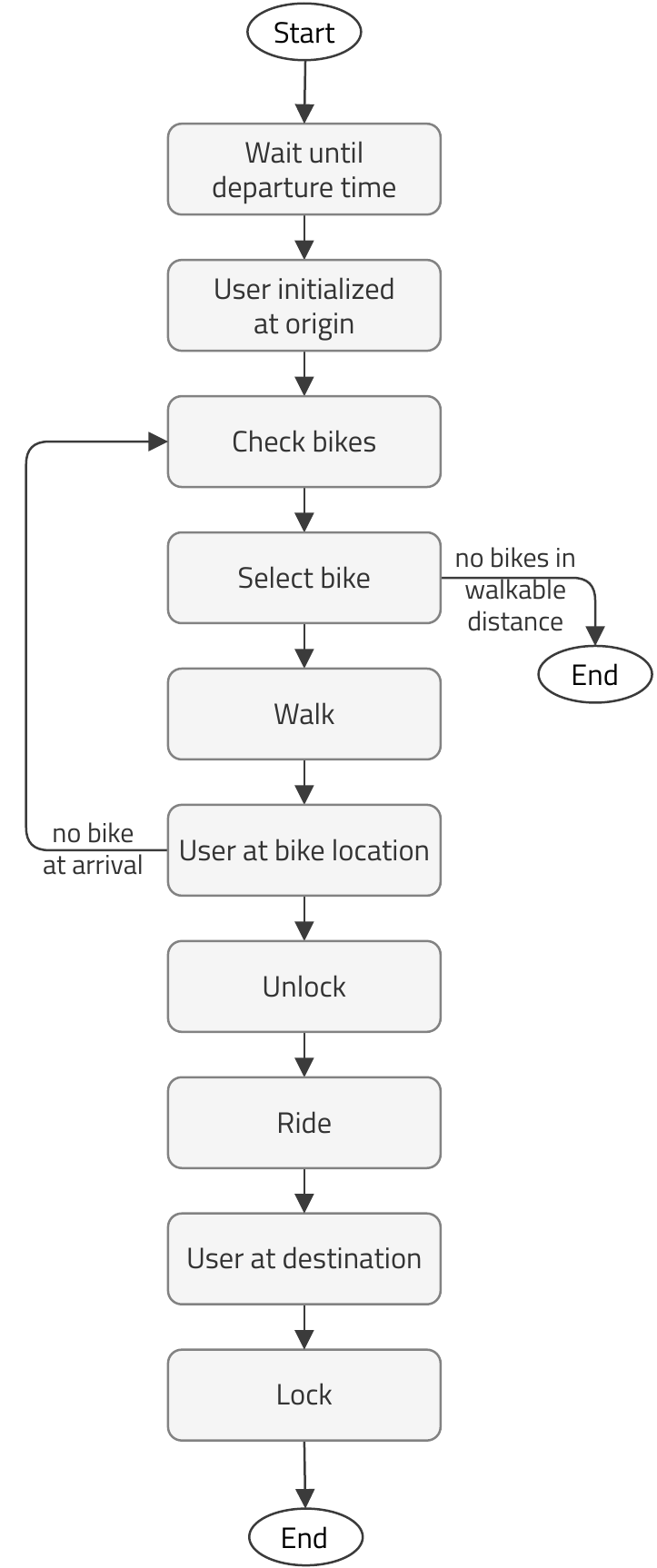} \label{fig:process_user_dockless} }
    \subfloat[Autonomous user]{\includegraphics[width=0.2226\linewidth,
    valign=b]{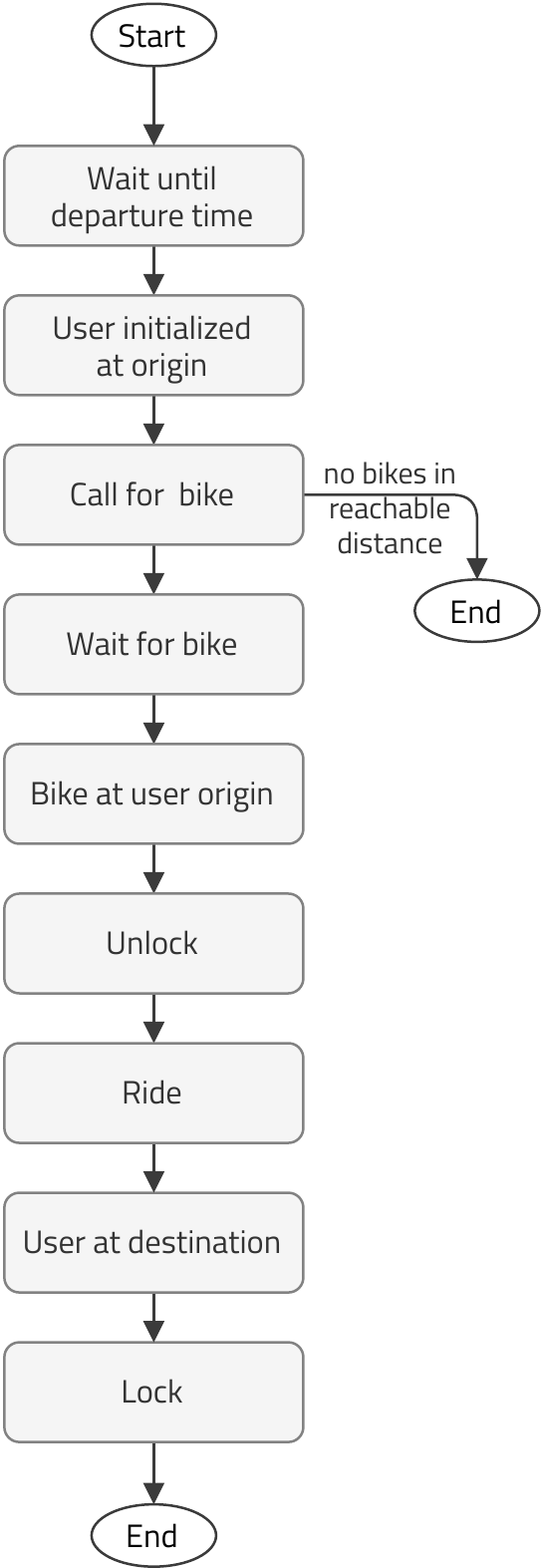} \label{fig:process_user_autonomous} }
    \caption{User life-cycle state diagrams for a) station-based, b) dockless and c) autonomous bike systems.} 
    \label{fig:user_processes}
\end{figure*}

\begin{figure*}[!htb]
    \centering
    \subfloat[Station-based bike]{\includegraphics[width=0.1878\linewidth, valign=b]{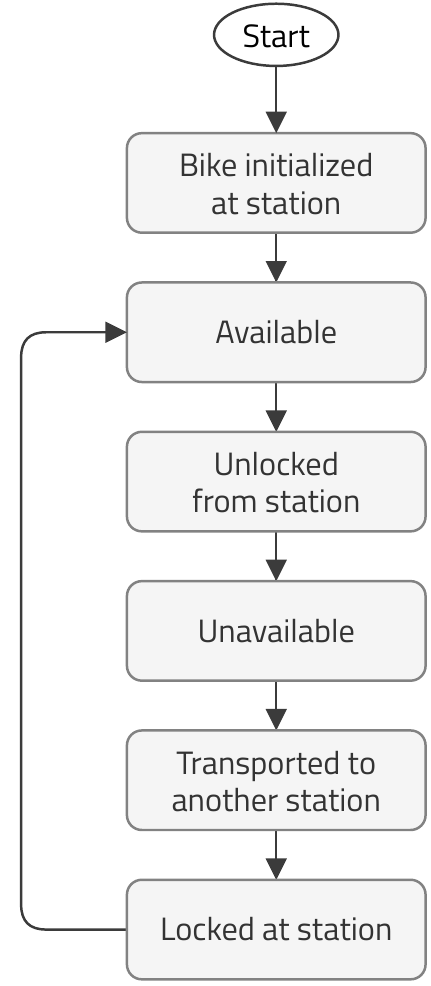} \label{fig:process_bike_station} }
    \hspace{4em}
    \subfloat[Dockless bike]{\includegraphics[width=0.1836\linewidth,
    valign=b]{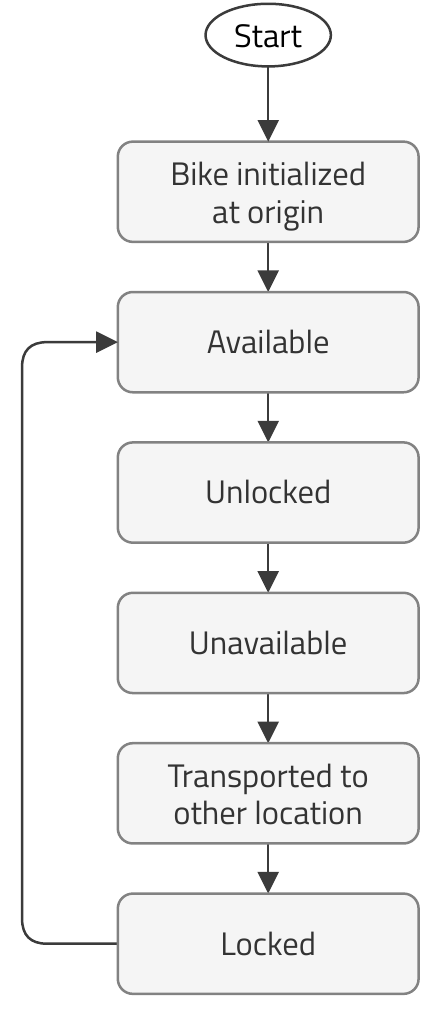} \label{fig:process_bike_dockless} }
    \hfill
    \subfloat[Autonomous bike. When SOC (State Of Charge) reaches zero, bike runs out of battery power.]{\includegraphics[width=0.8286\linewidth,
    valign=b]{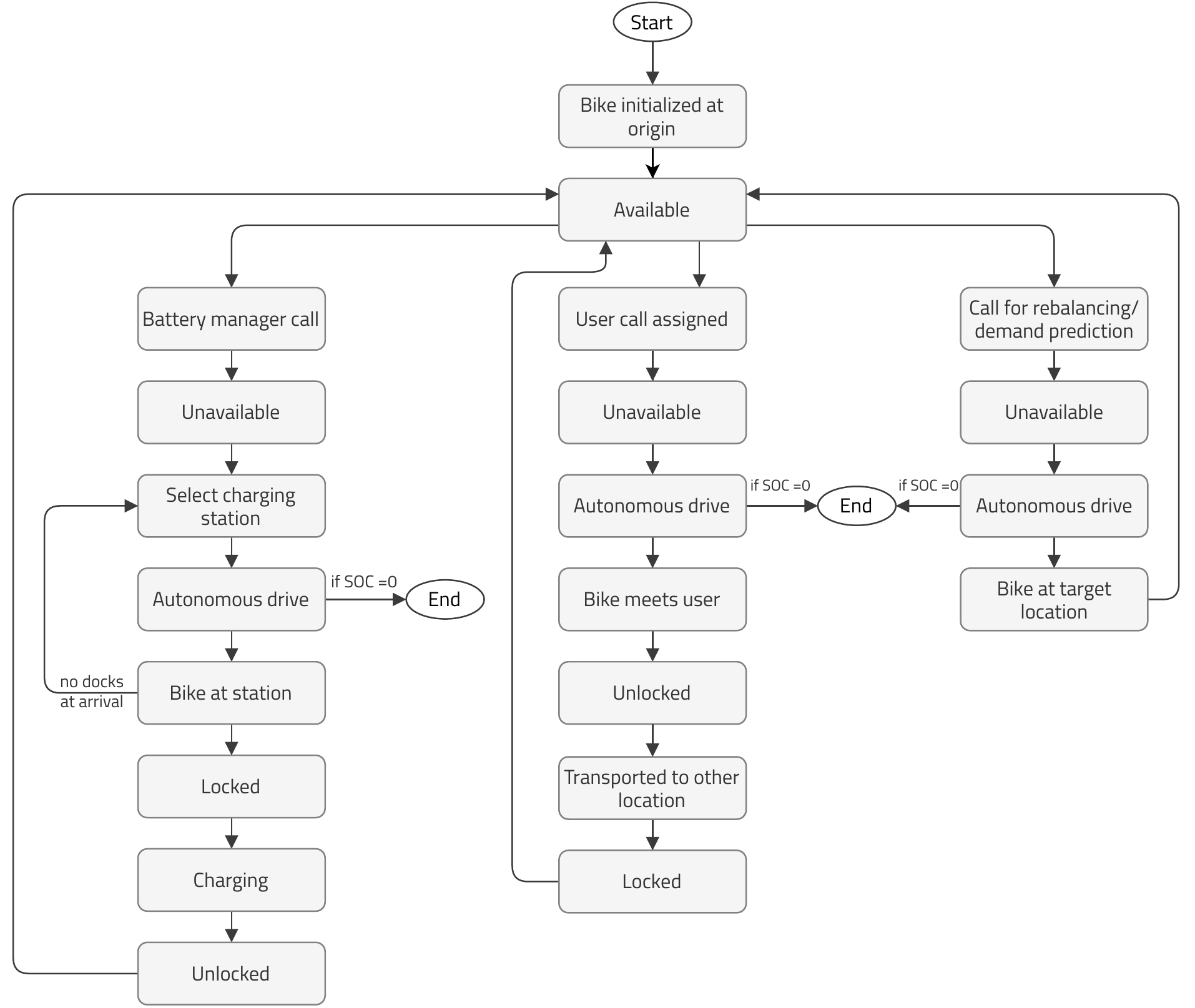} \label{fig:process_bike_autonomous} }
    \caption{Bike life-cycle state diagrams for a) station-based, b) dockless and c) autonomous bike systems.} 
    \label{fig:bike_processes}
\end{figure*}

\fi

\if\figformat0

\begin{figure}[!htb]
    \centering
    \includegraphics[width=0.60\linewidth]{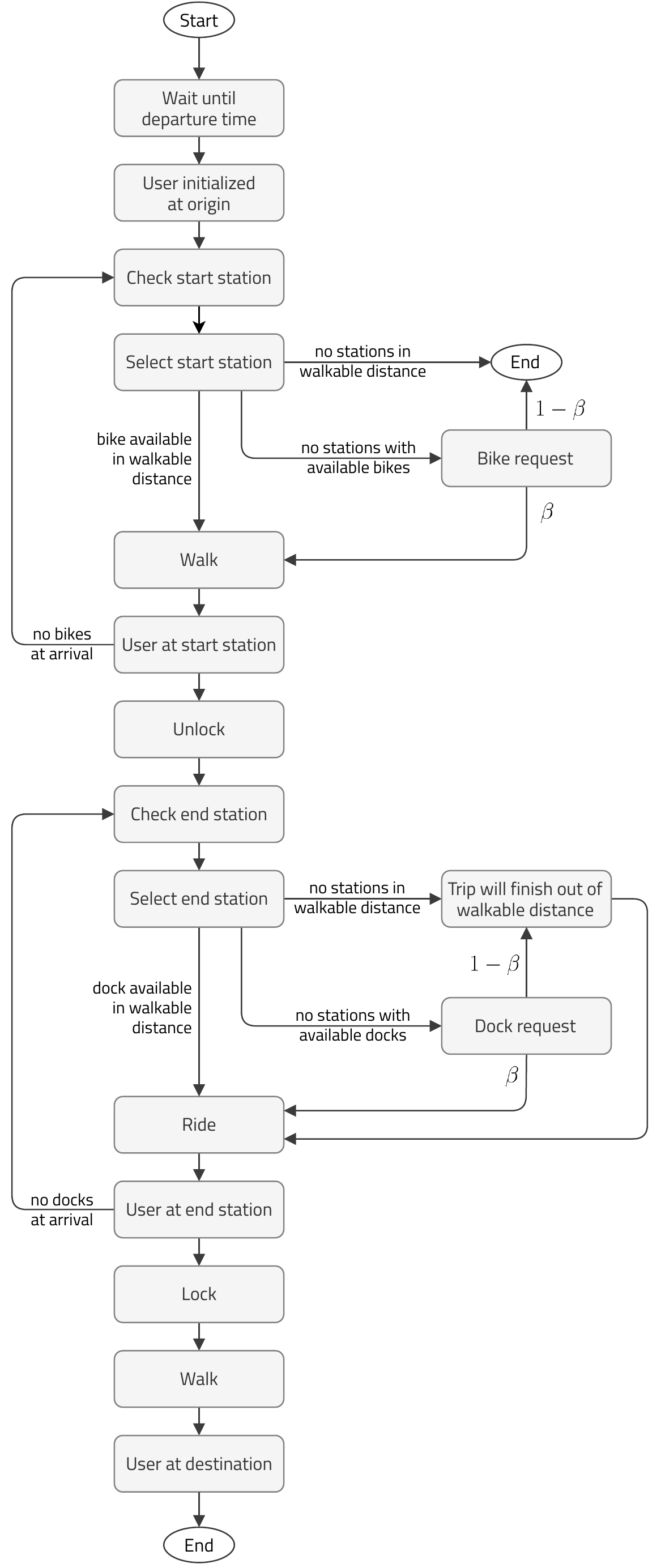} 
    \caption{User life-cycle state diagram in station-based bike system}
    \label{fig:process_user_station}
\end{figure}
\begin{figure}[!htb]
    \centering
    \includegraphics[width=0.22\linewidth]{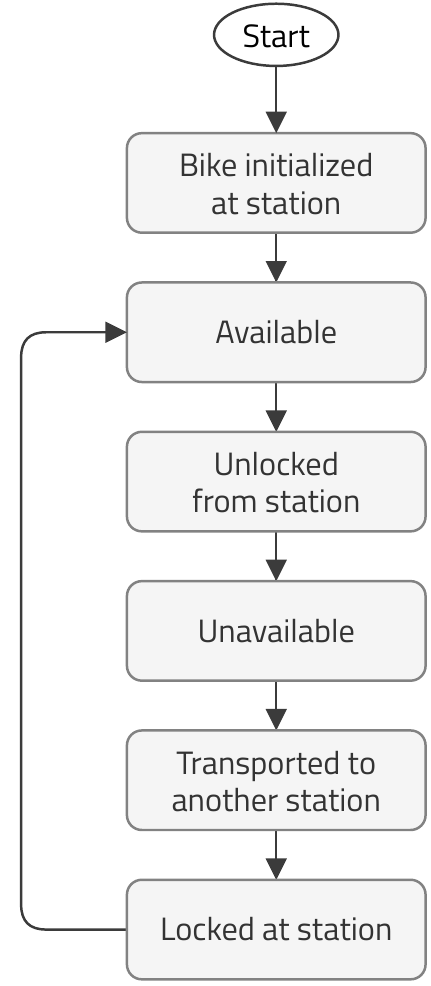} 
    \caption{Station-based bike life-cycle state diagram}
    \label{fig:process_bike_station}
\end{figure}

\begin{figure}[!htb]
    \centering
    \includegraphics[width=0.48\linewidth]{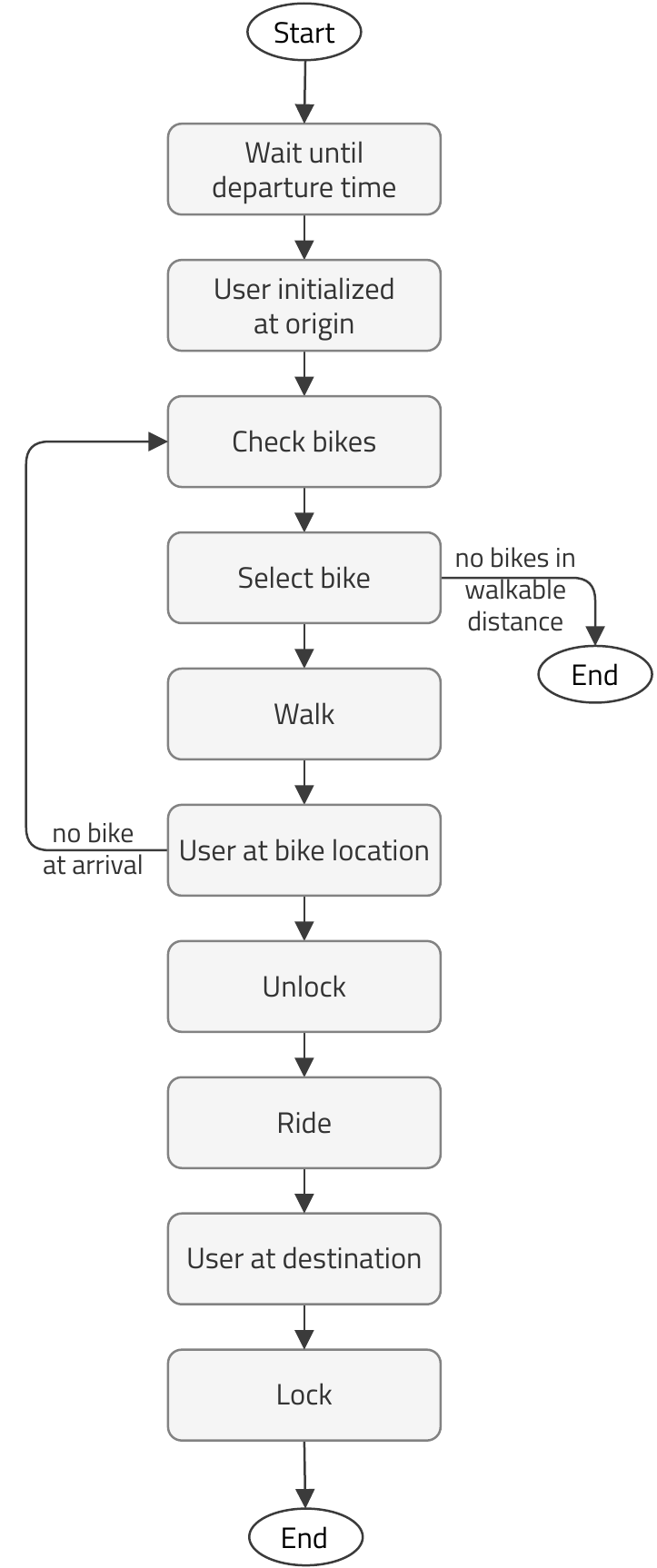} 
    \caption{User life-cycle state diagram in dockless bike system}
    \label{fig:process_user_dockless}
\end{figure}
\begin{figure}[!htb]
    \centering
    \includegraphics[width=0.22\linewidth]{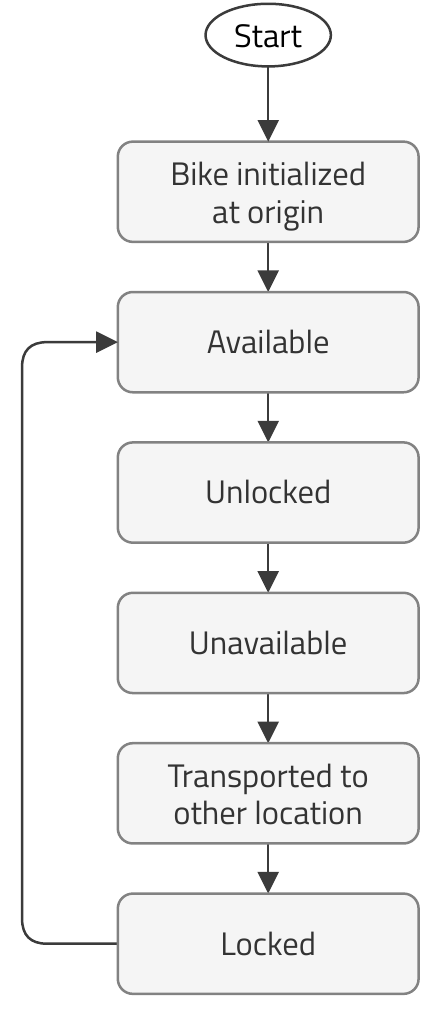} 
    \caption{Dockless bike life-cycle state diagram}
    \label{fig:process_bike_dockless}
\end{figure}

\begin{figure*}
\centering
\begin{minipage}{.28\textwidth}
    \centering
    \includegraphics[width=\linewidth]{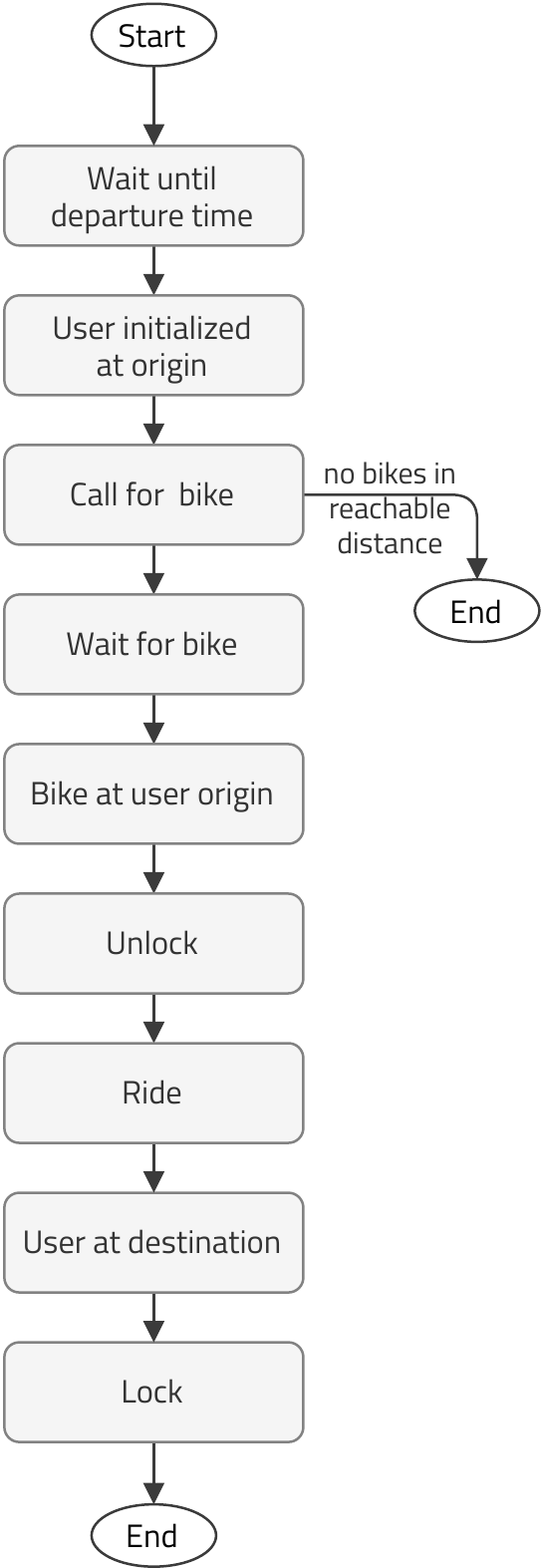}
    \caption{User life-cycle state diagram in autonomous bike system}
    \label{fig:process_user_autonomous}
\end{minipage}%
\begin{minipage}{.72\textwidth}
    \centering
    \includegraphics[width=\linewidth]{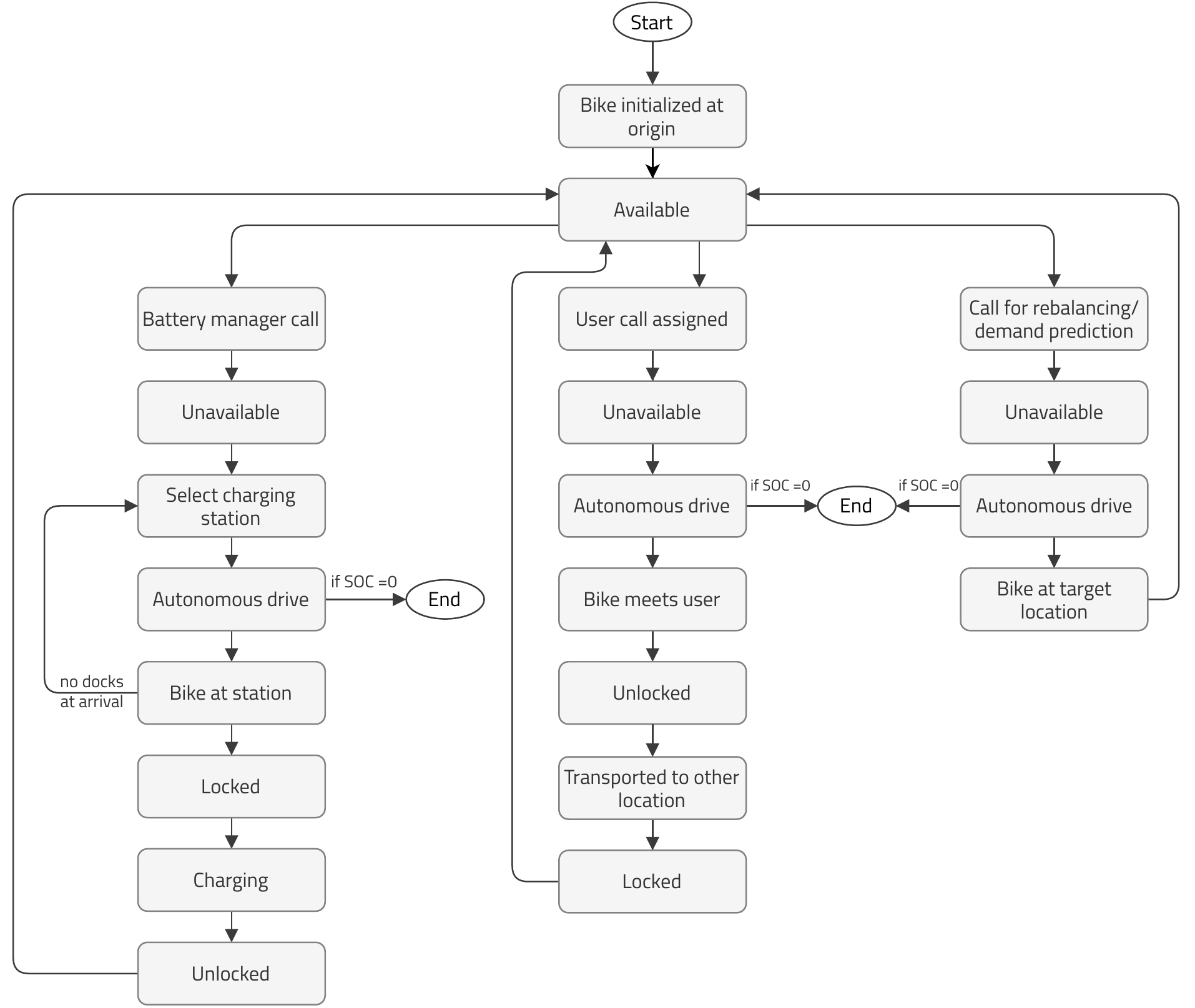}
    \caption{Autonomous bike life-cycle state diagram }
    \label{fig:process_bike_autonomous}
\end{minipage}
\end{figure*}

\fi

\subsubsection{Dockless system}

In dockless system, the users check for available bicycles in an app that shows the location of available bicycles. If there are no bicycles within a walkable distance, the users might choose to use another transportation mode. If there are bicycles within a walkable radius, the user will walk to the closest one. If the bike is no longer available at arrival because someone else took it already, the user will look for another bike and repeat the process. Finally, the user will ride the bike and drop it off right at the destination. 
Figure \ref{fig:process_user_dockless} illustrates the steps a user takes when using station-based BSS mode of transportation. 
The bike process is identical to the station-based bike process, as portrayed in Figure \ref{fig:process_bike_dockless}.

In the case of the dockless system, no rebalancing was considered. Most of the rebalancing in these systems occurs from the outskirts to the city center and, in the case of this simulation, the trips are restricted to be on the area around where the bike-sharing system operates. 
The configuration parameters for dockless systems are: 
\begin{itemize}
\itemsep 0em
\item \textit{Fleet size}: number of bikes in the system
\item \textit{Maximum walking radius}: the maximum distance a user is expected to walk to unlock a bike
\item \textit{Average riding speed}: average speed at which users ride the bikes including stopping times
\item \textit{Average walking speed}: average speed at which users walk
\end{itemize}


\subsubsection{Autonomous system}\label{bss_autonomous}

The autonomous bike process has three main sub-routines: a) user call request, b) battery management, and c) rebalancing management, as represented in Figure \ref{fig:process_bike_autonomous}.

The user call sub-routine (Figure \ref{fig:process_user_autonomous}) is very similar to the one in the station-based system, except that the user does not walk to the station, but rather the bike drives autonomously to the user location.
With autonomous bicycles, the user requests a bicycle, and the system assigns the nearest available bicycle to this user.
The user has to wait while the bicycle drives autonomously to pick them up.
If the wait time is too long, the system does not assign a bicycle, and the user has to choose another mode of transportation.
When the bike arrives at the user's location, they will ride it as if it was a regular bicycle before dropping it off at the destination location. 

Concerning battery management, the bike is sent to a charging station whenever the battery level falls below a predefined minimum value.
These charging stations are similar to station-based docking stations, with the main difference being that they are used to charge autonomous bikes. In the current simulation, charging stations are located in the exact same locations as docking stations. While it is beyond the scope of this paper, the number and location of the charging stations could be optimized based on bicycle usage. The battery level is checked right after a user locks the bike or in case the rebalancing system is activated after a bike has been rebalanced.

The third sub-routine corresponds to bike rebalancing management. In this regard, for this simulation, two extreme scenarios are implemented. One in which there is no rebalancing, and the bikes stay wherever the last user left them until they receive a call. The other scenario represents an ideal rebalancing in which there would be a perfect demand prediction model and routing algorithm. In this case, whenever a user calls for a bike, a bike immediately appears at the user's location. While this transition is supposed to be immediate, the distance traveled and the battery consumption are considered in the simulation. The real behavior is expected to be in-between scenarios; therefore, the scenario with no rebalancing and the scenario with ideal rebalancing provide, respectively, a lower- and upper- bound to the performance metrics. 

It should be noted that the autonomous drive sub-routine can be commanded by the rebalancing manager, the charging manager, or a user call. The battery level is discharged in proportion to the traveled distance.

The configuration parameters for autonomous systems are: 
\begin{itemize}
\itemsep 0em
\item \textit{Fleet size}: number of bikes in the system
\item \textit{Maximum autonomous radius}: maximum distance an autonomous bike is expected to move to pick-up a user
\item \textit{Average autonomous speed}: average speed at which autonomous bicycles drive including stopping times due to traffic interactions
\item \textit{Average riding speed}: average speed at which users ride the bikes including stopping times
\item \textit{Battery parameters}: kilometers of autonomy, total recharge time and minimum battery level to operate
\item \textit{Rebalancing parameters}: input window $W$, prediction ahead $P$ and prediction period $T$, explained in Section \ref{rebalancing}
\end{itemize}

        
        
        

\subsection{Demand Prediction Module}\label{demandprediction}


In a fleet of autonomous bicycles, reaching a perfect system efficiency and service quality would require instantly arranging a bike at the location demanded by each user. This level of performance is not possible to achieve with limited cost and fleet sizes. However, bikes can move around the city by taking advantage of their autonomous technology, locating themselves closer to high-demand areas, and, consequently, balancing the system. In that way, users would receive better average service and faster access to autonomous bikes. 
There are two requirements for providing an autonomous BSS fleet with such rebalancing capabilities: a demand prediction module and a routing optimization algorithm. In this section, the approach followed for the task of demand prediction is described.

The demand prediction module informs the rebalancing manager where users' demand will occur in the near future. Urban mobility patterns and transportation flows are predictable up to certain precision, as stated by Gonzalez et al. \cite{gonzalez2008understanding}, ``human trajectories show a high degree of temporal and spatial regularity, each individual being characterized by a time-independent characteristic travel distance and a significant probability to return to a few highly frequented locations.''.

The demand function $f$ is a mapping from the continuous 2D plane space to a scalar in the natural numbers: $f: \mathbb{R}^2 \rightarrow \mathbb{N}$. To facilitate the demand prediction, first the urban 2D space is discretized into a finite number of cells. Cells can be of any shape and size. For this implementation, Uber's hexagonal hierarchical spatial index (H3) \cite{brodsky_2019} with resolution level 8 was selected, yielding 180 hexagonal cells.

A demand prediction model for each cell separately fails to utilize hidden correlations between cells to enhance prediction performance. Therefore, for the task of demand prediction, a Graph Convolutional Neural Networks with Data-driven Graph Filter (GCNN-DDGF) \cite{lin2018predicting} was applied. The main limitation of a GCNN \cite{henaff2015deep} is that its performance relies on a pre-defined graph structure. The GCNN-DDGF model, on the contrary, can learn hidden heterogeneous pairwise correlations between grid cells to predict cell-level hourly demand in a large-scale bike-sharing network. The GCNN-DDGF model is enhanced with a Negative Binomial probabilistic neuron at the last layer of the neural topology.
    
The simulation experiments were applied to the city of Boston, as explained in Section \ref{setup}.
The bike-sharing demand dataset includes over 4.2 million bike-sharing transactions between 01/01/2018 and 31/12/2019, which are downloaded from Bluebikes Metro Boston's public bike share program \cite{bluebikes}. The dataset was split at date $31/09/2019$ into the train ($01/01/2018-31/09/2019$) and test (01/10/2019-21/12/2019) sets and was processed as follows: for each cell, 70080 (2 years x 365 days x 24 hours x 4) 15-minute bike demands were aggregated based on the bike check-out time and start station in transaction records. After preprocessing, as 180 hexagonal cells were considered in this study, a 180 by 70080 matrix was obtained. The input to the model is a window of $W$ data points (each point representing 15 minutes), and the output is computed $P$ data points ahead of time. Finally, the prediction model is invoked every $T$ data points. These three parameters, input window $W$, prediction ahead $P$, and prediction period $T$, were implemented on the simulation so that the rebalancing manager can customize this service. The model performance was evaluated using the Root Mean Square Error (RMSE) as the main criteria. The testing RMSE for the trained GCNN-DDGF model on Boston data was $2.69$. This performance metric is very close to the results obtained by the original implementation by Lin et al. \cite{lin2018predicting}, with a testing RMSE of $2.12$. The model training task was conducted using Tensorflow, an open-source deep learning neural network library, with Python 3.7.4 in a Ubuntu 18.04 Linux system with 64 GB RAM and GTX 1080 graphics card. 

\subsection{Bike Rebalancing Transportation Problem}\label{rebalancing}


The demand prediction module alone is not enough to generate an efficient rebalancing algorithm. A routing optimization algorithm is necessary to minimize the global transport costs between a set of supply points and a set of demand points. 
$T_{i,j}\geqslant0$ represents the number of bikes transported from supply point $i$ (with $B_{i}\geqslant0$ bikes available) to demand point $j$ (that requires $D_{j}\geqslant0$ bikes). Taking into account the number of bikes in each cell $B_{i}$, a unbalanced Hitchcock–Koopmans transportation problem is solved to yield the optimal flow of bikes between each pair of cells.

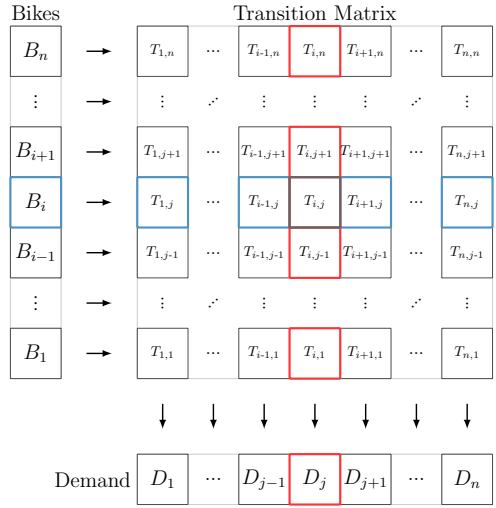
\begin{figure}[!htb]
    \centering
    \resizebox{0.4\linewidth}{!}{
\begin{tikzpicture}
    \pgfmathsetmacro{\n}{7}
    \pgfmathsetmacro{\a}{3}
    \pgfmathsetmacro{\b}{3}
    \def\constraints{0}
    \def\grid{1}
    \def\colorA{black!20}
    \def\colorB{black!70}
    
    \begin{scope}[shift={(-2.5,0)}]
        \if\grid1
            \draw[\colorA, ultra thin] (0,0) grid (1,\n);
        \fi
        \draw[\colorB, ultra thin] (0,0) grid (1,1);
        \draw[\colorB, ultra thin] (0,2) grid (1,5);
        \draw[\colorB, ultra thin] (0,6) grid (1,7);
        
        \node[above] at (0.5,\n) {\large Bikes};
        \draw[blue!80, very thick] (0,\b) grid (1, \b+1);

        \if\constraints1
            \draw[-latex, thick] (0, 3+0.5) -- (-1, 3+0.5);
            \node[left, draw=black!50] at (-1, 3 + 0.5) {\large $\sum_{j=1}^{n} t_{j,i} \leq B_{j}, \forall i \in {1,...,n}$};
        \fi

        \node[] at (0.5, 0 + 0.5) {\large $B_1$};
        \node[rotate=90] at (0.5, 1 + 0.5) {\large $...$};
        \node[] at (0.5, 2 + 0.5) {\large $B_{i-1}$};
        \node[] at (0.5, 3 + 0.5) {\large $B_{i}$};
        \node[] at (0.5, 4 + 0.5) {\large $B_{i+1}$};
        \node[rotate=90] at (0.5, 5 + 0.5) {\large $...$};
        \node[] at (0.5, \n - 0.5) {\large $B_n$};

        \foreach \i in {1,...,\n}
            \draw[-latex, thick] (1.5, \i-0.5) -- (2, \i-0.5);
    \end{scope}
    
    \begin{scope}[shift={(0,-2.5)}]
        \if\grid1
            \draw[\colorA, ultra thin] (0,0) grid (\n,1);
        \fi
        \draw[\colorB, ultra thin] (0,0) grid (1,1);
        \draw[\colorB, ultra thin] (2,0) grid (5,1);
        \draw[\colorB, ultra thin] (6,0) grid (7,1);

        \node[left] at (0,0.5) {\large Demand};
        \draw[red!80, very thick] (\a,0) grid (\a+1,1);
        
        \if\constraints1
        \draw[-latex, thick] (3+0.5,0) -- (3+0.5, -1);
        \node[below, draw=black!50] at (3 + 0.5, -1) {\large $\sum_{i=1}^{n} t_{i,j}  + s_{j} \geq D_{j}, \forall j \in {1,...,n}$};
        \fi

        \node[] at (0 + 0.5, 0.5) {\large $D_1$};
        \node[] at (1 + 0.5, 0.5) {\large $...$};
        \node[] at (2 + 0.5, 0.5) {\large $D_{j-1}$};
        \node[] at (3 + 0.5, 0.5) {\large $D_{j}$};
        \node[] at (4 + 0.5, 0.5) {\large $D_{j+1}$};
        \node[] at (5 + 0.5, 0.5) {\large $...$};
        \node[] at (\n - 0.5, 0.5) {\large $D_n$};

        \foreach \i in {1,...,\n}
            \draw[-latex, thick] (\i-0.5, 2) -- (\i-0.5, 1.5);
    \end{scope}
    
    \node[above] at (\n/2, \n) {\large Transition Matrix};

    \if\grid1
        \draw[\colorA, ultra thin] (0,0) rectangle (\n,\n);
    \fi
    \def\u{{0,2,6}}
    \def\v{{1,5,7}}
    \foreach \i in {0,1,2}{
        \draw[\colorB, ultra thin] (\u[\i],0) grid (\v[\i],1);
        \draw[\colorB, ultra thin] (\u[\i],2) grid (\v[\i],5);
        \draw[\colorB, ultra thin] (\u[\i],6) grid (\v[\i],7);
    }
    \foreach \i in {0,1,2}{
        \draw[blue!80, very thick] (\u[\i],\b) grid (\v[\i],\b+1);
        \draw[red!80, very thick] (\a,\u[\i]) grid (\a+1,\v[\i]);
    }
    \draw[black!60, very thick] (\a,\b) grid (\a+1, \b+1);


    \def\namesX{{1,...,"i\text{-}1","i","i+1",...,"n"}}
    \def\namesY{{1,...,"j\text{-}1","j","j+1",...,"n"}}
    \foreach \x in {0,...,6}
        \foreach \y in {0,...,6}{
            \pgfmathsetmacro{\p}{\namesX[\x]}
            \pgfmathsetmacro{\q}{\namesY[\y]}

            \ifthenelse{\equal{\p}{...} \AND \equal{\q}{...}}{
                \node[rotate=45] at (\x+0.5, \y+0.5) {$...$};
            }{
                \if\p...
                    \node[] at (\x+0.5, \y+0.5) {$...$};
                \else
                    \if\q...
                        \node[rotate=90] at (\x+0.5, \y+0.5) {$...$};
                    \else
                        \node[scale=0.8] at (\x+0.5, \y+0.5) {$T_{\p,\q}$};
                    \fi
                \fi
            }
        }

\end{tikzpicture}
}
    \caption{Bike rebalancing transportation problem: minimize the global transport costs between a set of supply points and a set of demand points. $T_{i,j}\geqslant0$ represents the number of bikes transported from supply point $i$ (with $B_{i}\geqslant0$ bikes available) to demand point $j$ (that requires $D_{j}\geqslant0$ bikes).
    }
    \label{fig:transportation}
\end{figure}

The urban area is subdivided into a grid of $n$ cells, label these $i=1,...,n$. The grid can be rectangular, square, or hexagonal. For this application, Uber's Hexagonal Hierarchical Spatial Index (H3) \cite{brodsky_2019} with resolution level 8 was selected as illustrated on Figure \ref{fig:boston_h3_cells}. The transportation cost between two cells is denoted as $C_{i,j}$ and it is proportional to the road distance from cell $i$ to cell $j$. The rebalancing transportation problem is to get bikes from supply cells to demand cells. The goal is to minimize the total cost:

\begin{equation}
\begin{aligned}
\min \quad & \sum_{i=1}^{n}\sum_{j=1}^{n} C_{i,j} \cdot T_{i,j} + \sum_{i=1}^{n} \lambda_{i} \cdot S_{i} \\
\textrm{s.t.} \quad &  \sum_{j=1} T_{i,j} \leq B_{i} \; \forall i \in {1,...,n} \\
& \sum_{i=1} T_{j,i} + S_{j} \geq D_{j} \; \forall j \in {1,...,n} \\
& T_{i,j} \geq 0 \\
\end{aligned}
\end{equation}

where
$n$ is number of grid cells.
$C_{i,j}$ is road distance from cell $i$ to cell $j$, computed using their position (longitude, latitude) and the road network graph.
$T_{i,j}$ represents the number of transported bikes from cell $i$ to cell $j$.
$B_{i}$ is number of autonomous bikes available (not busy and with enough battery) on cell $i$, and
$D_{i}$ is number of autonomous bikes demanded at cell $i$. 
The first constraint limits the number of bikes supplied by each cell $i$ and the second constraint tries to match the expected demand on each cell $j$. The $S_{i}$ represents the slack variable for cell $i$ and $\lambda_i$ is the cost associated with not reaching the expected demand $D_i$. Figure \ref{fig:transportation} illustrates the transition matrix $T$, and both the bikes $B$ and demand $D$ vectors.
The rebalancing transportation problem was solved using SciPy linear programming implementation along with the HiGHS solvers.

    


\section{Simulation setup} \label{setup}

The simulation was applied to the Boston (USA) metropolitan area, comprising the cities of Arlington, Boston, Brookline, Cambridge, Chelsea, Everett, Newton, Revere, Salem, Somerville, and Watertown. A satellite image of the area under study is included in Figure \ref{fig:boston_satellite}.
Geospatial data was obtained using OpenStreetMap \cite{OpenStreetMap} services: the road graph was queried from the \textit{highway} tag of OpenStreetMap (Figure \ref{fig:boston_road_network}) and the buildings were obtained using the Overpass API (Figure \ref{fig:boston_buildings}). 

\def\bigsatellite{1}

\if\bigsatellite1
\newcommand{\fh}{0.40\linewidth}
\begin{figure*}[!htb]
    \centering
    \subfloat[Satellite imagery]{\includegraphics[height=\fh]{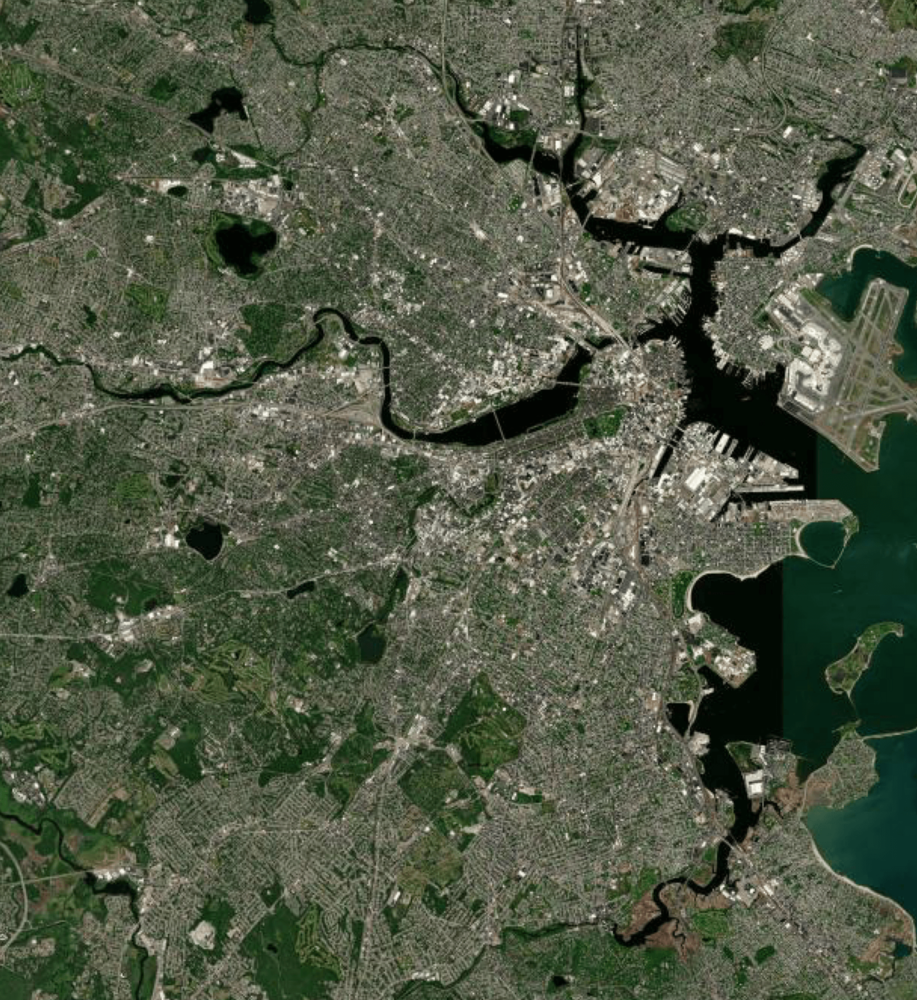} \label{fig:boston_satellite} }
    \subfloat[Buildings]{\includegraphics[height=\fh]{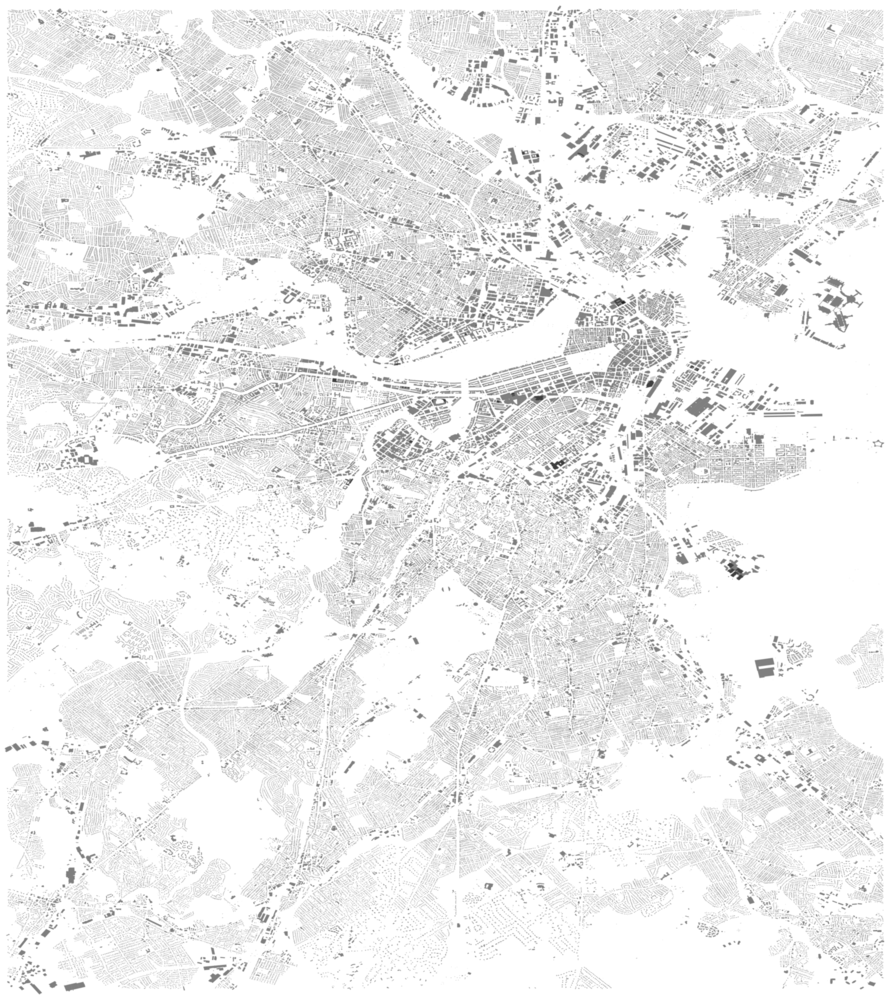} \label{fig:boston_buildings} } \\
    \subfloat[Road network]{\includegraphics[height=\fh]{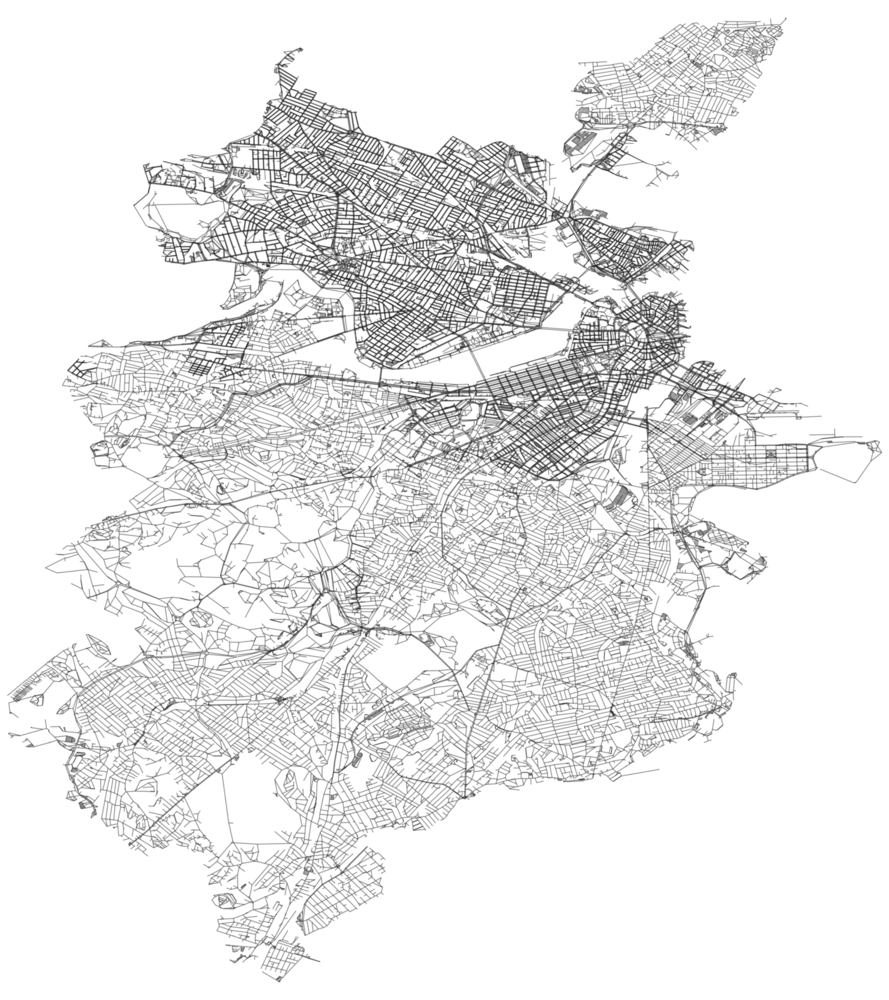} \label{fig:boston_road_network} }
    \subfloat[H3 indexing \cite{brodsky_2019}]{\includegraphics[height=\fh]{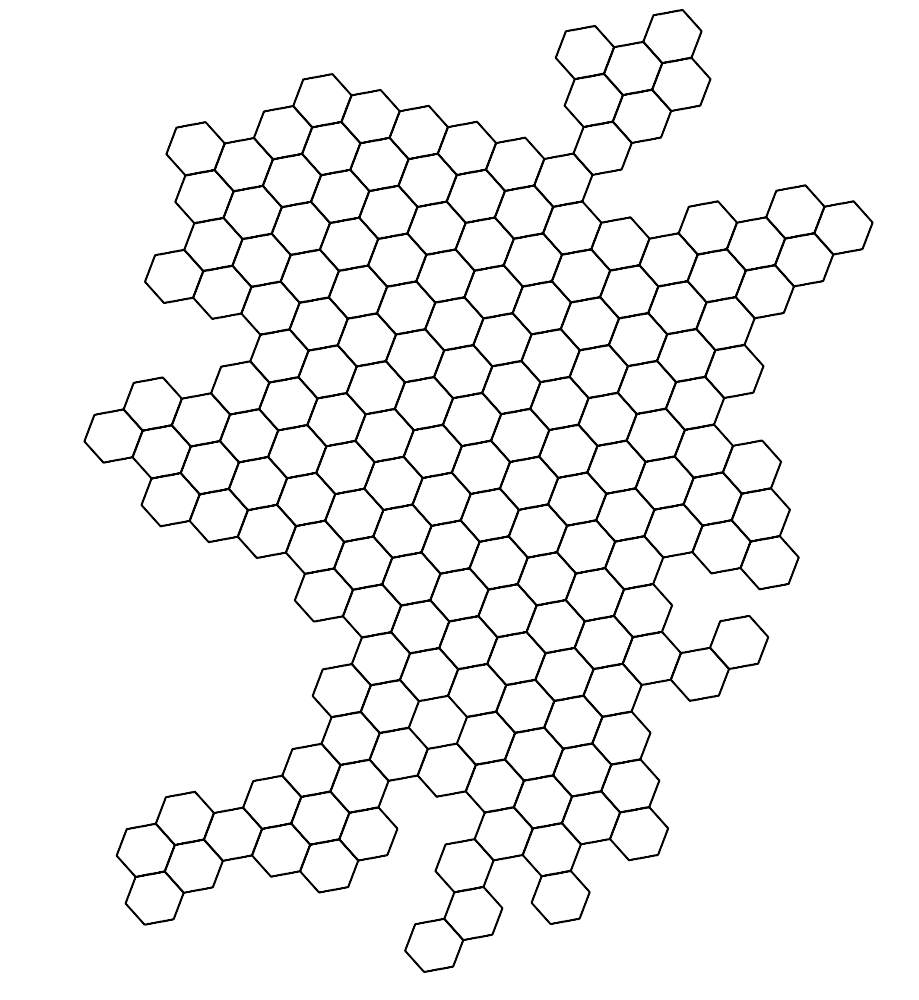} \label{fig:boston_h3_cells} }
    \caption{Data for Metro Boston Region: a) satellite imagery b) buildings c) road network d) H3 hexagonal hierarchical geospatial indexing system \cite{brodsky_2019} 
    }
    \label{fig:boston-spatial-1}
\end{figure*}

\else
\newcommand{\fw}{0.28\linewidth}
\begin{figure*}[!htb]
    \centering
    \subfloat[Satellite imagery]{\includegraphics[height=\fw]{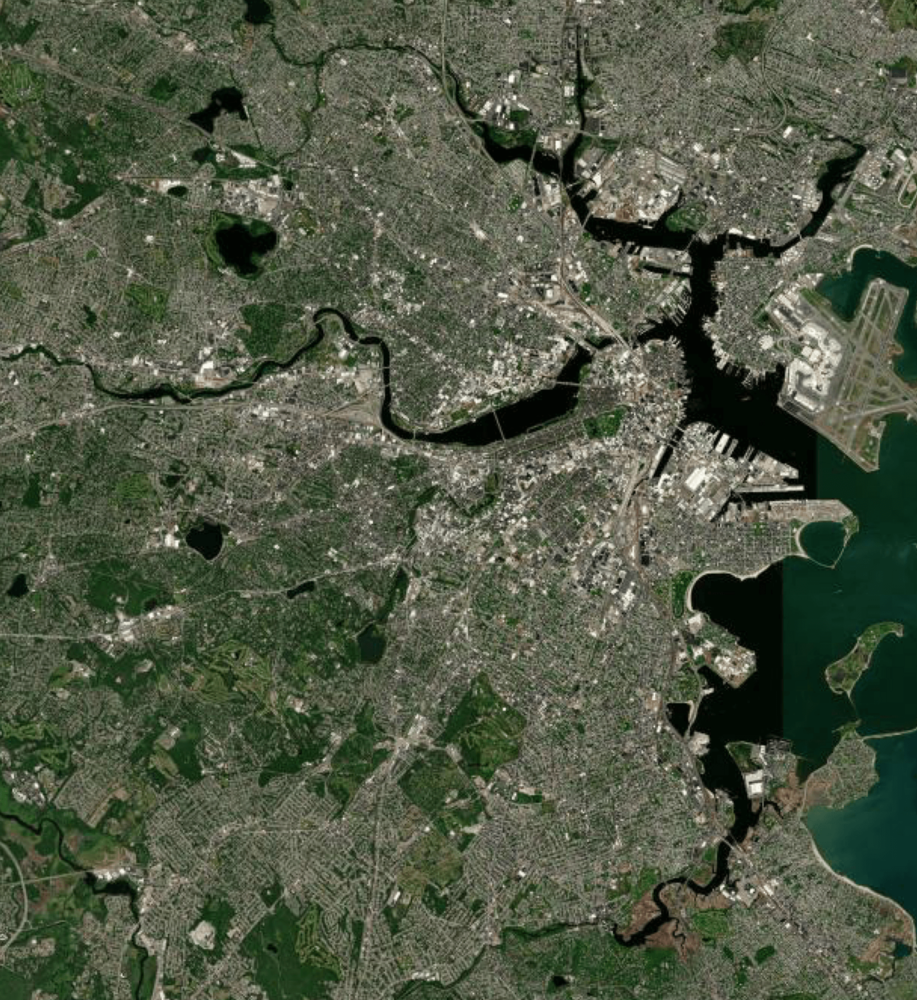} \label{fig:boston_satellite} }
    \subfloat[Buildings]{\includegraphics[height=\fw]{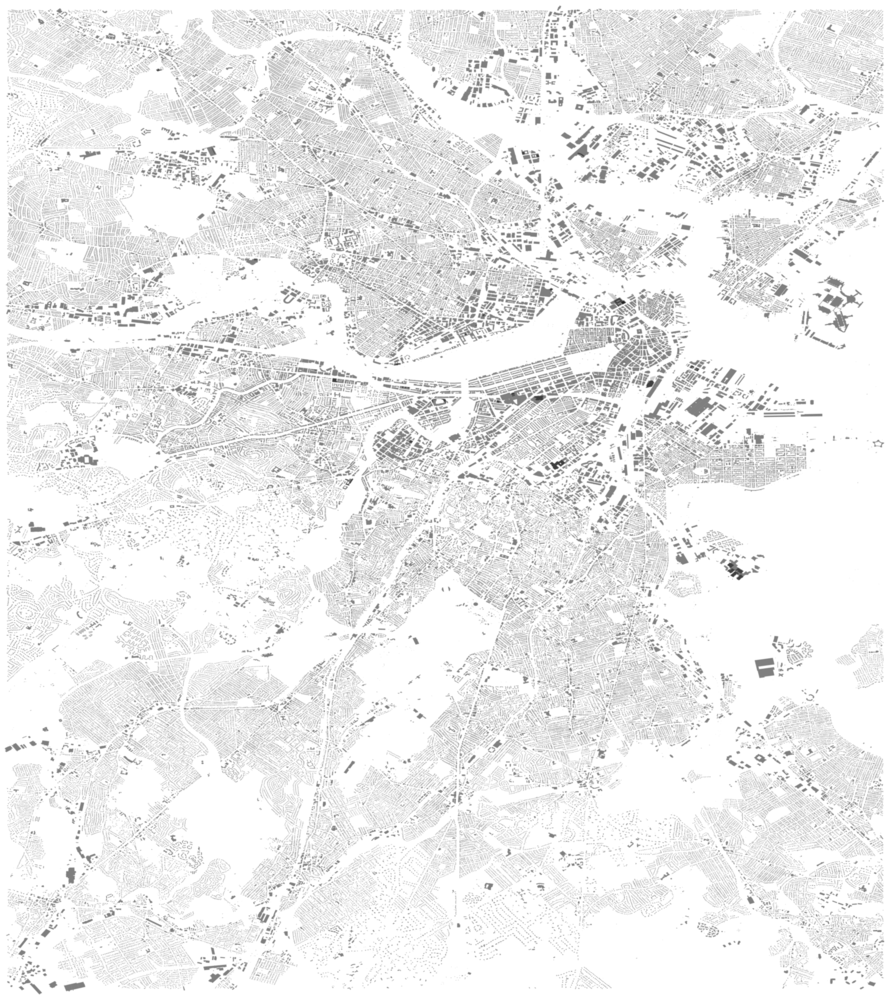} \label{fig:boston_buildings} }
    \subfloat[Road network]{\includegraphics[height=\fw]{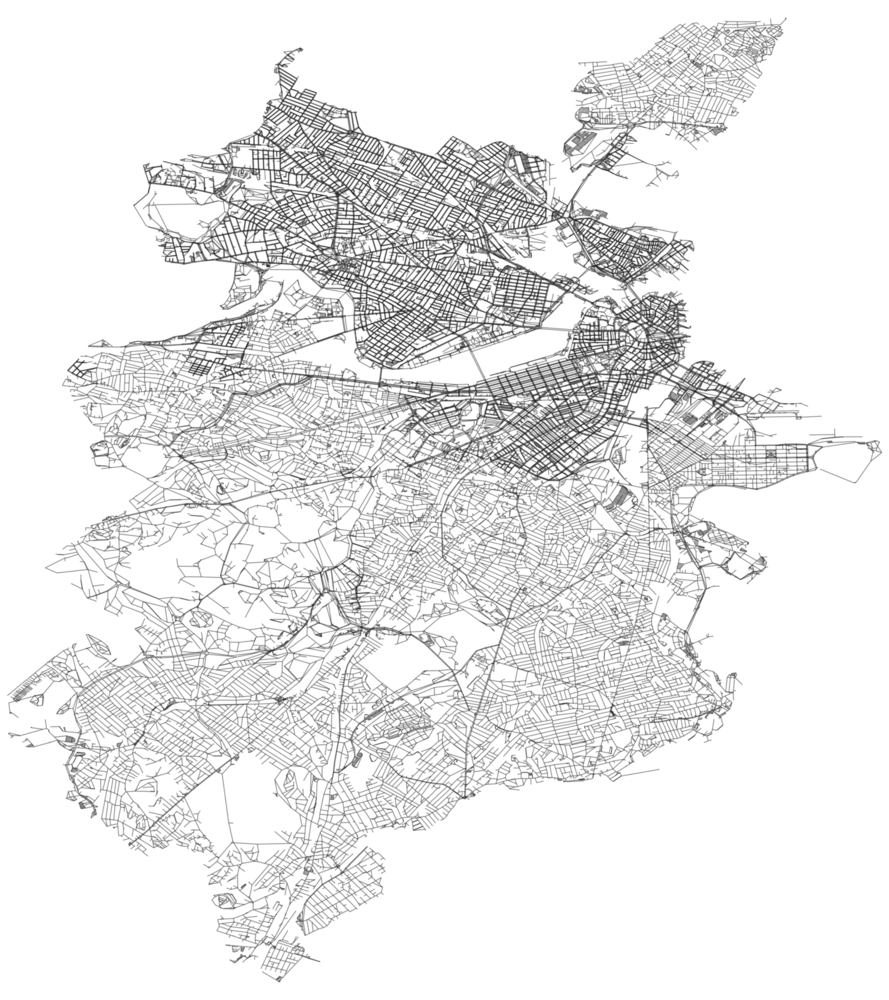} \label{fig:boston_road_network} }
    \subfloat[H3 indexing]{\includegraphics[height=\fw]{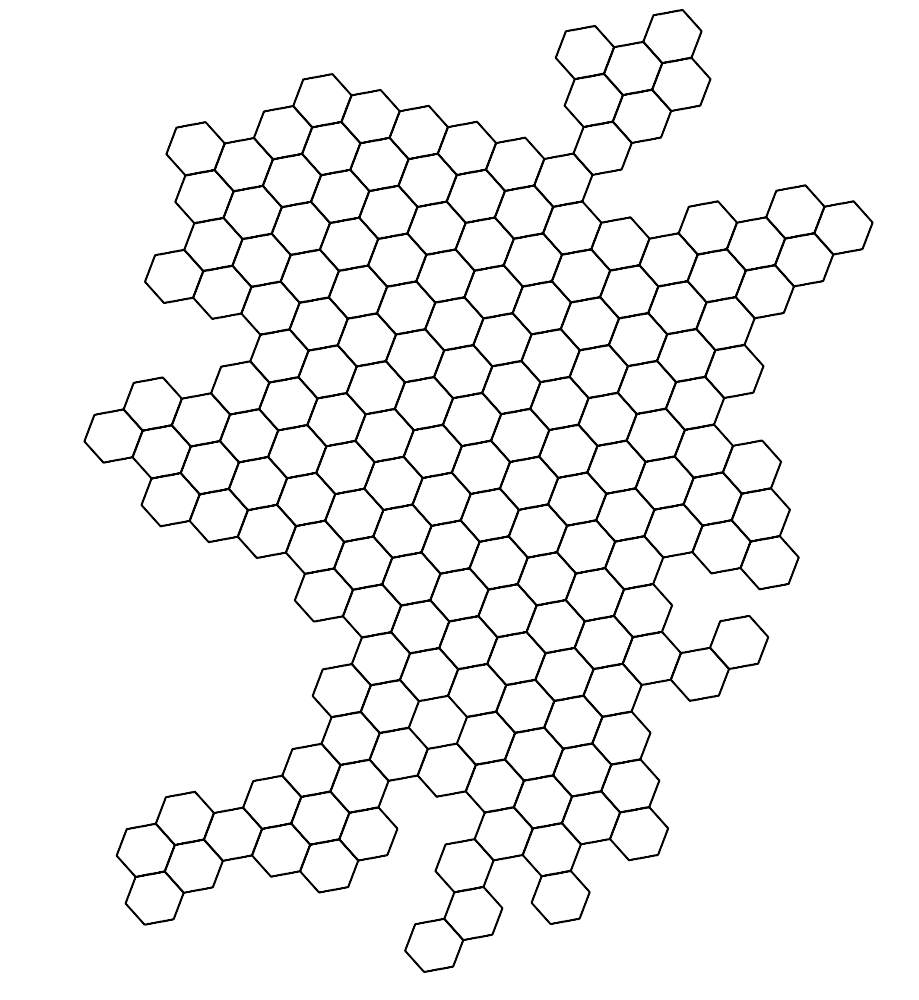} \label{fig:boston_h3_cells} }
    \caption{Data for Boston metropolitan area: a) satellite imagery b) buildings c) road network d) H3 hexagonal hierarchical geospatial indexing system 
    }
    \label{fig:boston-spatial-2}
\end{figure*}
\fi

The demand considered for the simulation is based on Bluebikes public bike-sharing system usage data \cite{bluebikes}. The user generation process takes advantage of this historical usage data and the buildings' spatial data. Buildings data is used to generate users' origin and destination locations inside the buildings.  This process yields realistic locations and avoids geographical obstacles such as highways or rivers. The impact of variations in the demand was analyzed by repeating the simulations with a randomized distribution of the scattering of the origins and destinations in buildings within 300 m around stations.
 
\begin{figure*}[!htb]
    \centering
    \subfloat[]{
    \includegraphics[width=0.5\linewidth]{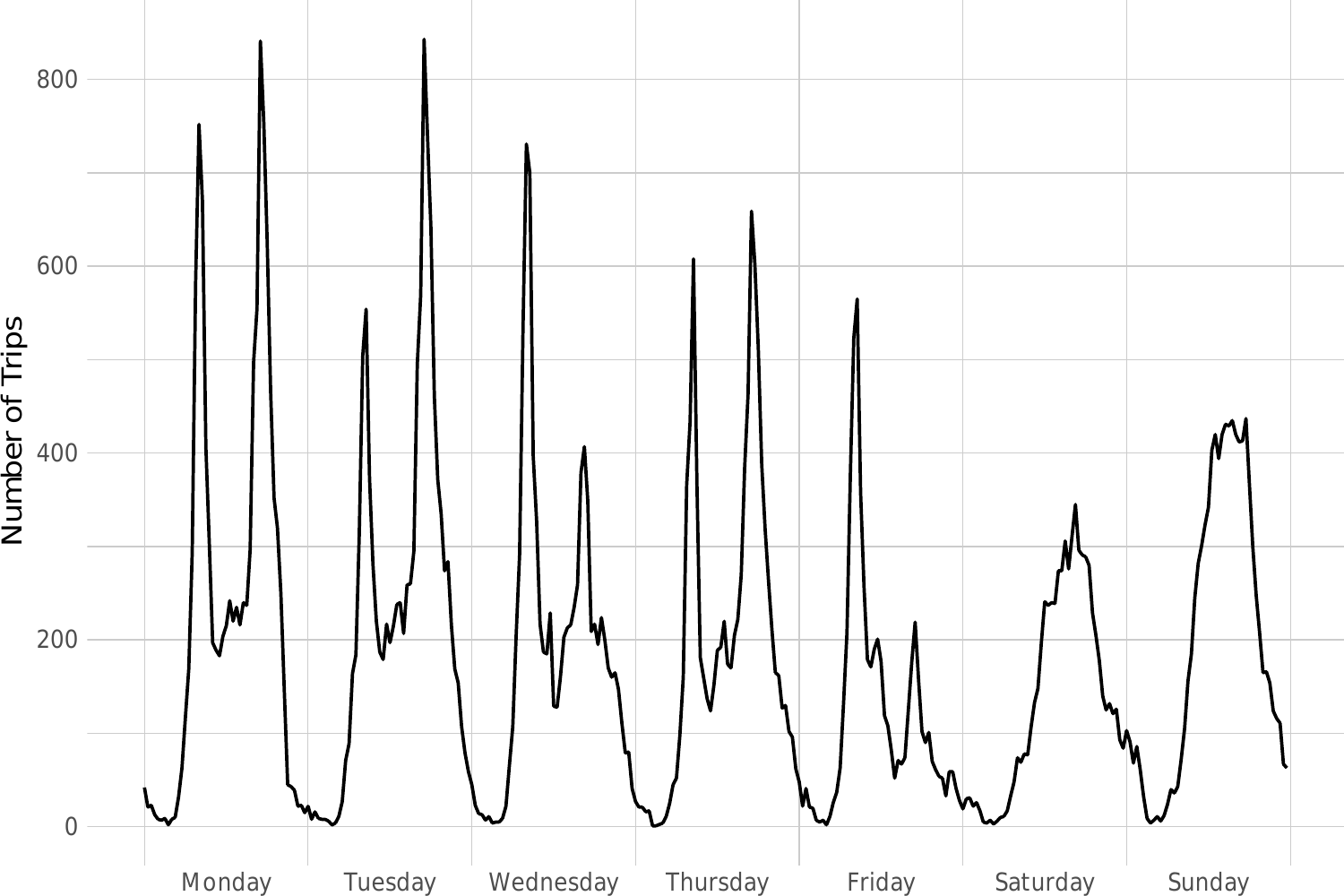}
    \label{fig:demand-1}}
    \subfloat[]{
    \includegraphics[width=0.5\linewidth]{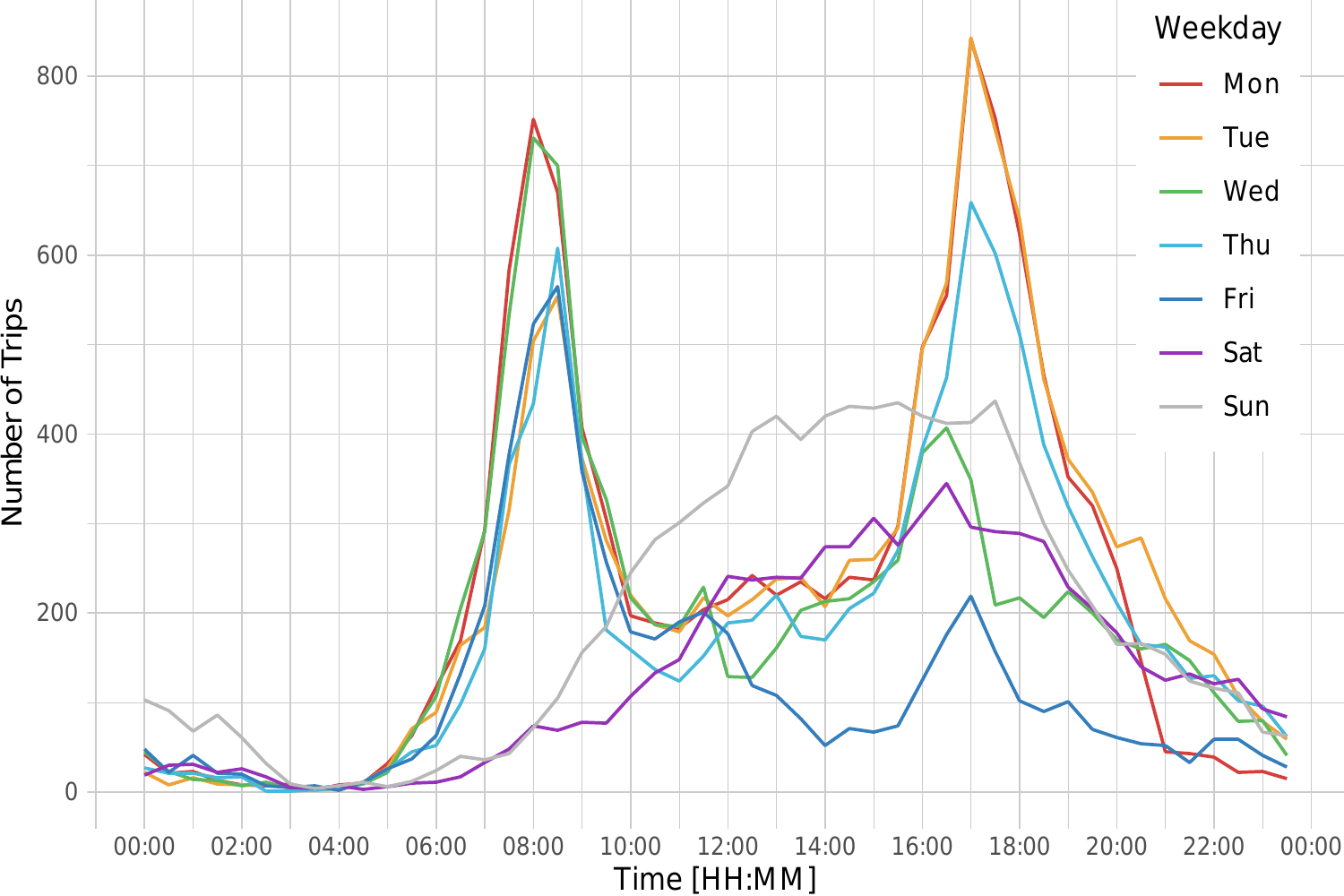}
    \label{fig:demand-2}} \\
    \subfloat[]{
    \includegraphics[width=0.5\linewidth]{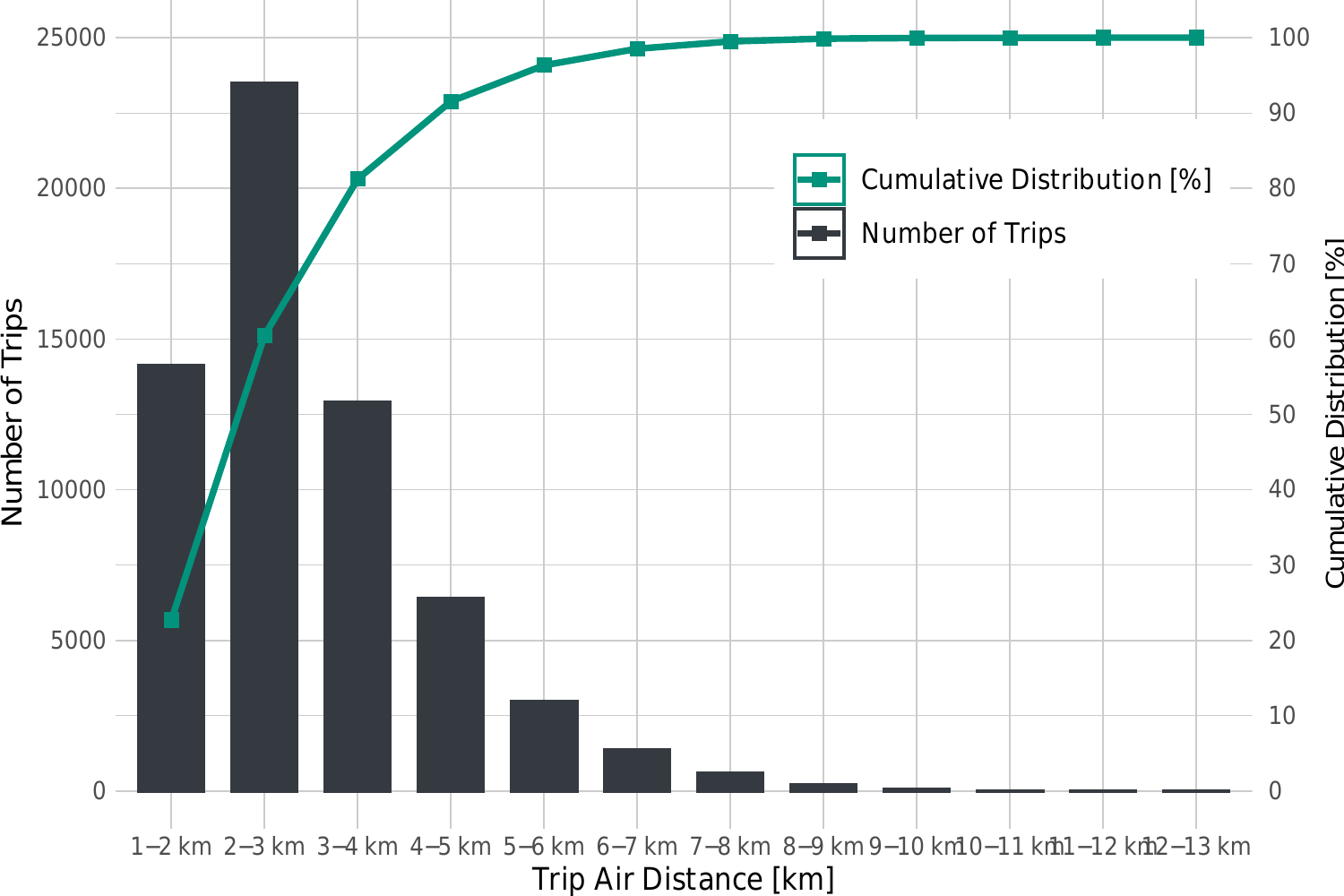}
    \label{fig:demand-3}}
    \caption{Characterization of the demand: a) Evolution of the demand during the week, b) Comparison of the demand at the same hour in the different days of the week, c) Histogram of the trip lengths measured from the air distance between origin and destination.}
    \label{fig:demand}
\end{figure*}

Figures \ref{fig:demand-1} and \ref{fig:demand-2} represent the evolution of demand during the week under study (07/10/2019 - 14/10/2019), Monday to Sunday. 
Figure \ref{fig:demand-3} displays a histogram of the trip lengths measured from the air distance between origin and destination. 90\% of the trips are less than 5km long.
With regards to the demand prediction module and the rebalancing manager, the urban area is subdivided into a grid of 180 H3 \cite{brodsky_2019} cells, as illustrated in Figure \ref{fig:boston_h3_cells}. The station-based system's location and capacity are also based on the Bluebikes data, and bikes are initially distributed proportionally to stations' capacity. The specific values considered for the configuration parameters in each system and scenario are depicted in Section \ref{results}.



        
\section{Results} \label{results}

This section gathers the main results on the performance of the station-based, dockless, and autonomous systems. As explained in Section \ref{bss_autonomous}, the case of the autonomous system is represented by two scenarios: one with no rebalancing, which provides the lower bound for the performance, and a second one with ideal rebalancing, which provides the upper-bound.

First, we examine the main performance metrics in the nominal state for each system separately, providing a detailed overview of each system's characteristics. The performance of the three BSS is then compared from different perspectives: a) in the nominal state, b) providing the same level of service, and c) with the same fleet size. Analyzing the differences in performance under various scenarios provides a thorough understanding of each system's strengths and weaknesses in comparison to the others.  

Intending to provide results for the autonomous system that are more precise than the upper and lower bound defined by the scenarios of no rebalancing and ideal rebalancing, this section includes an analysis of the performance of an autonomous system with a rebalancing mechanism based on demand prediction and routing.

Subsequently, we conduct several analyses to see how different configuration parameters affect the performance metrics of each system. This analysis is particularly interesting for the autonomous BSS case due to the uncertainty surrounding how these systems will function in complex real-world conditions. At the same time, it provides a great understanding of how variations in parameters are reflected on the system's performance and the most relevant parameters for each performance metric. Moreover, this analysis allowed us to assess the impact of the values considered for the nominal state.

\begin{figure*}[!htb]
    \centering
    \subfloat[User activities timeline]{
    \includegraphics[width=0.5\linewidth]{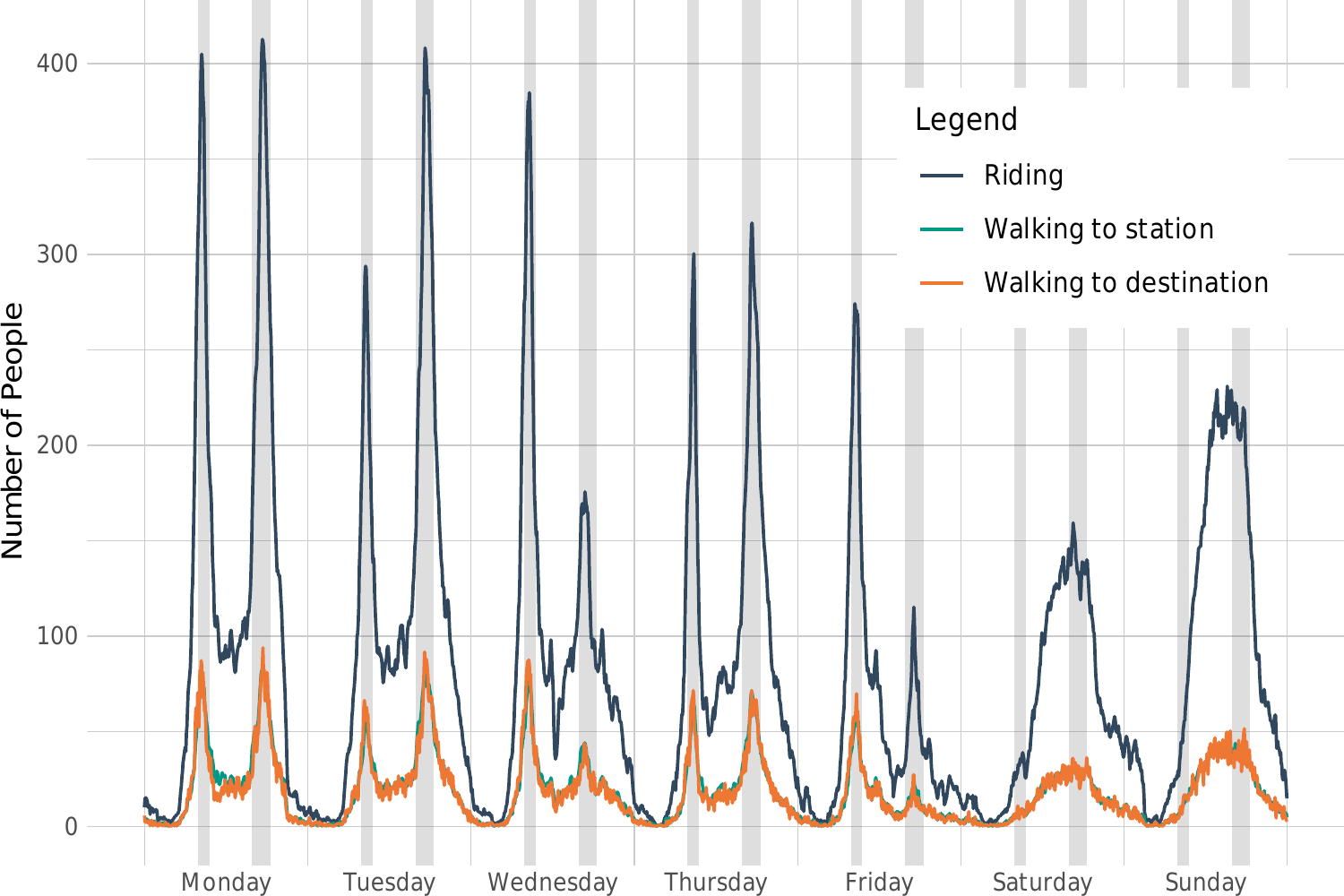}
    \label{fig:SB-nominal-1}}
    \subfloat[Bike activities timeline: dashed line indicates fleet size]{
    \includegraphics[width=0.5\linewidth]{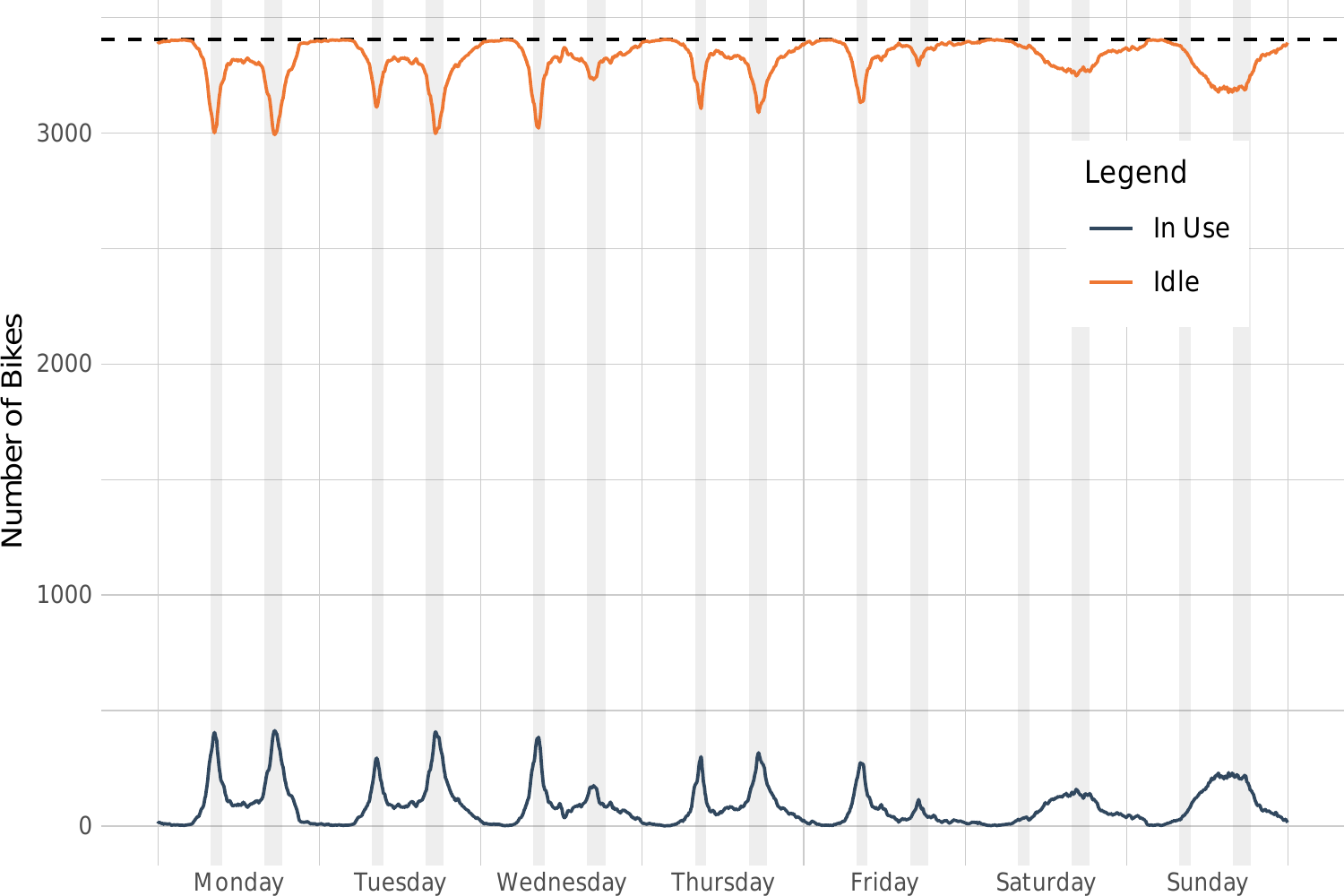}
    \label{fig:SB-nominal-2}} \\
    \subfloat[Served and unserved trips timeline]{
    \includegraphics[width=0.5\linewidth]{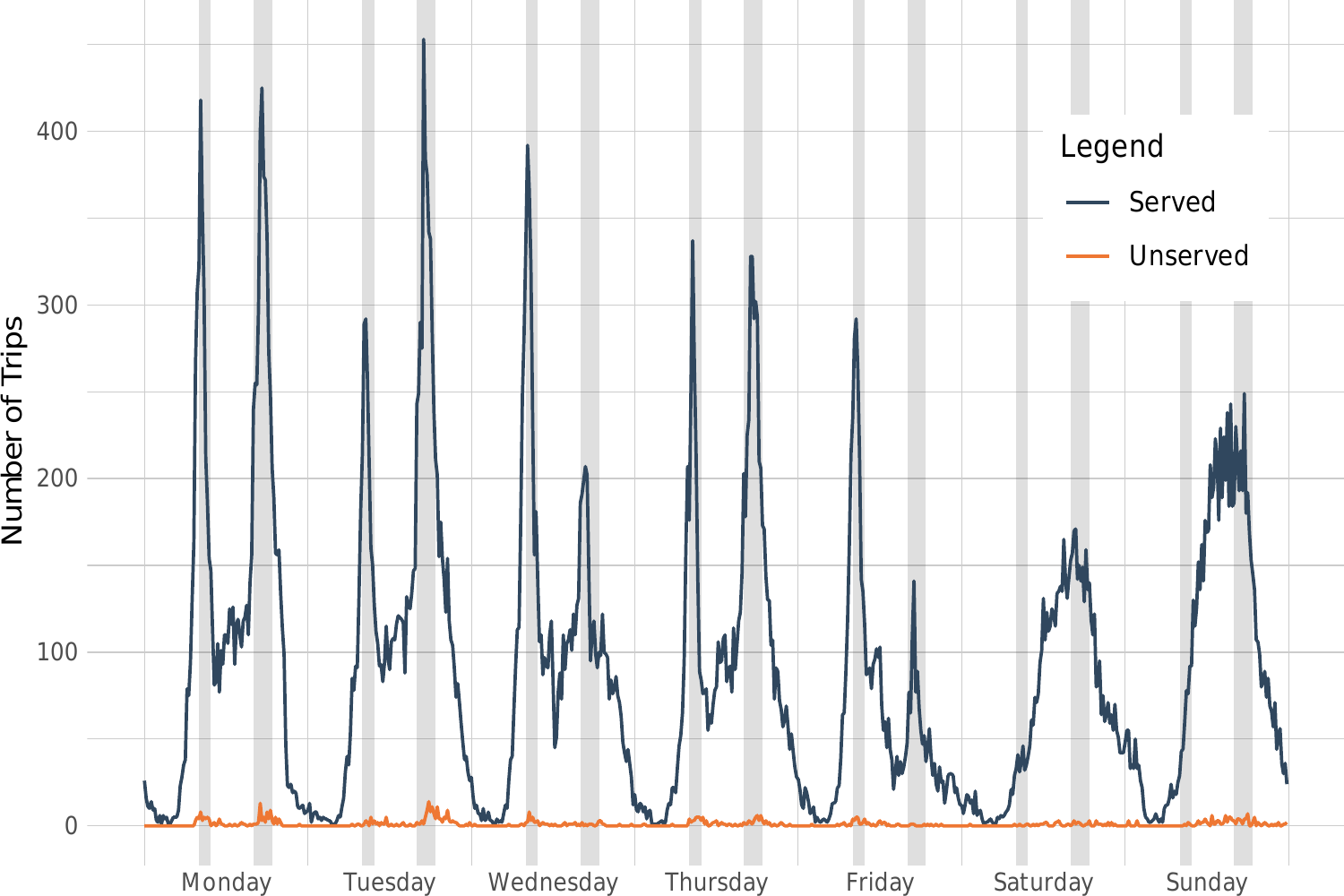}
    \label{fig:SB-nominal-3}}
    \subfloat[User activities mean duration with shaded std. dev.: $\mu \pm \sigma$ ]{
    \includegraphics[width=0.5\linewidth]{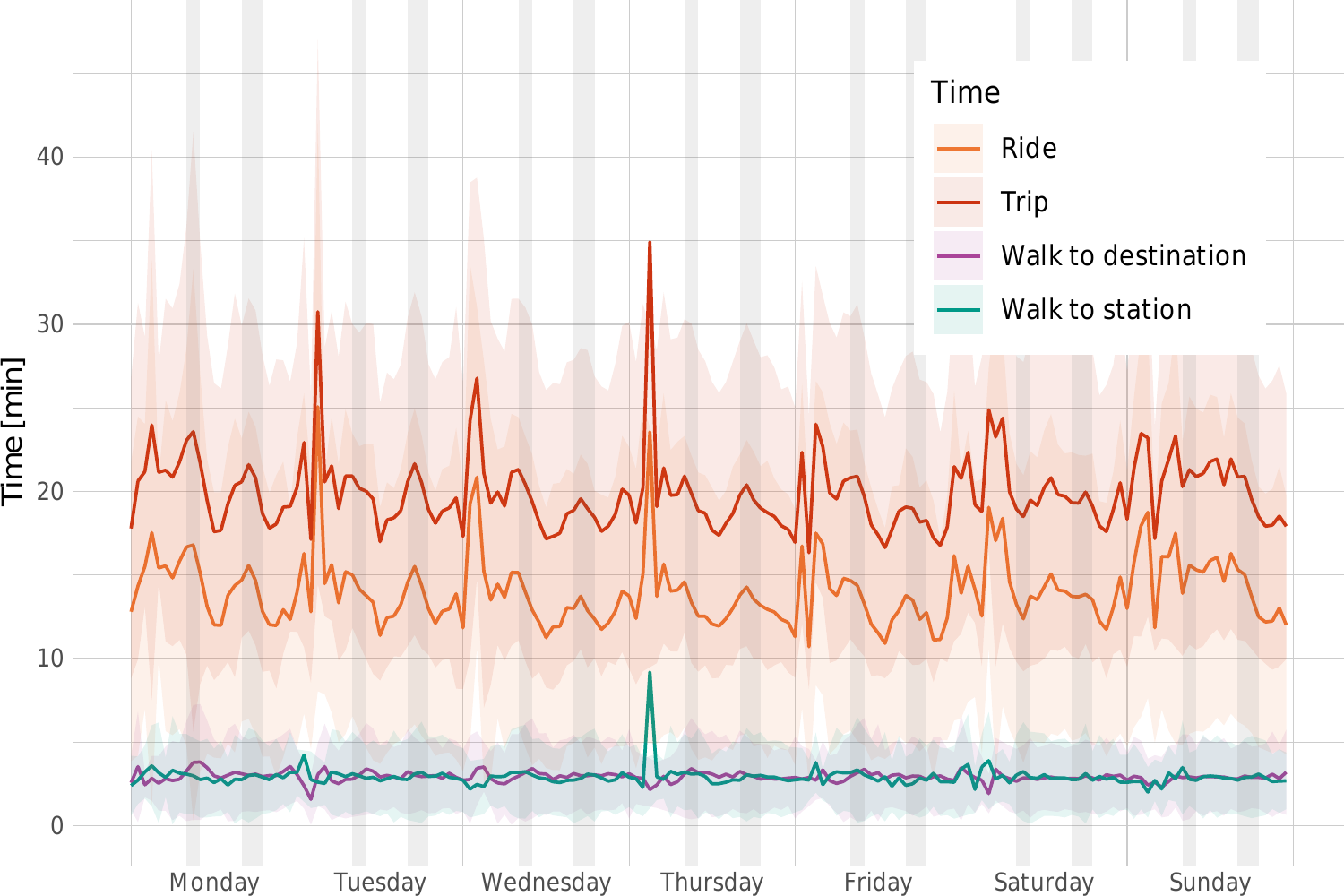}
    \label{fig:SB-nominal-4}}
    \caption{Station-based system nominal state: a) Timeline of the number of users performing each activity; b) Timeline of the number of users performing each activity; c) Timeline of the number of served and unserved trips; d) Timeline of the mean duration of different user activities. Shaded areas indicate morning and evening periods of peak demand.
    }
    \label{fig:SB-nominal}
\end{figure*}

\subsection{Nominal state analysis}

This section summarizes the parameters that define each system's nominal state, as well as the performance metrics for this configuration. The values assigned to the parameters in the nominal state are those that are most likely to accurately represent a realistic behavior. These values are defined in Tables \ref{tb:sb-nominal}, \ref{tb:dl-nominal}, and \ref{tb:aut-nominal}.  The performance metrics, instead, can be found on Tables \ref{tb:sb-nominal-results}, \ref{tb:dl-nominal-results}, and \ref{tb:aut-nominal-results}. 

In the case of the station-based system, the fleet size was determined considering the number of daily trips in the simulation and the number of trips per bike and day from \cite{sanchez2020autonomous}. In the case of the dockless system, the average number of trips per bike and day was extracted from \cite{kondor2021estimating}. Lastly, in the case of the autonomous system, the fleet size was estimated iterating over the results from \cite{sanchez2020autonomous} and validated using the results from Section \ref{level-of-service}.

As for other parameters that shared by two or more systems, the maximum distance that a user is willing to walk to at the beginning or end of the trip was considered to be 300 m \cite{kabra2020bike}. Furthermore, the average walking speed was considered to be 5 km/h \cite{browning2006effects}, and the average riding speed were considered to be 10.2 km/h \cite{jensen2010}.

\subsubsection{Station-based}
    
In addition to the configuration parameters shared among the three systems, the configuration of the station-based system requires some additional parameters that represent the bicycle rebalancing process (see Table \ref{tb:sb-nominal}). The rebalancing parameter represents the likelihood of a user getting a bicycle or dock in the cases when, without rebalancing, a station would have been either empty or full. For the rebalancing system to move bicycles to and from balanced stations, the minimum number of bikes and docks per station were set to three, which is approximately 20\% of the capacity of the average docking station. 

\begin{table}[!htb]
  \centering
  \scalebox{\tablescale}{
  \begin{tabular}{p{5.5cm}p{2cm}}
    \toprule
    \multicolumn{2}{c}{\textbf{Station-based system nominal parameters}}  \\ \midrule
    Number of bikes & 3500 \\ 
    Maximum walking radius & 300 m \\ 
    Average walking speed & 5 km/h \\  
    Average riding speed & 10.2 km/h \\ 
    Rebalancing parameter & 90 \% \\  
    Minimum bikes/docks per station  & 3 \\ \bottomrule
    \end{tabular}}
    \caption{Nominal values of the configuration parameters in the station-based system}
    \label{tb:sb-nominal}
\end{table}

\begin{table}[!htb]
\centering
\scalebox{\tablescale}{
\begin{tabular}{l|lr}
\toprule
\multicolumn{3}{c}{\textbf{Station-based system nominal metrics}} \\ \midrule
\multicolumn{2}{l}{User demand}                 & 62192     \\ 
\multicolumn{2}{l}{Average trip time}           & 19.87 min \\ 
 & Walk time at origin                            & 2.95 min  \\ 
 & Walk time at origin   \textgreater 10 min      & 0.71 \%    \\ 
 & Walk time at origin   \textgreater 15 min      & 0.25 \%    \\ 
 & Ride time                                      & 13.96 min \\ 
 & Walk time at   destination                     & 2.97 min  \\ 
 & Walk time at   destination \textgreater 10 min & 0.62 \%    \\ 
 & Walk time at   destination \textgreater 15 min & 0.17 \%    \\ 
\multicolumn{2}{l}{Served trips}                & 99.00 \%   \\ 
\multicolumn{2}{l}{Unserved trips}              & 1.00 \%    \\ 
 & No walkable   stations                       & 0.26 \%    \\ 
 & No bikes in walkable   distance              & 0.73 \%    \\ 
\multicolumn{2}{l}{Bikes used}                  & 91.59 \%   \\ 
\multicolumn{2}{l}{Trips/bike/day}              & 2.51      \\ 
\multicolumn{2}{l}{Rebalanced bikes}          & 9193    \\ 
\multicolumn{2}{l}{v.k.t total}                 & 146068 km \\ 
\multicolumn{2}{l}{v.k.t. per bike}           & 41.73     \\ 
 & v.k.t. in use                                & 100 \%     \\ 
\multicolumn{2}{l}{Total time}                & 7 days    \\
 & Time in use                                  & 2.44 \%    \\ 
 & Time idling                                 & 97.56 \%   \\ \bottomrule
\end{tabular}}
    \caption{Performance metrics for station-based system in nominal state}
    \label{tb:sb-nominal-results}
\end{table}

The performance metrics of the station-based system in the nominal state can be found in Table \ref{tb:sb-nominal-results}. The evolution of the different metrics over time can be observed in Figure \ref{fig:SB-nominal}. The average total duration of the trip in the station-based system is 19.87 minutes, including 2.95 minutes of walking at the beginning and 2.97 minutes of walking at the end of each trip. 

The system is capable of serving 99\% of the demand. It can be observed that most unserved trips occur during high-demand times, especially on days in which the rush hour peak is high. The average number of trips per bike and day is 2.51, which is close to the 2.62 trips/bike/day reported in \cite{sanchez2020autonomous}. Throughout the week, 92\% of the bikes are used at some point. On average, the bikes are in use for 2.44\% of the time and, as shown in Figure \ref{fig:SB-nominal-2}, even during high demand periods, the number of bikes that are being used at the same time is relatively low.

In terms of the rebalancing rate, the ratio is one rebalanced bike for every 6.08 trips. In real-world station-based systems rebalancing rates are around a 1:6 ratio, and, therefore, this result validates the values set for the rebalancing configuration parameters.

\subsubsection{Dockless}

The parameters for the dockless system are the same that are common for the three BSS and can be found in Table \ref{tb:dl-nominal}.  

\begin{table}[!htb]
  \centering
  \scalebox{\tablescale}{
  \begin{tabular}{p{4cm}p{2cm}}
    \toprule
    \multicolumn{2}{c}{\textbf{Dockless system nominal parameters}}  \\ \midrule
    Number of bikes & 8000 \\ 
    Maximum walking radius & 300 m \\ 
    Average walking speed & 5 km/h \\ 
    Average riding speed & 10.2 km/h \\ \bottomrule
    \end{tabular}}
    \caption{Nominal values of the configuration parameters in the dockless system}
    \label{tb:dl-nominal}
\end{table}

\begin{figure*}[!htb]
    \centering
    \subfloat[User activities timeline]{
    \includegraphics[width=0.5\linewidth]{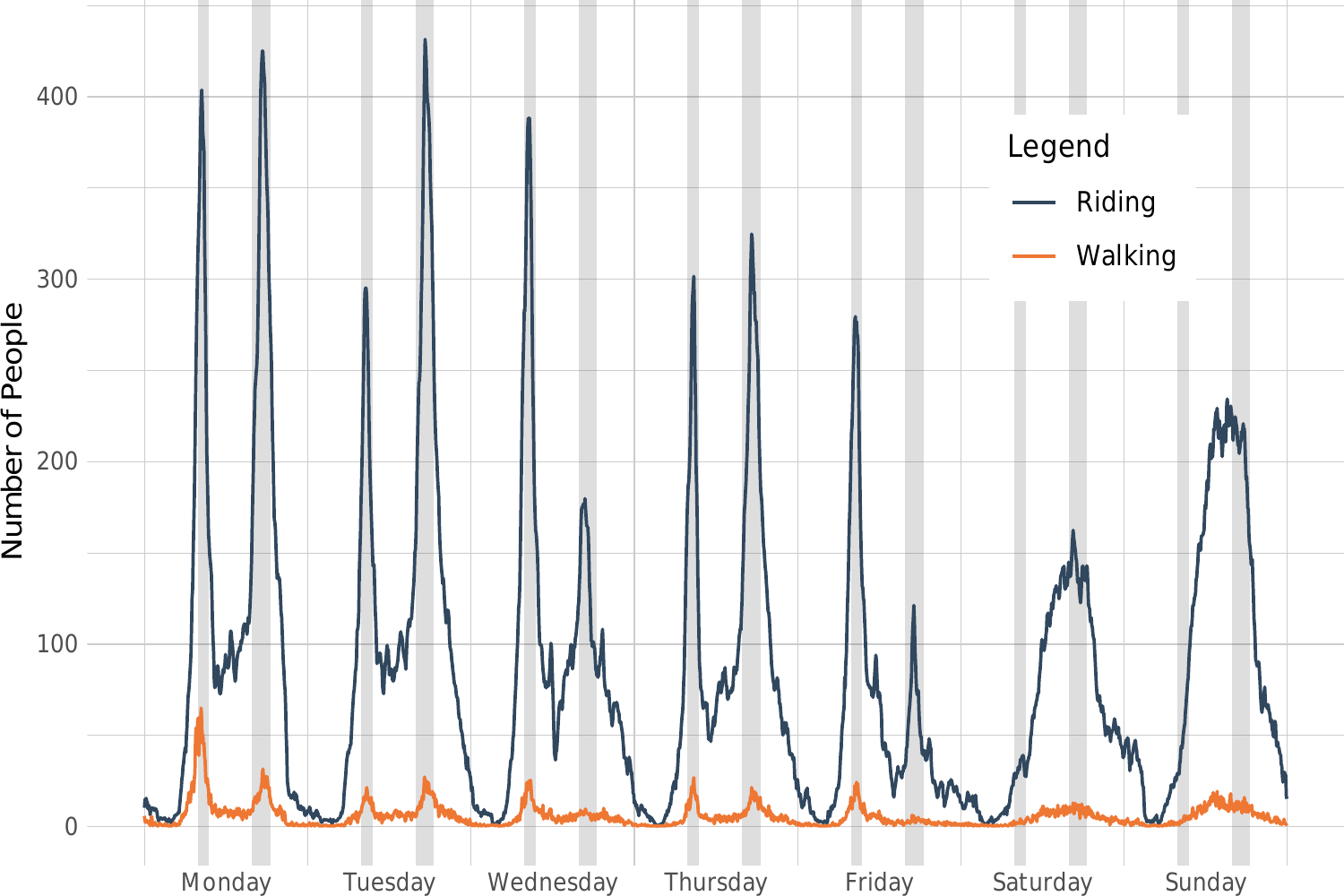}
    \label{fig:DK-nominal-1}}
    \subfloat[Bike activities timeline: dashed line indicates fleet size]{
    \includegraphics[width=0.5\linewidth]{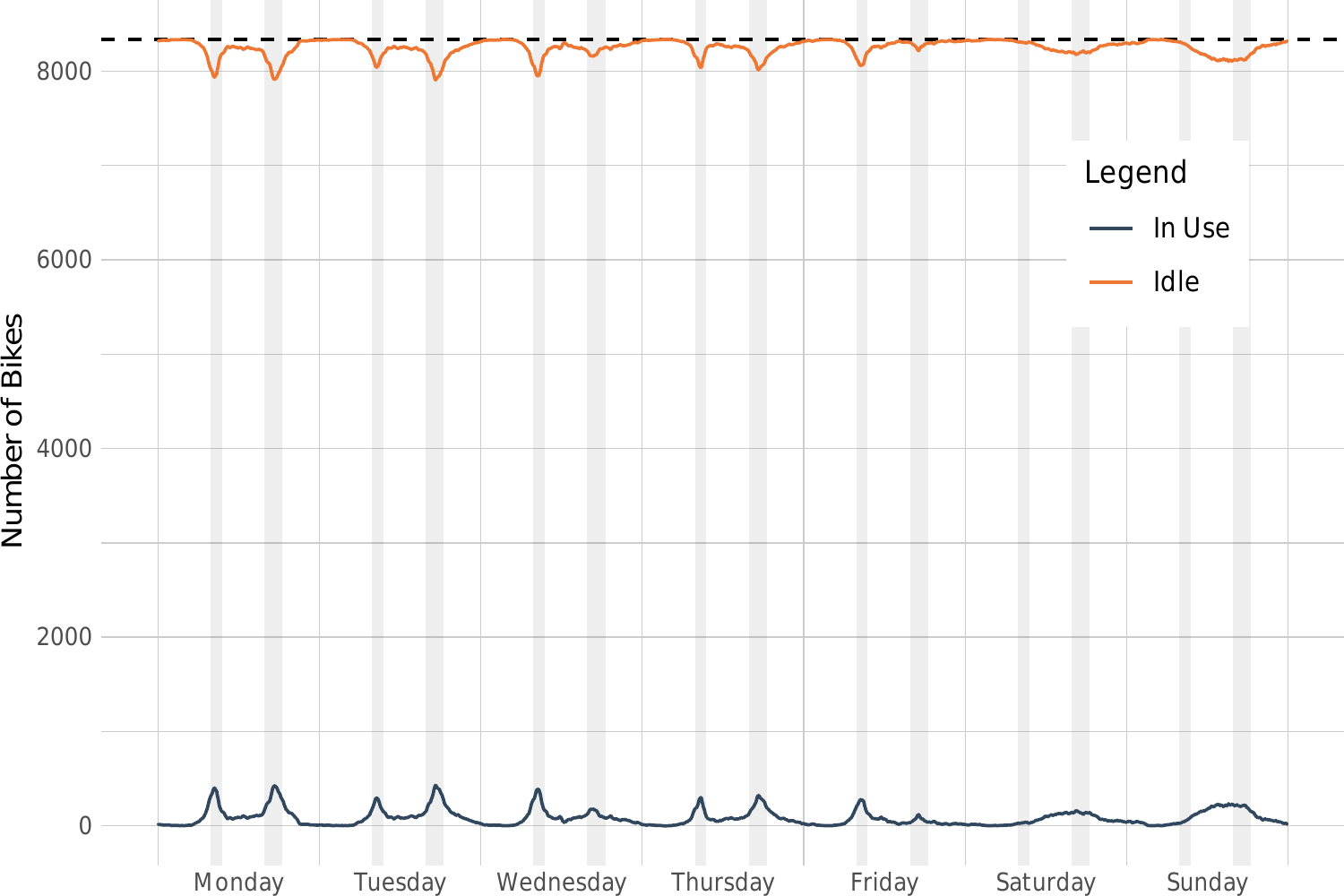}
    \label{fig:DK-nominal-2}} \\
    \subfloat[Served and unserved trips timeline]{
    \includegraphics[width=0.5\linewidth]{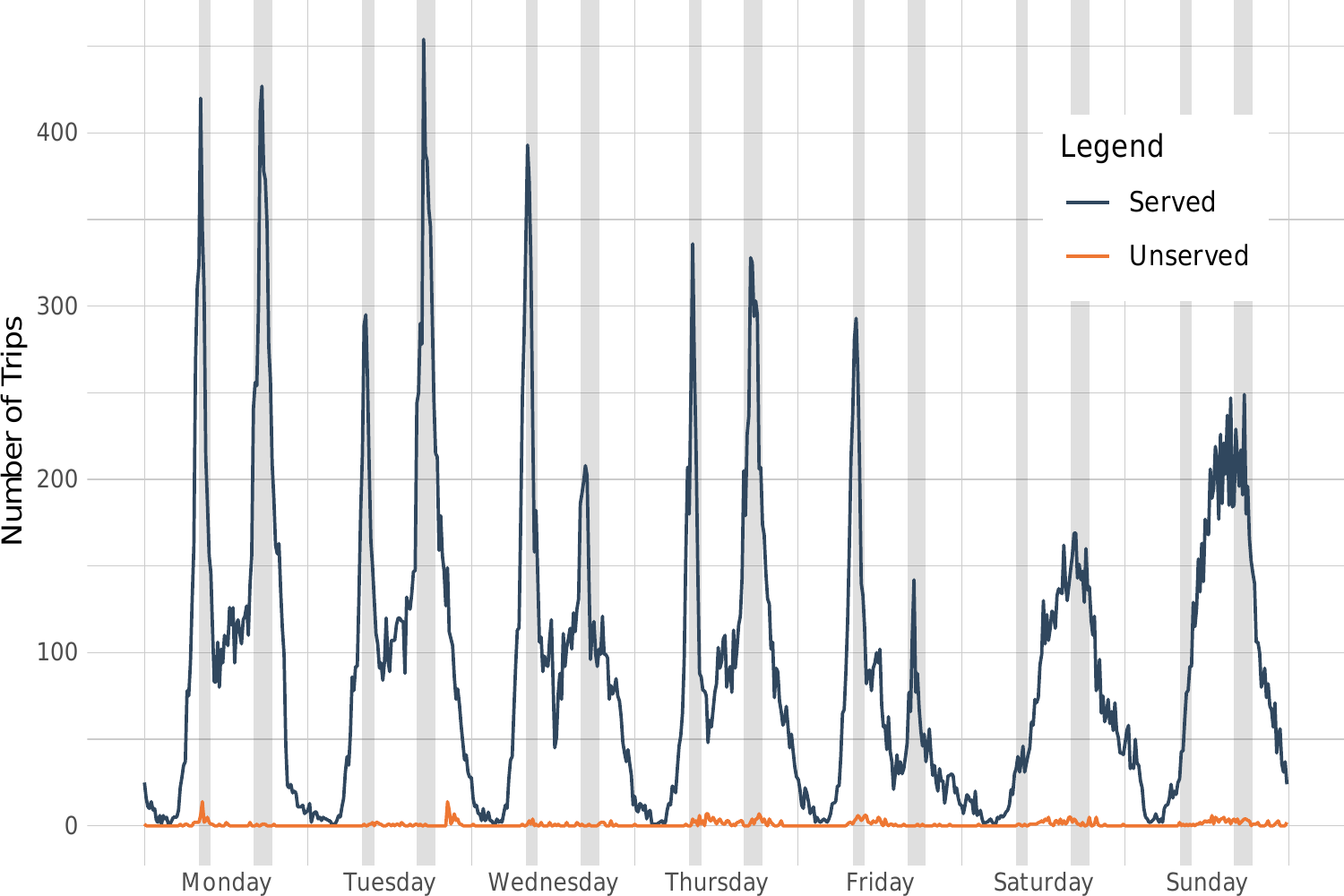}
    \label{fig:DK-nominal-3}}
    \subfloat[User activities mean duration with shaded std. dev.: $\mu \pm \sigma$]{
    \includegraphics[width=0.5\linewidth]{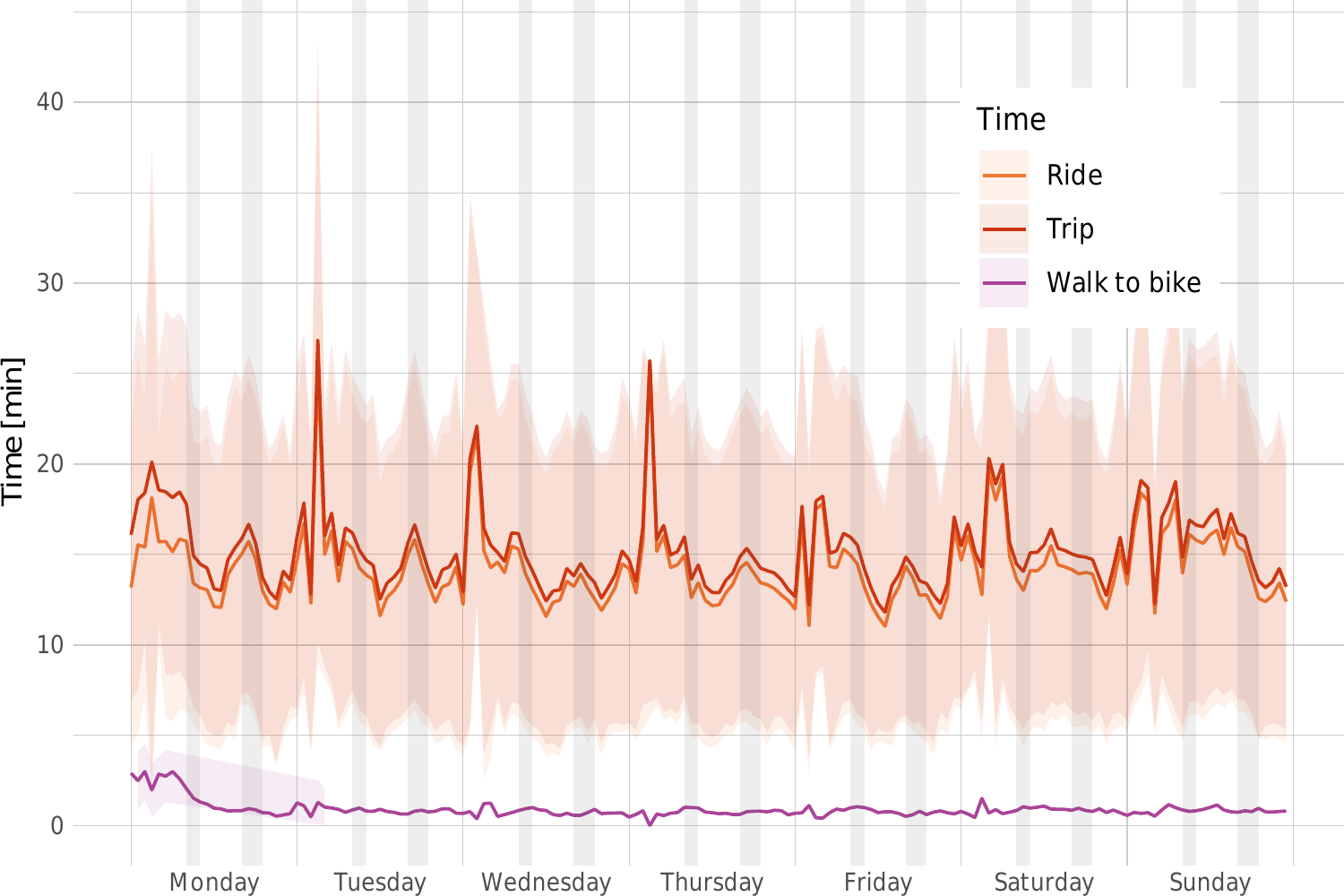}
    \label{fig:DK-nominal-4}}
    \caption{Dockless system nominal state: a) Timeline of the number of users performing each activity; b) Timeline of the number of users performing each activity; c) Timeline of the number of served and unserved trips; d) Timeline of the mean duration of different user activities. Shaded areas indicate morning and evening periods of peak demand.
    }
    \label{fig:DK-nominal}
\end{figure*}

The performance of the dockless system in the nominal state can be found in Table \ref{tb:dl-nominal-results} and the evolution of the performance metrics over time can be found in Figure \ref{fig:DK-nominal}. The average trip time is 15.14 minutes, which includes 1.07 minutes of walking at the beginning of the trip. The system is able to serve 99.02\% of the demand. Since the use of bicycles is not restricted to the stations' capacity, the number of unserved trips is less related to peaks in demand than in the case of the station-based system. While peaks of demand do also lead to an increase in unserved trips, on some days (e.g., Thursday), the unserved trips are more evenly distributed in time than the demand, while on others (e.g., Tuesday), the peak of unserved trips happens after the demand-peak.

The average number of trips per bike and day is 1.10, which is very close to the 1.09 reported by Kondor et al. \cite{kondor2021estimating}. During the week, 70.93\% of the bikes are used. On average, bikes are used for 1.07\% of the time. As can be observed in Figure \ref{fig:DK-nominal}, only a small fraction of the bicycles are used simultaneously.

\begin{table}[!htb]
\centering
\scalebox{\tablescale}{
\begin{tabular}{l|lr}
\toprule
\multicolumn{3}{c}{\textbf{Dockless system nominal metrics}}                  \\ \midrule
\multicolumn{2}{l}{User demand}            & 62192     \\ 
\multicolumn{2}{l}{Average trip time}      & 15.14 min \\ 
 & Walk time at origin                       & 1.07 min  \\ 
 & Walk time at origin   \textgreater 10 min & 0.15 \%    \\ 
 & Walk time at origin   \textgreater 15 min & 0.06 \%    \\ 
 & Ride time                                 & 14.07 min \\ 
\multicolumn{2}{l}{Served   trips}         & 99.02 \%   \\ 
\multicolumn{2}{l}{Unserved   trips}       & 0.98 \%    \\ 
\multicolumn{2}{l}{Bikes used}             & 70.93 \%   \\ 
\multicolumn{2}{l}{Trips/bike/day}         & 1.10      \\ 
\multicolumn{2}{l}{Total   v.k.t}          & 147327 km \\ 
\multicolumn{2}{l}{v.k.t. per   bike}      & 18 km     \\ 
 & v.k.t. in use                             & 100 \%     \\ 
\multicolumn{2}{l}{Total time}             & 7 days    \\ 
 & Time in use                               & 1.07 \%    \\ 
 & Time idling                              & 98.93 \%   \\ \bottomrule
\end{tabular}}
    \caption{Performance metrics for dockless system in nominal state}
    \label{tb:dl-nominal-results}
\end{table}

\subsubsection{Autonomous}

The configuration parameters for the autonomous system in the nominal state can be found in Table \ref{tb:aut-nominal}. The average speed in autonomous mode was determined to be 8 km/h; while this is a relatively low value, it aims to account for the uncertainty of autonomous systems behavior in complex real-world conditions. The maximum autonomous radius was set at 2000 m, which, at a speed of  8 km/h, is equivalent to a maximum wait time of 15 min.
The autonomy of the battery was considered to be 70 km with a charge time of 4.5 h based on the range of commercially available bicycles in throttle only mode \cite{juiced}. 

\begin{table}[!htb]
  \centering
  \scalebox{\tablescale}{
  \begin{tabular}{p{5.5cm}p{2cm}}
     \toprule
    \multicolumn{2}{c}{\textbf{Autonomous system nominal parameters}}  \\ \midrule
    Number of bikes & 1000 \\  
    Maximum autonomous radius & 2000 m \\  
    Average autonomous driving speed & 8 km/h \\ 
    Average riding speed & 10.2 km/h \\ 
    Minimum battery level & 15 \% \\ 
    Battery autonomy & 70 km \\ 
    Battery recharge time & 4.5 h \\ \bottomrule
    \end{tabular}}
    \caption{Nominal values of the configuration parameters in the autonomous system}
    \label{tb:aut-nominal}
\end{table}

\begin{figure*}[!htb]
    \centering
    \subfloat[User activities timeline]{
    \includegraphics[width=0.5\linewidth]{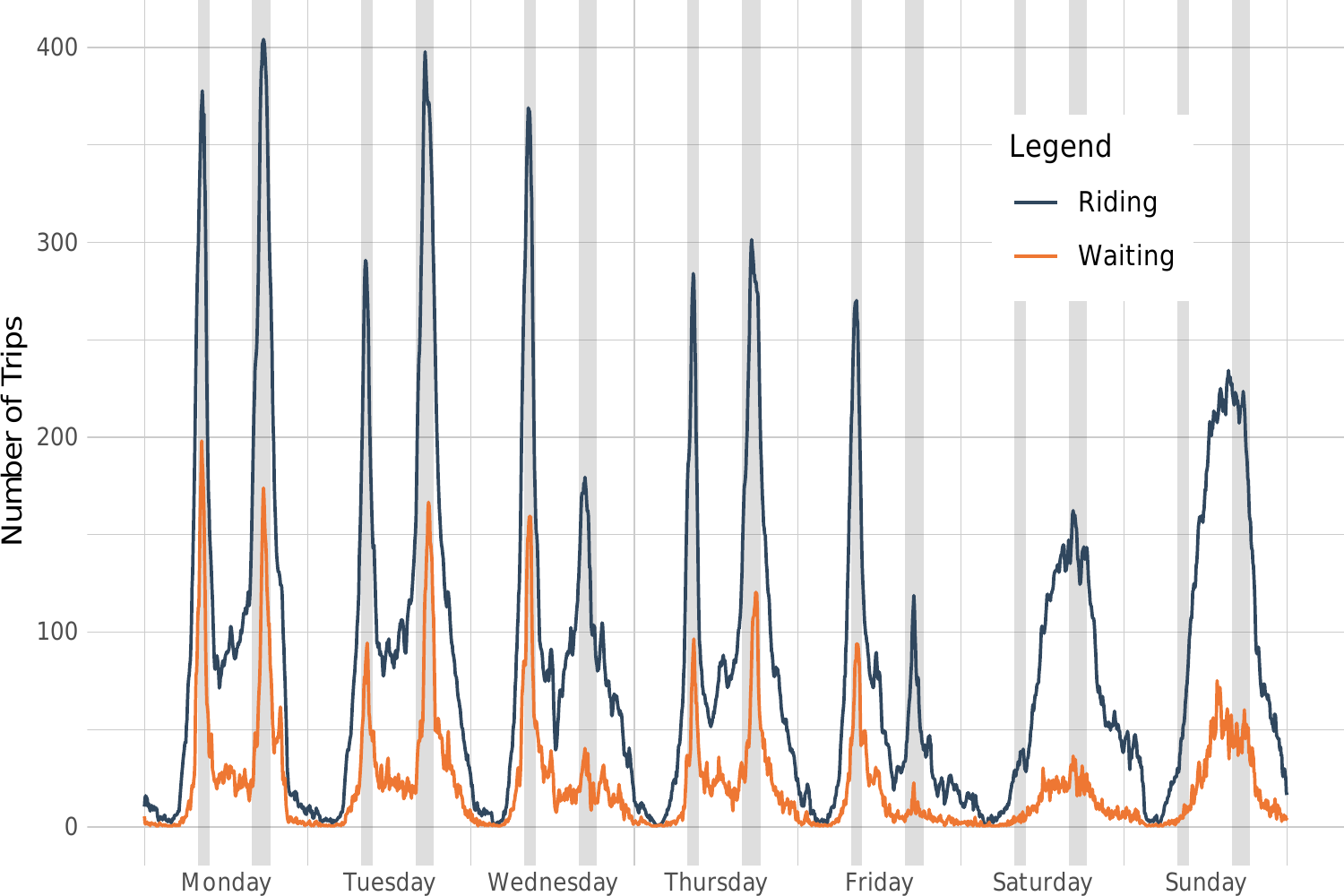}
    \label{fig:AU-nominal-1}}
    \subfloat[Bike activities timeline: dashed line indicates fleet size]{
    \includegraphics[width=0.5\linewidth]{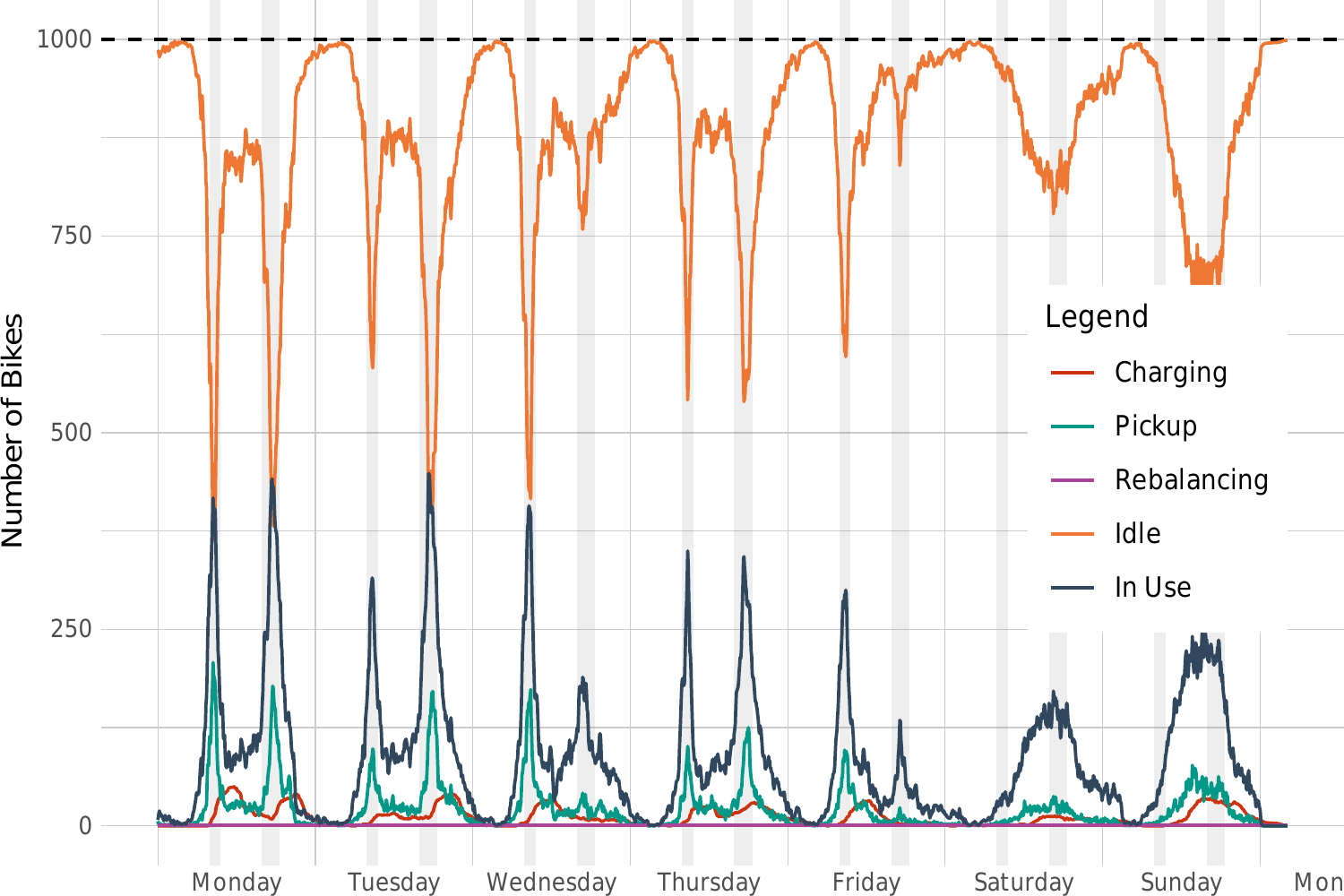}
    \label{fig:AU-nominal-2}} \\
    \subfloat[Served and unserved trips timeline]{
    \includegraphics[width=0.5\linewidth]{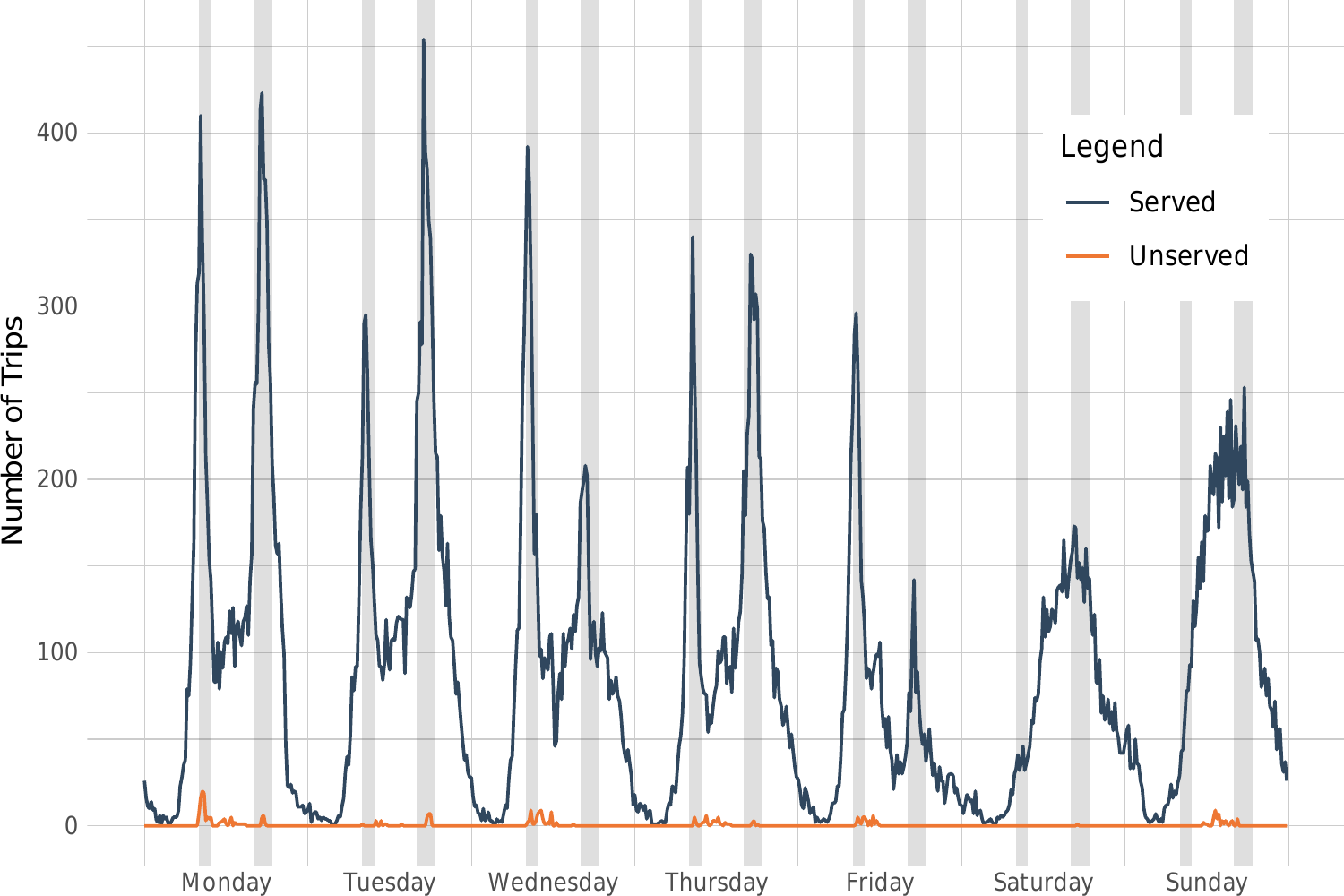}
    \label{fig:AU-nominal-3}}
    \subfloat[User activities mean duration with shaded std. dev.: $\mu \pm \sigma$]{
    \includegraphics[width=0.5\linewidth]{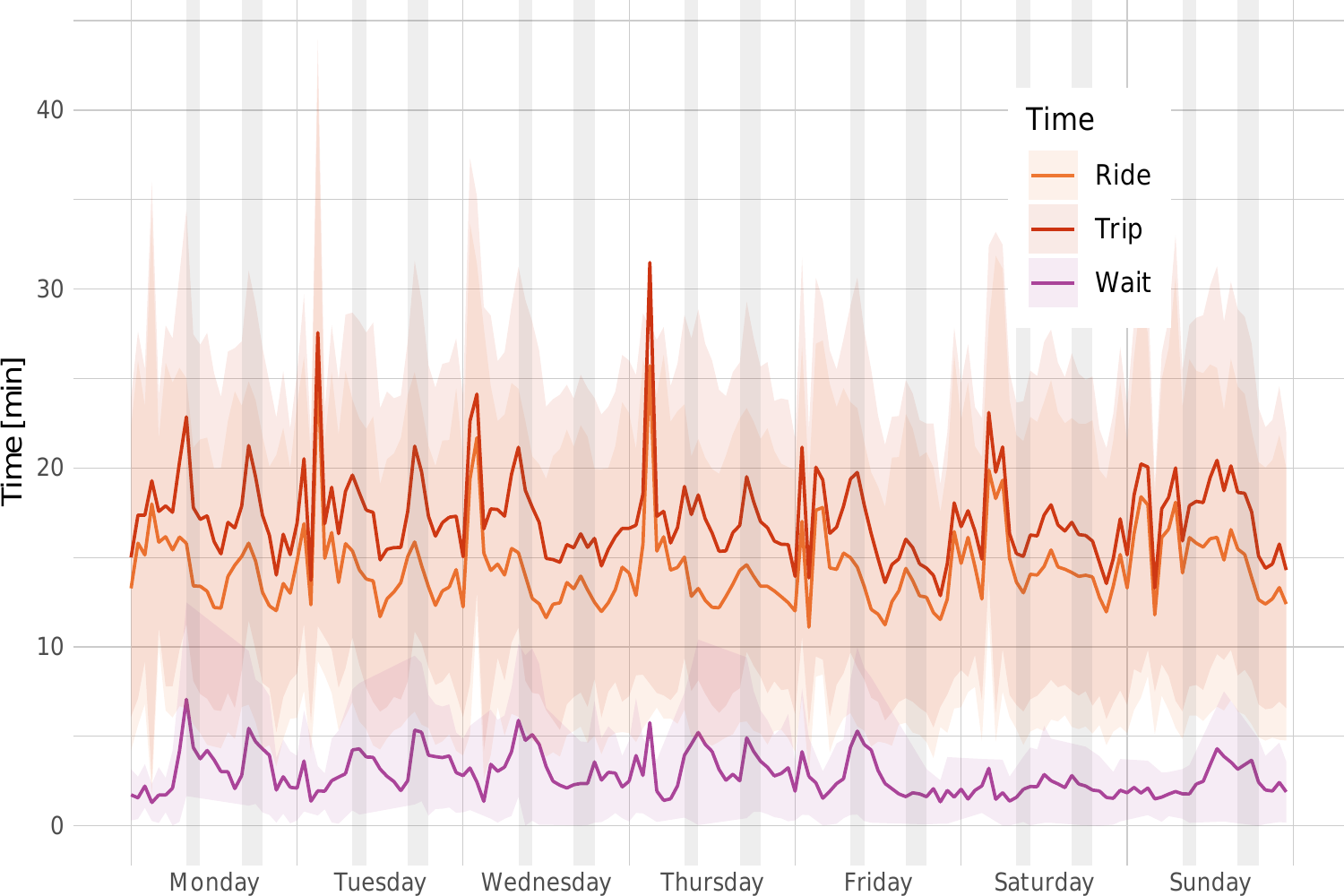}
    \label{fig:AU-nominal-4}}
    \caption{Autonomous system nominal state: a) Timeline of the number of users performing each activity; b) Timeline of the number of users performing each activity; c) Timeline of the number of served and unserved trips; d) Timeline of the mean duration of different user activities. Shaded areas indicate morning and evening periods of peak demand.
    }
    \label{fig:AU-nominal}
\end{figure*}

The performance metrics for the autonomous system can be found in Table \ref{tb:aut-nominal-results} and their evolution over time in Figure \ref{fig:AU-nominal}. In the absence of rebalancing, the average trip time is 17.62 minutes, which includes an average of 3.53 minutes of wait time for the autonomous bike to arrive. With ideal rebalancing, the average trip time is 14.1 minutes with no wait time. The wait times vary more than the walk time in station-based and dockless systems, and  they seem to have a higher dependency on the level of the demand at each moment.

With ideal rebalancing, all trips are served, and, even without rebalancing, the system can serve 99.46\% of the demand, with unserved trips happening mainly during rush hours.

In the scenario without rebalancing, the average number of trips per bike and day is 8.84, and 97.02\% of the bikes are used throughout the week. With ideal rebalancing, bicycles do 8.88 tips per bike and day, and the percentage of used bikes is 96.43\%. 

With no rebalancing, vehicles are used for 8.64\% of the time, they spend 2.18\% of the time picking up people, 1.14\% charging or going for a charge, and 88.04\% idling. In terms of traveled distance, 83.45\% of the kilometers traveled are in use, 16.48\% during pickup, and 0.07\% going for a charge. In the case of ideal rebalancing, vehicles are in use for 8.71\% of the time, charging or going for a charge for 1.15\% of the time, and idling for 90.14\% of the time. The distance traveled in use is 99.92\%, and 0.08\% for a charge. As can be observed in Figure \ref{fig:AU-nominal-2}, the percentage of bicycles that are being used simultaneously is much higher than for the other two systems.

\begin{table}[!htb]
\centering
\scalebox{\tablescale}{
\begin{tabular}{l|lrr}
\toprule
\multicolumn{4}{c}{\textbf{Autonomous system nominal metrics}} \\ \midrule
\multicolumn{2}{c}{\textbf{Metric}} & \textbf{NR} & \textbf{IR} \\ \midrule
\multicolumn{2}{l}{User demand} & 62192     & 62192     \\ 
\multicolumn{2}{l}{Average trip time} & 17.62 min   & 14.11 min \\ 
         & Wait time                          & 3.53 min  & 0 min     \\ 
         & Wait time \textgreater 10 min       & 6.16 \%    & 0.00 \%    \\ 
         & Wait time \textgreater 15 min      & 1.02 \%    & 0.00 \%    \\ 
         & Ride time                          & 14.09 min & 14.11 min \\ 
\multicolumn{2}{l}{Served trips}            & 99.46 \%   & 100 \%     \\ 
\multicolumn{2}{l}{Unserved trips}          & 0.54 \%    & 0 \%       \\ 
\multicolumn{2}{l}{Bikes used}              & 97.20 \%   & 96.43 \%   \\ 
\multicolumn{2}{l}{Trips/bike/day}          & 8.84      & 8.88      \\ 
\multicolumn{2}{l}{Total charges}           & 504       & 500       \\ 
\multicolumn{2}{l}{Total charges   per day} & 72        & 71        \\ 
\multicolumn{2}{l}{Total v.k.t}             & 177488 km & 149310 km \\ 
\multicolumn{2}{l}{v.k.t. per   bike}       & 178 km    & 149 km    \\ 
         & v.k.t. in use                      & 83.45 \%      & 99.92 \%   \\ 
         & v.k.t. pick-up                     & 16.48 \%   & 0.00 \%    \\ 
         & v.k.t. rebalancing                 & 0.00 \%    & 0.00 \%    \\ 
         & v.k.t. charge                      & 0.07 \%    & 0.08 \%    \\ 
\multicolumn{2}{l}{Total time}              & 7 days    & 7 days    \\ 
         & Time in use                        & 8.64 \%    & 8.71 \%    \\ 
         & Time pick-up                       & 2.18 \%    & 0.00 \%    \\ 
         & Time rebalancing                   & 0.00 \%    & 0.00 \%    \\ 
         & Time charge                        & 1.14 \%    & 1.15 \%    \\ 
         & Time idling                       & 88.04 \%   & 90.14 \%   \\ \bottomrule
\end{tabular}}
    \caption{Performance metrics for autonomous system in nominal state with no rebalancing (NR) and with ideal rebalancing (IR). }
    \label{tb:aut-nominal-results}
\end{table}

\subsection{Comparative analysis}\label{comparative-analysis}

The behavior of the three types of BSS considered was studied in three different scenarios. First, the nominal state performance of the systems is compared. Second, we analyze the fleet size required to provide a certain level of service, which is defined by a minimum percentage of served demand of 99\%. Third, we compare the performance of the three systems with identical fleet sizes. Specifically, the studied fleet sizes were of 2000 and 3000 bicycles. For all cases, the comparative results are included in the two rightmost columns of the tables.

\subsubsection{Nominal state} \label{comparative-nominal}
As can be observed in Table \ref{tb:comparative-nominal}, with a fleet size of just 1000 bikes, even with no rebalancing, most performance metrics improve for the autonomous system compared to a station-based system with 3500 bikes and a dockless system with 8000 bikes. 

Regarding the user experience, the autonomous system can serve between 0.47\% and 1.01\% more of the demand than the station-based and dockless systems. This increase in served trips is a significant advantage because it represents the reliability of the system, and reliability is a very relevant factor in the mode choice of commuting trips \cite{bhat2006impact}.
 
The wait time in the case of no rebalancing, which represents the upper bound to this value, is 3.53 minutes, so it is expected that the real-world wait time would be below that value. This wait time is well below the 5.91 minutes of walk in station-based systems. Compared to the walk time in the dockless system, the wait time in the autonomous mode could be longer if there was no rebalancing in the autonomous system. However, it can also be lower depending on the performance of the predictive rebalancing. In addition, it must be taken into account that, while waiting for the autonomous bike to arrive, users will be able to use their time for other purposes and, therefore, it seems likely that the tolerance to waiting will be higher than it is to walking. Increased reliability and convenience could incentivize more people to use bicycles as their primary mode of transportation for commuting.

In terms of efficiency, the levels of use of the bicycles are greatly improved in the autonomous system; autonomous bicycles make an average of 8.84-8.88 trips per day, while the station-based bikes do 2.51 the dockless bicycles 1.10. On average, the autonomous bikes are used for 8.64 - 8.71\% of the time, as opposed to 2.44\% in station-based systems and 1.07\% in dockless systems. Finally, the percentage of the fleet used during a week is 96.43 - 97.20\%  while in the case of the station-based system is 91.59\% and 70.93\% in the case of the dockless system.

Overall, it can be concluded that with a fleet size three and a half times smaller than the station-based system and eight times smaller than the dockless system, autonomous bicycles are be able to provide the same - or better - service. These results indicate that even if the cost of the autonomous bicycles is higher than that of the current systems, the smaller needed fleet size can compensate for this cost, and the total cost of an autonomous bicycle-sharing system can even be below the cost of the current station-based and dockless systems \cite{sanchez2020autonomous}.

\begin{table*}[]
\centering
\scalebox{\tablescale}{
\begin{tabular}{lrrR{1.5cm}lrr}
\toprule
\textbf{Metric} & \textbf{SB} & \textbf{DL} & \multicolumn{2}{c}{\textbf{Autonomous}} & \textbf{Aut. vs.\ SB} & \textbf{Aut. vs.\ DL} \\ \midrule
\multirow{2}{*}{{Number of bikes {[}-{]}}} & \multirow{2}{*}{3500} & \multirow{2}{*}{8000} & NR & 1000 & -71.43 \% & -87.50 \% \\ 
 &  &  & IR & 1000 & -71.43 \% & -87.50 \% \\ \midrule
\multirow{2}{*}{{Average trip time {[}min{]}}} & \multirow{2}{*}{19.87} & \multirow{2}{*}{15.14} & NR & 17.62 & -11.35 \% & 16.34 \% \\ 
 &  &  & IR & 14.11 & -28.99 \% & -6.62 \% \\ \midrule
\multirow{2}{*}{{Wait or walk time {[}min{]}}} & \multirow{2}{*}{5.92} & \multirow{2}{*}{1.07} & NR & 3.53 & -40.31 \% & 229.88 \% \\ 
 &  &  & IR & 0.00 & -100.00 \% & -100.00 \% \\ \midrule
\multirow{2}{*}{{Wait/walk \textgreater 10 min {[}\%{]}}} & \multirow{2}{*}{1.33 \%} & \multirow{2}{*}{0.15 \%} & NR & 6.16 \% & 361.46 \% & 4016.40 \% \\ 
 &  &  & IR & 0.00 \% & -100.00 \% & -100.00 \% \\ \midrule
\multirow{2}{*}{{Wait/walk \textgreater 15 min {[}\%{]}}} & \multirow{2}{*}{0.25 \%} & \multirow{2}{*}{0.06 \%} & NR & 1.02 \% & 306.21 \% & 1695.75 \% \\ 
 &  &  & IR & 0.00 \% & -100.00 \% & -100.00 \% \\ \midrule
\multirow{2}{*}{{Ride time {[}min{]}}} & \multirow{2}{*}{13.96} & \multirow{2}{*}{14.07} & NR & 14.09 & 0.93 \% & 0.09 \% \\ 
 &  &  & IR & 14.11 & 1.12 \% & 0.28 \% \\ \midrule
\multirow{2}{*}{{Served trips {[}\%{]}}} & \multirow{2}{*}{99.00 \%} & \multirow{2}{*}{99.02 \%} & NR & 99.46 \% & 0.47 \% & 0.45 \% \\ 
 &  &  & IR & 100.00 \% & 1.01 \% & 0.99 \% \\ \midrule
\multirow{2}{*}{{Unserved trips {[}\%{]}}} & \multirow{2}{*}{1.00 \%} & \multirow{2}{*}{0.98 \%} & NR & 0.54 \% & -46.45 \% & -45.39 \% \\ 
 &  &  & IR & 0.00 \% & -100.00 \% & -100.00 \% \\ \midrule
\multirow{2}{*}{{Bikes used {[}\%{]}}} & \multirow{2}{*}{91.59 \%} & \multirow{2}{*}{70.93 \%} & NR & 97.20 \% & 6.13 \% & 37.03 \% \\ 
 &  &  & IR & 96.43 \% & 5.29 \% & 35.95 \% \\ \midrule
\multirow{2}{*}{{Trips/bike/day {[}-{]}}} & \multirow{2}{*}{2.51} & \multirow{2}{*}{1.10} & NR & 8.84 & 251.64 \% & 703.60 \% \\ 
 &  &  & IR & 8.88 & 253.54 \% & 707.93 \% \\ \midrule
\multirow{2}{*}{{Total v.k.t. {[}km{]}}} & \multirow{2}{*}{146068} & \multirow{2}{*}{147327} & NR & 177488 & 21.51 \% & 20.47 \% \\ 
 &  &  & IR & 149310 & 2.22 \% & 1.35 \% \\ \midrule
\multirow{2}{*}{{v.k.t. per bike {[}km{]}}} & \multirow{2}{*}{41.73} & \multirow{2}{*}{18.42} & NR & 177.49 & 325.29 \% & 863.78 \% \\ 
 &  &  & IR & 149.31 & 257.77 \% & 710.77 \% \\ \midrule
\multirow{2}{*}{{Time in use {[}\%{]}}} & \multirow{2}{*}{2.44 \%} & \multirow{2}{*}{1.07 \%} & NR & 8.64 \% & 254.95 \% & 704.38 \% \\ 
 &  &  & IR & 8.71 \% & 257.56 \% & 710.77 \% \\ \midrule
\multirow{2}{*}{{Time idling {[}\%{]}}} & \multirow{2}{*}{97.56 \%} & \multirow{2}{*}{98.93 \%} & NR & 88.04 \% & -9.76 \% & -11.00 \% \\ 
 &  &  & IR & 90.14 \% & -7.61 \% & -8.88 \% \\ \bottomrule
\end{tabular}}
    \caption{Comparative of performance metrics in nominal state for the station-based, dockless and autonomous system (NR: No rebalancing, IR: Ideal rebalancing).}
    \label{tb:comparative-nominal}

\end{table*}

\subsubsection{Same level of service}\label{level-of-service}

As previously stated, the fleet sizes for the nominal state were determined based on available literature. However, from the point of view of a fleet operator, there is a trade-off between the fleet size and the level of service provided. Consequently, it seems reasonable to assume that the fleet sizing will be determined by the level of service that a specific operator might want to provide. Therefore, for this study, it was assumed that the desired level of service is 99\% of served trips.

The comparison of the fleet size and performance metrics of the three systems providing a percentage of served trips above 99\% can be found in Table \ref{tb:comparative-level-of-service}. The minimum fleet sizes required to give 99\% of served trips are 3500 bikes in the case of the station-based system, 8000 bicycles in the dockless system, and between 500 and 1000 for the autonomous system, depending on the performance of the rebalancing system. As these fleet sizes match with the nominal case, except in the case of ideal rebalancing, most of the results are the same as those analyzed in Section \ref{comparative-nominal}. 

These results prove that the fleet sizes assumed for the nominal case and their sizes relative to the other systems are coherent with a scenario in which the fleets would be sized to serve a certain percentage of the demand, in addition to being supported by literature.  Moreover, it also demonstrates that the autonomous system might further reduce fleet size depending on the rebalancing performance. 

\begin{table*}[]
\centering
\scalebox{\tablescale}{
\begin{tabular}{lrrR{1.5cm}lrr}
\toprule
\textbf{Metric} & \textbf{SB} & \textbf{DL} & \multicolumn{2}{c}{\textbf{Autonomous}} & \textbf{Aut. vs.\ SB} & \textbf{Aut. vs.\ DL} \\ \midrule
\multirow{2}{*}{{Number of bikes {[}-{]}}} & \multirow{2}{*}{3500} & \multirow{2}{*}{8000} & NR & 1000 & -71.43 \% & -87.50 \% \\ 
 &  &  & IR & 500 & -85.71 \% & -93.75 \% \\ \midrule
\multirow{2}{*}{{Average trip time {[}min{]}}} & \multirow{2}{*}{19.90} & \multirow{2}{*}{15.14} & NR & 17.62 & -11.28 \% & 16.34 \% \\ 
 &  &  & IR & 14.11 & -29.1 \% & -6.81 \% \\ \midrule
\multirow{2}{*}{{Wait or walk time {[}min{]}}} & \multirow{2}{*}{5.93} & \multirow{2}{*}{1.07} & NR & 3.53 & -40.39 \% & 229.88 \% \\ 
 &  &  & IR & 0.00 & -100.00 \% & -100.00 \% \\ \midrule
\multirow{2}{*}{{Wait/walk \textgreater 10 min {[}\%{]}}} & \multirow{2}{*}{1.37 \%} & \multirow{2}{*}{0.15 \%} & NR & 6.16 \% & 348.54 \% & 4016.40 \% \\ 
 &  &  & IR & 0.00 \% & -100.00 \% & -100.00 \% \\ \midrule
\multirow{2}{*}{{Wait/walk \textgreater 15 min {[}\%{]}}} & \multirow{2}{*}{0.26 \%} & \multirow{2}{*}{0.06 \%} & NR & 1.02 \% & 292.88 \% & 1695.75 \% \\ 
 &  &  & IR & 0.00 \% & -100.00 \% & -100.00 \% \\ \midrule
\multirow{2}{*}{{Ride time {[}min{]}}} & \multirow{2}{*}{13.98} & \multirow{2}{*}{14.07} & NR & 14.09 & 0.78 \% & 0.09 \% \\ 
 &  &  & IR & 14.11 & 0.97 \% & 0.28 \% \\ \midrule
\multirow{2}{*}{{Served trips {[}\%{]}}} & \multirow{2}{*}{99.03 \%} & \multirow{2}{*}{99.02 \%} & NR & 99.46 \% & 0.44 \% & 0.45 \% \\ 
 &  &  & IR & 100.00 \% & 0.98 \% & 0.99 \% \\ \midrule
\multirow{2}{*}{{Unserved trips {[}\%{]}}} & \multirow{2}{*}{0.97 \%} & \multirow{2}{*}{0.98 \%} & NR & 0.54 \% & -44.78 \% & -45.39 \% \\ 
 &  &  & IR & 0.00 \% & -100.00 \% & -100.00 \% \\ \midrule
\multirow{2}{*}{{Bikes used {[}\%{]}}} & \multirow{2}{*}{91.56 \%} & \multirow{2}{*}{70.93 \%} & NR & 97.20 \% & 6.16 \% & 37.03 \% \\ 
 &  &  & IR & 98.93 \% & 8.06 \% & 39.47 \% \\ \midrule
\multirow{2}{*}{{Trips/bike/day {[}-{]}}} & \multirow{2}{*}{2.51} & \multirow{2}{*}{1.10} & NR & 8.84 & 251.54 \% & 703.60 \% \\ 
 &  &  & IR & 17.77 & 606.84 \% & 1515.81 \% \\ \midrule
\multirow{2}{*}{{Total v.k.t. {[}km{]}}} & \multirow{2}{*}{146328} & \multirow{2}{*}{147327} & NR & 177487.64 & 21.29 \% & 20.47 \% \\ 
 &  &  & IR & 149382 & 2.09 \% & 1.39 \% \\ \midrule
\multirow{2}{*}{{v.k.t. per bike {[}km{]}}} & \multirow{2}{*}{41.81} & \multirow{2}{*}{18.42} & NR & 177.49 & 324.53 \% & 863.78 \% \\ 
 &  &  & IR & 298.76 & 614.61 \% & 1522.32 \% \\ \midrule
\multirow{2}{*}{{Time in use {[}\%{]}}} & \multirow{2}{*}{2.44 \%} & \multirow{2}{*}{1.07 \%} & NR & 8.64 \% & 254.29 \% & 704.31 \% \\ 
 &  &  & IR & 17.41 \% & 613.71 \% & 1520.27 \% \\ \midrule
\multirow{2}{*}{{Time idling {[}\%{]}}} & \multirow{2}{*}{97.56 \%} & \multirow{2}{*}{98.93 \%} & NR & 88.04 \% & -9.76 \% & -11.00 \% \\ 
 &  &  & IR & 78.71 \% & -19.32 \% & -20.44 \% \\ \bottomrule
\end{tabular}}
    \caption{Comparison of fleet size performance metrics of the station-based, dockless and autonomous system for providing a minimum level of service of 99 \% of served trips (NR= No rebalancing, IR= Ideal rebalancing).}
    \label{tb:comparative-level-of-service}
\end{table*}

\subsubsection{Same fleet size}\label{fleet-size}

Given that the fleet size has a significant impact on the performance of a bicycle-sharing system, this last scenario aims to compare the station-based, dockless, and autonomous systems with the same number of bikes. More specifically, two different fleet sizes were considered: 2000 bikes and 3000 bikes. 

As shown in Table \ref{tb:comparative-2000}, with 2000 bikes the shortest wait time corresponds to the autonomous system. The wait time is lower than the walk time in the station-based system by a greater percentage than in the nominal state. In this case, the wait time in the autonomous system is also less than the walk time for the dockless system.

The autonomous system also has the highest percentage of served trips, with a larger difference than in the nominal state. In this scenario, it is clear that a dockless system requires a large number of bicycles to perform properly. With 2000 bikes, the walk time is doubled compared to the nominal state, and the number of served trips falls to 87.41\%, with 12.59\% of the trips left unserved. 

The percentage of used bikes is higher than for the dockless systems but lower than for the station-based system because, in the autonomous system, there are more bikes than needed. However, the autonomous system is still the one with the highest number of trips per bike and day and the highest percentage of time in use, albeit with a minor difference from the nominal case.

\begin{table*}[]
\centering
\scalebox{\tablescale}{
\begin{tabular}{lrrR{1.5cm}lrr}
\toprule
\textbf{Metric} & \textbf{SB} & \textbf{DL} & \multicolumn{2}{c}{\textbf{Autonomous}} & \textbf{Aut. vs.\ SB} & \textbf{Aut. vs.\ DL} \\ \midrule
\multirow{2}{*}{{Number of bikes {[}-{]}}} & \multirow{2}{*}{2000} & \multirow{2}{*}{2000} & NR & 2000 & 0.00 \% & 0.00 \% \\ 
 &  &  & IR & 2000 & 0.00 \% & 0.00 \% \\ \midrule
\multirow{2}{*}{{Average trip time {[}min{]}}} & \multirow{2}{*}{19.61} & \multirow{2}{*}{16.15} & NR & 16.12 & -17.80 \% & -0.22 \% \\ 
 &  &  & IR & 14.11 & -28.03 \% & -12.64 \% \\ \midrule
\multirow{2}{*}{{Wait or walk time {[}min{]}}} & \multirow{2}{*}{5.87} & \multirow{2}{*}{2.07} & NR & 2.01 & -65.79 \% & -2.93 \% \\ 
 &  &  & IR & 0.00 & -100.00 \% & -100.00 \% \\ \midrule
\multirow{2}{*}{{Wait/walk \textgreater 10 min {[}\%{]}}} & \multirow{2}{*}{1.16 \%} & \multirow{2}{*}{0.73 \%} & NR & 1.40 \% & 21.49 \% & 91.19 \% \\ 
 &  &  & IR & 0.00 \% & -100.00 \% & -100.00 \% \\ \midrule
\multirow{2}{*}{{Wait/walk \textgreater 15 min {[}\%{]}}} & \multirow{2}{*}{0.25 \%} & \multirow{2}{*}{0.14 \%} & NR & 0.22 \% & -10.83 \% & 58.43 \% \\ 
 &  &  & IR & 0.00 \% & -100.00 \% & -100.00 \% \\ \midrule
\multirow{2}{*}{{Ride time {[}min{]}}} & \multirow{2}{*}{13.74} & \multirow{2}{*}{14.08} & NR & 14.11 & 2.71 \% & 0.18 \% \\ 
 &  &  & IR & 14.11 & 2.74 \% & 0.20 \% \\ \midrule
\multirow{2}{*}{{Served trips {[}\%{]}}} & \multirow{2}{*}{97.68 \%} & \multirow{2}{*}{87.41 \%} & NR & 99.96 \% & 2.34 \% & 14.36 \% \\ 
 &  &  & IR & 100.00 \% & 2.38 \% & 14.40 \% \\ \midrule
\multirow{2}{*}{{Unserved trips {[}\%{]}}} & \multirow{2}{*}{2.32 \%} & \multirow{2}{*}{12.59 \%} & NR & 0.04 \% & -98.23 \% & -99.67 \% \\ 
 &  &  & IR & 0.00 \% & -100.00 \% & -100.00 \% \\ \midrule
\multirow{2}{*}{{Bikes used {[}\%{]}}} & \multirow{2}{*}{95.29 \%} & \multirow{2}{*}{91.28 \%} & NR & 92.16 \% & -3.28 \% & 0.96 \% \\ 
 &  &  & IR & 91.95 \% & -3.51 \% & 0.73 \% \\ \midrule
\multirow{2}{*}{{Trips/bike/day {[}-{]}}} & \multirow{2}{*}{4.34} & \multirow{2}{*}{3.88} & NR & 4.44 & 2.34 \% & 14.36 \% \\ 
 &  &  & IR & 4.44 & 2.38 \% & 14.40 \% \\ \midrule
\multirow{2}{*}{{Total v.k.t. {[}km{]}}} & \multirow{2}{*}{141853} & \multirow{2}{*}{130151} & NR & 165878 & 16.94 \% & 27.45 \% \\ 
 &  &  & IR & 149257 & 5.22 \% & 14.68 \% \\ \midrule
\multirow{2}{*}{{v.k.t. per bike {[}km{]}}} & \multirow{2}{*}{70.93} & \multirow{2}{*}{65.08} & NR & 82.94 & 16.94 \% & 27.45 \% \\ 
 &  &  & IR & 74.63 & 5.22 \% & 14.68 \% \\ \midrule
\multirow{2}{*}{{Time in use {[}\%{]}}} & \multirow{2}{*}{4.14 \%} & \multirow{2}{*}{3.80 \%} & NR & 4.35 \% & 5.11 \% & 14.56 \% \\ 
 &  &  & IR & 4.35 \% & 5.18 \% & 14.63 \% \\ \midrule
\multirow{2}{*}{{Time idling {[}\%{]}}} & \multirow{2}{*}{95.86 \%} & \multirow{2}{*}{96.20 \%} & NR & 94.72 \% & -1.19 \% & -1.54 \% \\ 
 &  &  & IR & 95.34 \% & -0.54 \% & -0.89 \% \\ \bottomrule
\end{tabular}}
    \caption{Comparison of fleet size performance metrics of the station-based, dockless and autonomous system with a fleet size of 2000 bikes (NR:No rebalancing, IR: Ideal rebalancing).}
    \label{tb:comparative-2000}
\end{table*}

Analyzing the results with a fleet size of 3000 bikes, as shown in Table \ref{tb:comparative-3000}, it can be observed how a bigger fleet size increases the level of improvement of the autonomous system over the other two systems, especially in terms of user experience. The efficiency-related performance metrics are still better than in the station-based and dockless systems, but less so than with 2000 bicycles or in the nominal state. This minor improvement is the result of having more bikes for the same demand, which reduces efficiency.

Based on these two scenarios, it can be concluded that, for the same fleet size, the autonomous system outperforms the station-based and dockless systems in terms of user experience and efficiency. An increase in the number of bikes improves the user experience in the autonomous system, and it stands further from the other two systems than in the nominal state. However, having more bicycles reduces efficiency-related metrics such as the trips per bike and day and the percentage of time in use. 

\begin{table*}[!htb]
\centering
\scalebox{\tablescale}{
\begin{tabular}{lrrR{1.5cm}lrr}
\toprule
\textbf{Metric} & \textbf{SB} & \textbf{DL} & \multicolumn{2}{c}{\textbf{Autonomous}} & \textbf{Aut. vs.\ SB} & \textbf{Aut. vs.\ DL} \\ \midrule
\multirow{2}{*}{{Number of bikes {[}-{]}}} & \multirow{2}{*}{3000} & \multirow{2}{*}{3000} & NR & 3000 & 0.00 \% & 0.00 \% \\ 
 &  &  & IR & 3000 & 0.00 \% & 0.00 \% \\ \midrule
\multirow{2}{*}{{Average trip time {[}min{]}}} & \multirow{2}{*}{19.74} & \multirow{2}{*}{15.77} & NR & 15.22 & -21.23 \% & -1.41 \% \\ 
 &  &  & IR & 14.11 & -28.52 \% & -10.53 \% \\ \midrule
\multirow{2}{*}{{Wait or walk time {[}min{]}}} & \multirow{2}{*}{5.89} & \multirow{2}{*}{1.68} & NR & 1.44 & -75.57 \% & -14.47 \% \\ 
 &  &  & IR & 0.00 & -100.00 \% & -100.00 \% \\ \midrule
\multirow{2}{*}{{Wait/walk \textgreater 10 min {[}\%{]}}} & \multirow{2}{*}{1.26 \%} & \multirow{2}{*}{0.40 \%} & NR & 0.64 \% & -49.42 \% & 59.54 \% \\ 
 &  &  & IR & 0.00 \% & -100.00 \% & -100.00 \% \\ \midrule
\multirow{2}{*}{{Wait/walk \textgreater 15 min {[}\%{]}}} & \multirow{2}{*}{0.25 \%} & \multirow{2}{*}{0.09 \%} & NR & 0.11 \% & -57.42 \% & 22.69 \% \\ 
 &  &  & IR & 0.00 \% & -100.00 \% & -100.00 \% \\ \midrule
\multirow{2}{*}{{Ride time {[}min{]}}} & \multirow{2}{*}{13.85} & \multirow{2}{*}{14.09} & NR & 14.11 & 1.86 \% & 0.15 \% \\ 
 &  &  & IR & 14.11 & 1.87 \% & 0.15 \% \\ \midrule
\multirow{2}{*}{{Served trips {[}\%{]}}} & \multirow{2}{*}{98.71 \%} & \multirow{2}{*}{94.08 \%} & NR & 99.99 \% & 1.30 \% & 6.28 \% \\ 
 &  &  & IR & 100.00 \% & 1.31 \% & 6.29 \% \\ \midrule
\multirow{2}{*}{{Unserved trips {[}\%{]}}} & \multirow{2}{*}{1.29 \%} & \multirow{2}{*}{5.92 \%} & NR & 0.01 \% & -99.60 \% & -99.91 \% \\ 
 &  &  & IR & 0.00 \% & -100.00 \% & -100.00 \% \\ \midrule
\multirow{2}{*}{{Bikes used {[}\%{]}}} & \multirow{2}{*}{92.25 \%} & \multirow{2}{*}{88.27 \%} & NR & 88.31 \% & -4.28 \% & 0.05 \% \\ 
 &  &  & IR & 88.27 \% & -4.32 \% & 0.00 \% \\ \midrule
\multirow{2}{*}{{Trips/bike/day {[}-{]}}} & \multirow{2}{*}{2.92} & \multirow{2}{*}{2.79} & NR & 2.96 & 1.30 \% & 6.28 \% \\ 
 &  &  & IR & 2.96 & 1.31 \% & 6.29 \% \\ \midrule
\multirow{2}{*}{{Total v.k.t. {[}km{]}}} & \multirow{2}{*}{144571} & \multirow{2}{*}{140161} & NR & 161195 & 11.50 \% & 15.01 \% \\ 
 &  &  & IR & 149244 & 3.23 \% & 6.48 \% \\ \midrule
\multirow{2}{*}{{v.k.t. per bike {[}km{]}}} & \multirow{2}{*}{48.19} & \multirow{2}{*}{46.72} & NR & 53.73 & 11.50 \% & 15.01 \% \\ 
 &  &  & IR & 49.75 & 3.23 \% & 6.48 \% \\ \midrule
\multirow{2}{*}{{Time in use {[}\%{]}}} & \multirow{2}{*}{2.81 \%} & \multirow{2}{*}{2.73 \%} & NR & 2.90 \% & 3.19 \% & 6.44 \% \\ 
 &  &  & IR & 2.90 \% & 3.20 \% & 6.45 \% \\ \midrule
\multirow{2}{*}{{Time idling {[}\%{]}}} & \multirow{2}{*}{97.19 \%} & \multirow{2}{*}{97.27 \%} & NR & 96.67 \% & -0.54 \% & -0.62 \% \\ 
 &  &  & IR & 96.94 \% & -0.25 \% & -0.34 \% \\ \bottomrule
\end{tabular}}
    \caption{Comparison of fleet size performance metrics of the station-based, dockless and autonomous system with a fleet size of 3000 bikes (NR: No rebalancing, IR: Ideal rebalancing).}
    \label{tb:comparative-3000}
\end{table*}

\subsection{Autonomous system with predictive rebalancing}

While the previous section considered two extreme rebalancing scenarios (i.e., no rebalancing and ideal rebalancing) that provide an upper- and lower-bounds to the performance metrics, this scenario aims to provide more precise values by describing the performance of an autonomous system with a rebalancing model based on demand prediction and routing optimization. The performance metrics of the system with predictive rebalancing are compared to the cases of no rebalancing and ideal rebalancing for the same scenarios than in Section \ref{comparative-analysis}.

\subsubsection {Nominal}

As it can be observed in Table \ref{tb:predictive-nominal}, in the nominal state, the predictive rebalancing does not improve the performance over the scenario of no rebalancing. The wait time slightly increase, and the number of served trips are reduced by 1\%. In this scenario, the additional kilometers traveled by the bikes while rebalancing are not compensated by an improvement in performance. 

The reason for this is that, because the fleet size is only 1000 bikes in the nominal state, during rush hour, up to 60\% of the bicycles are busy even in the absence of rebalancing \ref{fig:AU-nominal-2}. Therefore, if bikes are being rebalanced, there are even fewer available bicycles at that critical time of the day, which deteriorates system performance. However, as evidenced by the examples in the following sections, larger fleet sizes improve performance. In addition, as shown in \ref{appendix-A}, higher autonomous driving speeds also lead to an improvement in the performance of the rebalancing system.

From this study, it can be concluded that there is the risk that a predictive rebalancing system might worsen the performance, especially if the fleet size is closely adjusted to the demand and the autonomous driving speed is relatively low. However, it must be kept in mind that these results highly depend on the demand prediction and routing models. Theoretically, the results of an effective predictive routing system could fall anywhere between the performance with no rebalancing and the ideal rebalancing case. Therefore, rebalancing systems must be appropriately designed considering these conclusions. 

\begin{table*}[]
\centering
\scalebox{\tablescale}{
\begin{tabular}{l|lrrrrr}
\toprule
\multicolumn{2}{l}{\textbf{Metric}} & \textbf{NR} & \textbf{PR} & \textbf{IR} & \multicolumn{1}{l}{\textbf{PR vs.\ NR}} & \multicolumn{1}{l}{\textbf{PR vs.\ IR}} \\ \midrule
\multicolumn{2}{l}{User demand {[}-{]}} & 62192 & 62192 & 62192 & 0.00 \% & 0.00 \% \\ 
\multicolumn{2}{l}{Total trip time   {[}min{]}} & 17.62 & 17.88 & 14.11 & 1.47 \% & 26.68 \% \\ 
 & Wait time {[}min{]} & 3.53 & 3.79 & 0.00 & 7.18 \% & - \\ 
 & Wait time \textgreater 10min   {[}\%{]} & 6.16 \% & 6.95 \% & 0.00 \% & 12.98 \% & - \\ 
 & Wait time \textgreater 15 min   {[}\%{]} & 1.02 \% & 1.51 \% & 0.00 \% & 48.46 \% & - \\ 
 & Ride time {[}min{]} & 14.09 & 14.09 & 14.11 & 0.03 \% & -0.15 \% \\ 
\multicolumn{2}{l}{Served trips  {[}\%{]}} & 99.46 \% & 98.55 \% & 100 \% & -0.92 \% & -1.45 \% \\ 
\multicolumn{2}{l}{Unserved trips  {[}\%{]}} & 0.54 \% & 1.45 \% & 0.00 \% & 170.16 \% & - \\ 
\multicolumn{2}{l}{Bikes used {[}\%{]}} & 97.20 \% & 98.80 \% & 96.43 \% & 1.65 \% & 2.45 \% \\ 
\multicolumn{2}{l}{Trips/bike/day {[}-{]}} & 8.84 & 8.76 & 8.88 & -0.92 \% & -1.45 \% \\ 
\multicolumn{2}{l}{Total charges {[}-{]}} & 504 & 967 & 500 & 91.77 \% & 93.40 \% \\ 
\multicolumn{2}{l}{Total charges per day {[}-{]}} & 72 & 138 & 71 & 91.77 \% & 93.40 \% \\ 
\multicolumn{2}{l}{Total v.k.t  {[}km{]}} & 177488 & 203724 & 149310 & 14.78 \% & 36.44 \% \\ 
\multicolumn{2}{l}{v.k.t. per bike {[}km{]}} & 177.49 & 203.72 & 149.31 & 14.78 \% & 36.44 \% \\ 
 & v.k.t. in use {[}\%{]} & 83.45 \% & 72.06 \% & 99.92 \% & -13.65 \% & -27.88 \% \\ 
 & v.k.t. pick-up {[}\%{]} & 16.48 \% & 15.35 \% & 0.00 \% & -6.88 \% & - \\ 
 & v.k.t. rebalancing   {[}\%{]} & 0.00 \% & 12.43 \% & 0.00 \% & - & - \\ 
 & v.k.t. charge {[}\%{]} & 0.07 \% & 0.16 \% & 0.08 \% & 144.01 \% & 110.89 \% \\ 
 
\multicolumn{2}{l}{Total time {[}days{]}} & 7 days & 7 days & 7 days & 0 \% & 0 \% \\ 
& Time in use {[}\%{]} & 8.64 \% & 8.57 \% & 8.71 \% & -0.87 \% & -1.59 \% \\ 
& Time pick-up {[}\%{]} & 2.18 \% & 2.33 \% & 0.00 \% & 6.90 \% & - \\
& Time rebalancing {[}\%{]} & 0.00 \% & 1.88 \% & 0.00 \% & - & - \\
& Time charge {[}\%{]} & 1.14 \% & 2.27 \% & 1.15 \% & 99.43 \% & 97.31 \% \\ 
& Time idling {[}\%{]} & 88.04 \% & 84.95 \% & 90.14 \% & -3.51 \% & -5.76 \% \\ \bottomrule
\end{tabular}}
    \caption{Performance metrics for the autonomous system with predictive rebalancing (PR) compared to no rebalancing (NR) and ideal rebalancing (IR) in nominal state}
    \label{tb:predictive-nominal}
\end{table*}

\subsubsection{Level of service}

The fleet sizes required to provide a minimum level of service of 99 \% of served trips can be observed in Table \ref{tb:predictive-99}. In this case, the fleet size needed to rebalance the bicycles according demand prediction is 50 \% greater than if no rebalancing is done. 

With this fleet size, predictive rebalancing reduces wait time and increases the percentage of served trips compared to the scenario with no rebalancing, adding 3.27\% of total vehicle kilometers traveled. Bicycles spend a smaller fraction of their time in pick-up and more time rebalancing or idling. Because the fleet is larger, bikes make fewer trips per bike and day, and a smaller portion of the bikes are used during the week. 

The results in this section confirm that to provide a good level of service --- depending on the fleet size, demand, and characteristics of the predictive rebalancing model ---  a system with predictive bicycle rebalancing may require bicycles so that, even if some bicycles are rebalancing during rush hour, there are still enough bikes available for users. 

\begin{table*}[]
\centering
\scalebox{\tablescale}{
\begin{tabular}{l|lrrrrr}
\toprule
\multicolumn{2}{l}{\textbf{Metric}} & \textbf{NR} & \textbf{PR} & \textbf{IR} & \textbf{PR vs.\ NR} & \textbf{PR vs.\ IR} \\ \midrule
\multicolumn{2}{l}{User demand {[}-{]}} & 62192 & 62192 & 62192 & 0.00 \% & 0.00 \% \\ 
\multicolumn{2}{l}{Number of bikes {[}-{]}} & 1000 & 1500 & 500 & 50.00 \% & 200.00 \% \\ 
\multicolumn{2}{l}{Total trip time {[}min{]}} & 17.62 & 16.78 & 14.11 & -4.78 \% & 18.88 \% \\ 
 & Wait time {[}min{]} & 3.53 & 2.67 & 0.00 \% & -24.56 \% & - \\ 
 & Wait time \textgreater 10min   {[}\%{]} & 6.16 \% & 2 \% & 0.00 \% & -69.88 \% & - \\ 
 & Wait time \textgreater 15 min   {[}\%{]} & 1.02 \% & 0 \% & 0.00 \% & -87.71 \% & - \\ 
 & Ride time {[}min{]} & 14.09 & 14.11 & 14.11 & 0.19 \% & 0.00 \% \\ 
\multicolumn{2}{l}{Served trips  {[}\%{]}} & 99.46 \% & 100.00 \% & 100 \% & 0.54 \% & 0.00 \% \\ 
\multicolumn{2}{l}{Unserved trips  {[}\%{]}} & 0.54 \% & 0.00 \% & 0.00 \% & -99.70 \% & - \\ 
\multicolumn{2}{l}{Bikes used {[}\%{]}} & 97.20 \% & 71.30 \% & 98.93 \% & -26.65 \% & -27.93 \% \\ 
\multicolumn{2}{l}{Trips/bike/day {[}-{]}} & 8.84 & 5.92 & 17.77 & -32.97 \% & -66.67 \% \\ 
\multicolumn{2}{l}{Total charges {[}-{]}} & 504 & 563 & 821 & 11.65 \% & -31.45 \% \\ 
\multicolumn{2}{l}{Total charges per day {[}-{]}} & 72 & 80 & 117 & 11.65 \% & -31.45 \% \\ 
\multicolumn{2}{l}{Total v.k.t  {[}km{]}} & 177488 & 183293 & 149382 & 3.27 \% & 22.70 \% \\ 
\multicolumn{2}{l}{v.k.t. per bike {[}km{]}} & 177.49 & 122.20 & 298.76 & -31.15 \% & -59.10 \% \\ 
 & v.k.t. in use {[}\%{]} & 83 \% & 81 \% & 99.97 \% & -2.46 \% & -18.50 \% \\ 
 & v.k.t. pick-up {[}\%{]} & 16.48 \% & 12 \% & 0.00 \% & -26.29 \% & - \% \\ 
 & v.k.t. rebalancing   {[}\%{]} & 0.00 \% & 6 \% & 0.00 \% & - & - \\ 
 & v.k.t. charge {[}\%{]} & 0.07 \% & 0 \% & 0.13 \% & 40.96 \% & -27.33 \% \\ 
 \multicolumn{2}{l}{Total time {[}days{]}} & 7 days & 7 days & 7 days & 0 \% & 0 \% \\ 
& Time in use {[}\%{]} & 8.64 \% & 5.80 \% & 17.41 \% & -32.85 \% & -66.67 \% \\
& Time pick-up {[}\%{]} & 2.18 \% & 1.10 \% & 0.00 \% & -49.25 \% &  - \\
& Time rebalancing {[}\%{]} & 0.00 \% & 0.58 \% & 0.00 \% & - & - \\
& Time charge {[}\%{]} & 1.14 \% & 0.88 \% & 3.82 \% & -23.13 \% & -77.09 \% \\
& Time idling {[}\%{]} & 88.04 \% & 91.62 \% & 78.71 \% & 4.07 \% & 16.40 \% \\ \bottomrule
\end{tabular}}
    \caption{Fleet size and performance metrics for the autonomous system with predictive rebalancing (PR) compared to no rebalancing (NR) and ideal rebalancing (IR) for providing a minimum level of service of 99 \% of served trips}
    \label{tb:predictive-99}
\end{table*}

\subsubsection{Fleet size}
This section compares the performance of the autonomous system with different rebalancing scenarios for a fleet size of 2000 and 3000 bicycles. As shown in Table \ref{tb:predictive-2000}, with a fleet of 2000 bicycles, predictive rebalancing slightly reduces wait times while increasing served trips when compared to the case of no rebalancing. In order to provide this improvement, the rebalancing adds 4.82\% of total vehicle miles traveled and increases the need for charging. 
In all three cases, bicycles serve roughly the same number of trips per day. Bicycles also spend the same amount of time in use and pick-up, but with predictive rebalancing, more time is spent in rebalancing and charging and less time idling.   

\begin{table*}[]
\centering
\scalebox{\tablescale}{
\begin{tabular}{l|lrrrrr}
\toprule
\multicolumn{2}{l}{\textbf{Metric}} & \textbf{NR} & \textbf{PR} & \textbf{IR} & \textbf{PR vs.\ NR} & \textbf{PR vs.\ IR} \\ \midrule
\multicolumn{2}{l}{User demand {[}-{]}} & 62192 & 62192 & 62192 & 0.00 \% & 0.00 \% \\ 
\multicolumn{2}{l}{Total trip time   {[}min{]}} & 16.12 & 16.09 & 14.11 & -0.14 \% & 14.05 \% \\ 
 & Wait time {[}min{]} & 2.01 & 1.98 & 0.00 \% & -1.29 \% & - \\ 
 & Wait time \textgreater 10min   {[}\%{]} & 1.40 \% & 1 \% & 0.00 \% & -52.73 \% & - \\ 
 & Wait time \textgreater 15 min   {[}\%{]} & 0.22 \% & 0 \% & 0.00 \% & -68.44 \% & - \\ 
 & Ride time {[}min{]} & 14.11 & 14.11 & 14.11 & 0.02 \% & 0.00 \% \\ 
\multicolumn{2}{l}{Served trips  {[}\%{]}} & 99.96 \% & 100.00 \% & 100 \% & 0.04 \% & 0.00 \% \\ 
\multicolumn{2}{l}{Unserved trips  {[}\%{]}} & 0.04 \% & 0.00 \% & 0.00 \% & -92.16 \% & - \\ 
\multicolumn{2}{l}{Bikes used {[}\%{]}} & 92.16 \% & 94.25 \% & 91.95 \% & 2.27 \% & 2.50 \% \\ 
\multicolumn{2}{l}{Trips/bike/day {[}-{]}} & 4.44 & 4.44 & 4.44 & 0.04 \% & 0.00 \% \\ 
\multicolumn{2}{l}{Total charges {[}-{]}} & 272 & 438 & 265 & 61.33 \% & 65.08 \% \\ 
\multicolumn{2}{l}{Total charges per day {[}-{]}} & 39 & 63 & 38 & 61.33 \% & 65.08 \% \\ 
\multicolumn{2}{l}{Total v.k.t  {[}km{]}} & 165878 & 173876 & 149257 & 4.82 \% & 16.50 \% \\ 
\multicolumn{2}{l}{v.k.t. per bike {[}km{]}} & 82.94 & 86.94 & 74.63 & 4.82 \% & 16.50 \% \\ 
 & v.k.t. in use {[}\%{]} & 89.89 \% & 85.80 \% & 99.96 \% & -4.54 \% & -14.16 \% \\ 
 & v.k.t. pick-up {[}\%{]} & 10.08 \% & 9.53 \% & 0.00 \% & -5.39 \% & - \\ 
 & v.k.t. rebalancing   {[}\%{]} & 0.00 \% & 4.59 \% & 0.00 \% & - & - \\ 
 & v.k.t. charge {[}\%{]} & 0.04 \% & 0.07 \% & 0.04 \% & 104.52 \% & 91.01 \% \\ 
\multicolumn{2}{l}{Total time {[}days{]}} & 7 days & 7 days & 7 days & 0.00 \% & 0.00 \% \\ 
& Time in use {[}\%{]} & 4.35 \% & 4.35 \% & 4.35 \% & 0.06 \% & 0.00 \% \\ 
& Time pick-up {[}\%{]} & 0.62 \% & 0.62 \% & 0.00 \% & -0.83 \% & - \\
& Time rebalancing {[}\%{]} & 0.00 \% & 0.30 \% & 0.00 \% & - & - \\ 
& Time charge {[}\%{]} & 0.30 \% & 0.51 \% & 0.30 \% & 68.62 \% & 69.46 \% \\
& Time idling {[}\%{]} & 94.72 \% & 94.22 \% & 95.34 \% & -0.53 \% & -01.18 \% \\ \bottomrule
\end{tabular}}
    \caption{Performance metrics for the autonomous system with predictive rebalancing (PR) compared to no rebalancing (NR) and ideal rebalancing (IR) with a fleet size of 2000 bikes}
    \label{tb:predictive-2000}
\end{table*}

In the case of having a fleet size of 3000 bicycles (see Table \ref{tb:predictive-3000}, the improvement in wait time thanks to the predictive rebalancing is greater than with 2000 bikes. The average wait time is reduced by 1.29\% with 2000 bikes and by -3.73\% with 3000 bikes. The difference in terms of served trips is minor because the system with predictive rebalancing was already covering 100\% of the demand with 2000 bikes. 

Interestingly, the total vehicle miles traveled and the total charges are increased in a lower proportion than with 2000 bikes. This is because an increase in bicycles does not lead to a proportional increase in the number of rebalancing trips; more bicycles are available to rebalance, but there is less need for rebalancing. 

\begin{table*}[]
\centering
\scalebox{\tablescale}{
\begin{tabular}{l|lrrrrr}
\toprule
\multicolumn{2}{l}{\textbf{Metric}} & \textbf{NR} & \textbf{PR} & \textbf{IR} & \textbf{PR vs.\ NR} & \textbf{PR vs.\ IR} \\ \midrule
\multicolumn{2}{l}{User demand {[}-{]}} & 62192 & 62192 & 62192 & 0.00 \% & 0.00 \% \\ 
\multicolumn{2}{l}{Total trip time   {[}min{]}} & 15.55 & 15.50 & 14.11 & -0.34 \% & 9.81 \% \\ 
 & Wait time {[}min{]} & 1.44 & 1.38 & 0.00 \% & -3.73 \% & - \\ 
 & Wait time \textgreater 10min   {[}\%{]} & 0.64 \% & 0 \% & 0.00 \% & -60.23 \% & - \\ 
 & Wait time \textgreater 15 min   {[}\%{]} & 0.11 \% & 0 \% & 0.00 \% & -50.19 \% & - \\ 
 & Ride time {[}min{]} & 14.11 & 14.11 & 14.11 & 0.00 \% & 0.00 \% \\ 
\multicolumn{2}{l}{Served trips  {[}\%{]}} & 99.99 \% & 100.00 \% & 100 \% & 0.01 \% & 0.00 \% \\ 
\multicolumn{2}{l}{Unserved trips  {[}\%{]}} & 0.01 \% & 0.00 \% & 0 \% & -100.00 \% & - \\ 
\multicolumn{2}{l}{Bikes used {[}\%{]}} & 88.31 \% & 88.10 \% & 88.27 \% & -0.24 \% & -0.19 \% \\ 
\multicolumn{2}{l}{Trips/bike/day {[}-{]}} & 2.96 & 2.96 & 2.96 & 0.01 \% & 0.00 \% \\ 
\multicolumn{2}{l}{Total charges {[}-{]}} & 186 & 247 & 199 & 32.97 \% & 24.33 \% \\ 
\multicolumn{2}{l}{Total charges per day {[}-{]}} & 27 & 35 & 28 & 32.97 \% & 24.33 \% \\ 
\multicolumn{2}{l}{Total v.k.t  {[}km{]}} & 161195 & 165331 & 149244 & 2.57 \% & 10.78 \% \\ 
\multicolumn{2}{l}{v.k.t. per bike {[}km{]}} & 53.73 & 55.11 & 49.75 & 2.57 \% & 10.78 \% \\ 
 & v.k.t. in use {[}\%{]} & 93 \% & 90.24 \% & 99.97 \% & -2.49 \% & -9.73 \% \\ 
 & v.k.t. pick-up {[}\%{]} & 7.42 \% & 6.99 \% & 0.00 \% & -5.87 \% & - \\ 
 & v.k.t. rebalancing   {[}\%{]} & 0.00 \% & 2.73 \% & 0.00 \% & - & - \\ 
 & v.k.t. charge {[}\%{]} & 0.03 \% & 0.04 \% & 0.03 \% & 70.46 \% & 42.84 \% \\ 
\multicolumn{2}{l}{Total time {[}days{]}} & 7 days & 7 days & 7 days & 0.00 \% & 0.00 \% \\ 
 & Time in use {[}\%{]} & 2.90 \% & 2.90 \% & 2.90 \% & 0.01 \% & 0.00 \% \\ 
 & Time pick-up {[}\%{]} & 0.30 \% & 0.29 \% & 0.00 \% & -3.45 \% & - \\
 & Time rebalancing {[}\%{]} & 0.00 \% & 0.11 \% & 0.00 \% & - & - \\ 
 & Time charge {[}\%{]} & 0.13 \% & 0.19 \% & 0.15 \% & 44.43 \% & 24.70 \% \\  
 & Time idling {[}\%{]} & 96.67 \% & 96.51 \% & 96.94 \% & -0.17 \% & -0.45 \% \\ \bottomrule
\end{tabular}}
    \caption{Performance metrics for the autonomous system with predictive rebalancing (PR) compared to no rebalancing (NR) and ideal rebalancing (IR) with a fleet size of 3000 bikes}
    \label{tb:predictive-3000}
\end{table*}

The results in this section show that the benefits of having a predictive rebalancing system become more significant as the fleet size increases. Furthermore, the additional costs of rebalancing are lower, such as the extra vehicle miles traveled and increased need for charging. Therefore, for predictive rebalancing to be most effective, there seems to be a trade-off between improving user experience and increasing the fleet size. 

\subsection{Parameters influence on performance metrics}

We performed simulations for a range of values other than the nominal state to quantify the impact of configuration parameters on system performance. These batch simulations also helped to quantify the impact of the configuration parameters assumed for the nominal state. While this section highlights some of the most relevant findings from this analysis, a more detailed version of the results can be found in \ref{appendix-A}. 

The fleet size was considered to be the most important factor. Therefore, for each value of the configuration parameters, results were calculated for all the fleet sizes, with the rest of the parameters set to their nominal values. These specific values can be found on Tables \ref{tb:sb-parameters}, \ref{tb:dl-parameters}, and \ref{tb:aut-parameters}. The combination of all these values for the three systems has resulted in 1870 simulated scenarios, with a total of approximately 116M simulated trips. This large number of simulations provides an excellent foundation for studying the potential impacts of uncertainties related to how these systems might perform in real-world conditions and analyze many what-if scenarios once the technologies and regulations mature. 

\begin{table}[!htb]
  \centering
  \scalebox{\tablescale}{
  \begin{tabular}{ll}
    \toprule
    \multicolumn{2}{c}{\textbf{Station-based system parameters sets}}  \\ \midrule
    Number of bikes [-] & \{1000,1500,2000,2500,3000,3500,4000,4500,5000,5500\} \\ 
    Maximum walking radius [m] & \{100,300,500,750,1000,1500\} \\ 
    Average walking speed [km/h] & \{3,4,5,6,7,8\} \\  
    Average riding speed [km/h] & \{5,8,10,12,15,20\} \\ 
    Rebalancing parameter[\%] & \{0,50,80,90,98,100\} \\  
    Minimum bikes/docks per station [-] & \{0,1,2,3,4,5\} \\ \bottomrule
    \end{tabular}}
    \caption{Variations of the configuration parameters for the station-based system}
    \label{tb:sb-parameters}
\end{table}

\begin{table}[!htb]
  \centering
  \scalebox{\tablescale}{
  \begin{tabular}{ll}
    \toprule
    \multicolumn{2}{c}{\textbf{Dockless system parameters sets}}  \\ \midrule
    Number of bikes [-] & \{2000,3000,4000,5000,6000,7000,8000,9000,10000,11000\} \\ 
    Maximum walking radius [m] & \{100,300,500,750,1000,1500\} \\ 
    Average walking speed [km/h] & \{3,4,5,6,7,8\} \\  
    Average riding speed [km/h] & \{5,8,10,12,15,20\} \\  \bottomrule
    \end{tabular}}
    \caption{Variations of the configuration parameters for the dockless system}
    \label{tb:dl-parameters}
\end{table}

\begin{table}[!htb]
  \centering
  \scalebox{\tablescale}{
  \begin{tabular}{ll}
     \toprule
    \multicolumn{2}{c}{\textbf{Autonomous system parameters sets}}  \\ \midrule
    Number of bikes [-] &
    \{300,500,600,700,800,1000,1500,2000,2500,3000\} \\  
    Maximum autonomous radius [m] & \{500,1000,1500,2000,2500,3000\}\\  
    Average autonomous driving speed [km/h] & \{1,2.5,5,10,15,20\}\\ 
    Average riding speed [km/h] & \{5,8,10,12,15,20\} \\
    Minimum battery level [\%] & \{5,10,15,20,25,30\} \\ 
    Battery autonomy [km] & \{30,50,70,90,110,130\} \\ 
    Battery recharge time [h] & \{0.5,1,2,4,6,8\} \\ \bottomrule
    \end{tabular}}
    \caption{Variations of the configuration parameters for the autonomous system}
    \label{tb:aut-parameters}
\end{table}

Starting with the fleet size, an increase in the number of bikes directly influences the reduction of the walk and wait times and the number of unserved trips, significantly improving user experience. On the contrary, efficiency-related metrics, such as the number of trips per bike and day or the average time in use per bike, are worsened by an increase in the number of bikes. In the station-based system, there is a particular effect that must be noted. Because stations' capacity is considered to be fixed, an increase in fleet size over the nominal state leads to difficulty finding available docks, increasing ride times, walk times at destination, and the need for rebalancing.

The riding speed has a similar effect on the three systems; a decrease in riding speed increases the average ride time and the average trip time. Consequently, bicycles are also used for longer periods of time. In the autonomous system, especially in the cases of small fleet size, as bicycles are kept busy for a longer period of time, the number of available bikes decreases, while the number of unserved trips and wait times increase.

Similarly to the effect of the riding speed parameter, walking speed in station-based and dockless systems directly impacts walk times and, thus, average trip times. In the case of the station-based system, a slower walking pace also leads to more visited stations; users check the status of the stations before heading there, so the longer they spend on the way, the more likely it is that someone else might take the bike or dock. An analogous but smaller effect is also generated by a decrease in riding speed when finding a dock at the destination. 

Regarding the maximum walking radius, lower values decrease the number of served trips because it is less likely to find an available bicycle or dock in a smaller area. In this case, only trips with a nearby station or bike are made, thus reducing the average walk and trip times. At the same time, idle times increase, while the need for rebalancing and the average number of stations visited decrease. On the contrary, a larger walk radius increases the number of served trips because it is more likely to find a bicycle within walking distance. For the same reason, the average walk time increases and the need for rebalancing decreases. 

The parameters examined up to this point affected two or more of the three systems. When it comes to station-based system specific parameters, the rebalancing parameter has a significant impact on performance. The higher the rebalancing parameter, the more bikes are rebalanced, which increases the number of served trips. In addition, users visit fewer stations in search of bikes or docks, spending also less time walking. In the case of small fleet sizes, the main issue is the lack of bicycles, so a reduction in rebalancing has the greatest impact at the start of the trip. On the contrary, the main issue in the case of large fleet sizes is a lack of available docks. In this case, a lower level of rebalancing also increases the average ride time because users have to travel further to find a station with available docks.

As for parameters specific to the autonomous systems, the autonomous driving speed is an especially relevant one. In the scenario with no rebalancing, driving speed has a direct impact on wait times and served trips; the faster you go, the shorter the wait times and the higher the level of served trips. Interestingly, the difference between 1 km/h and 2.5 km/h is significant, but the difference from 5 km/h to 10 km/h is negligible. Instead, the difference between 10 km/h, 15 km/h, and 20 km/h is much smaller at this scale. In light of this, we would argue that a speed of 8 km/h, as was considered for the nominal case of this paper, probably represents a good trade-off between the provided service and the cost, complexity, and consumption of having the bicycles travel faster.  
In the case of the ideal rebalancing, most metrics are unaffected by the autonomous driving speed because the rebalancing and pick-up are supposed to be instantaneous. 

Regarding the autonomous radius, it has an impact on the autonomous system that is very similar to the effect of the walking radius on the station-based and dockless systems. A reduction in the autonomous radius decreases the number of served trips but, as the trips that are served are those that had autonomous bicycles nearby, the average wait times and trip times are reduced. Vehicles travel shorter distance for pick-ups and, reducing the need for charging. 

Finally, in terms of battery parameters, battery autonomy has an evident impact on the number of recharges required and, as a consequence, on the time spent going for a charge and charging, and the distance traveled to charging stations. The difference between having a battery autonomy of 30 km and one with 50 km is greater than having a battery of 70 km instead of 50 km. The difference between having a battery autonomy of 90 km, 110km, or 130 km is, at this scale, much smaller. Therefore, we believe that a battery autonomy of around 70 km represents a good trade-off between performance, environmental impact, and cost. The time spent charging is primarily affected by the charging speed of the battery. In contrast, the minimum battery primarily influences the number of charges and the distance traveled for recharging. However, it should be noted that the scenarios of no rebalancing and ideal rebalancing may underestimate the impact of the batteries because, in both cases, there are fewer kilometers traveled in autonomous mode than there would be with a predictive rebalancing model.  

\section{Conclusions} \label{conclusions}

Cities urgently need innovative solutions to global challenges such as urban population growth, inequality, and climate change. Mobility is one of the fields that requires a radical transformation; by redesigning urban mobility systems, we can move towards a future in which cities are more livable, equitable, sustainable, and resilient. 

Combining transformative vehicle sharing, electrification, autonomy, and micro-mobility can lead to compelling mobility solutions. An autonomous bicycle-sharing system would efficiently combine the benefits of all these new mobility trends. On the one hand, it can drastically improve the system efficiency by addressing issues found in the current BSS, such as the rebalancing problem or the bicycle oversupply. On the other hand, it could bring the convenience of mobility-on-demand systems into bike-share, incentivizing more people to use bicycles as their preferred trip mode, allowing more people to travel around their cities in a more enjoyable and environmentally friendly way. 

Due to the radical novelty and uniqueness of introducing autonomous driving technology into BSS and the inherent complexities of these systems, an agent-based simulation was required in order to quantify its possible impacts. The proposed simulation framework is highly configurable and flexible, can be easily transferred to other cities and works with geospatial data of high resolution and precision. It provides an in-depth understanding of the performance of a fleet of autonomous bicycles and compares its performance to the existing bicycle-sharing schemes that are station-based and dockless systems.  

In the first place, the systems have been analyzed separately. This analysis included the average values of the performance metrics and their evolution over time. The values of the performance metrics provide a very detailed overview of each system's characteristics in terms, for example, of average trip times, level of served trips, or the utilization of the bicycles. The timelines of the parameters also provide valuable insights; for instance, it can be observed that the percentage of bicycles being used simultaneously is significantly higher for the autonomous system than for the other two. 

Subsequently, the three systems have been compared under different scenarios to obtain a thorough understanding of each system's strengths and weaknesses compared to the other two. From the perspective of system efficiency, the most relevant conclusion from these comparative studies is that in the nominal state, autonomous bicycles do, on average, 8.84-8.88 trips per day, while station-based bikes do 2.51 and the dockless bikes do 1.10. As a consequence of having a much higher level of use of the bicycles, with a fleet size of 1000 bicycles, the autonomous system performs better than a station-based system with 3500 bicycles and a dockless system with 8000 bicycles for most performance metrics, even for the case of not considering any rebalancing in the autonomous system. These results indicate that the remarkable efficiency of an autonomous bicycle-sharing system could compensate for the additional cost of autonomous bicycles. 

In all the cases studied, the autonomous system could serve a higher percentage of the demand, which is very relevant from the perspective of the user experience because the system's reliability is vital in the mode choice for commuting trips. Additionally, even without rebalancing, the wait time in the autonomous system is lower than the walk time in the station-based system. Compared to the dockless systems, the time can be higher or lower depending on the rebalancing performance in the autonomous system and the fleet size. Another important aspect to note is that wait times vary more than the walk time in station-based and dockless systems; they seem to have a higher dependency on the level of the demand at each moment. Nevertheless, since the users will be able to use their time for other purposes while waiting for the autonomous bicycle to arrive, it seems likely that the tolerance to waiting will be higher than it is to walking.  Increased reliability and convenience could incentivize more people to use bicycles as their preferred mode of transportation for commuting.

With the intention of providing results for the autonomous system that are more precise than the upper and lower bound defined by the scenarios of no rebalancing and ideal rebalancing, the performance of a rebalancing system based on demand prediction was analyzed. The results showed that, depending on the performance of the predictive rebalancing model, a larger fleet size might be required to have a positive impact. Since the fleet sizes are smaller in the autonomous system, rebalancing bicycles reduces the available fleet size, which is detrimental to the performance during peak times. Therefore, to improve the performance, the fleet size may need to be slightly increased, and the demand prediction and routing models that govern the rebalancing behavior should be adequately designed with these learnings in mind.  

Finally, we quantified the impact of the different configuration parameters on each system's performance and validated the impact of the nominal state values by running batch simulations with a range of values for each parameter. These results answer one of the most central questions in shared autonomous systems: the relationship between the wait times and the fleet size. Other relevant takeaways from this analysis are related to the impacts of metrics related to the bicycles, such as the autonomous speed and battery autonomy, since these results can guide the vehicles' design. 

To conclude, we would like to highlight that the results presented in this paper can provide valuable insights for many stakeholders. First and foremost, it provides fleet operators with guidelines for designing, implementing, and operating an autonomous bicycle-sharing system. It also provides insights that can assist engineers in defining the vehicle's design requirements. In addition, it can be used by city planners to understand the potential urban impacts of an autonomous bike-sharing system. Finally, it can support governments in defining the regulations and incentive mechanisms for these new mobility modes.

In the future, we would like to include additional scenarios, such as the impact of adding a function that allows users to make a trip reservation in advance, to study the impact of the number and location of the charging stations, or having users charge or consume battery while pedaling based on incentive systems. Furthermore, we would like to integrate these findings into a mode choice model to learn how different people would use autonomous shared bicycles as an alternative to or combined with other mobility modes.


\clearpage


\section*{Funding}

This research did not receive any specific grant from funding agencies in the public, commercial, or not-for-profit sectors.

\section*{Declarations of interest: none}


\bibliographystyle{elsarticle-num.bst}
\bibliography{references}


\appendix
\section{Extended results of parameters influence on performance metrics} \label{appendix-A}

\subsection{Station-based system}

\begin{figure}[!htb]
    \centering
    \caption{Station-based system: influence of walk radius} 
    \includegraphics[width=\linewidth]{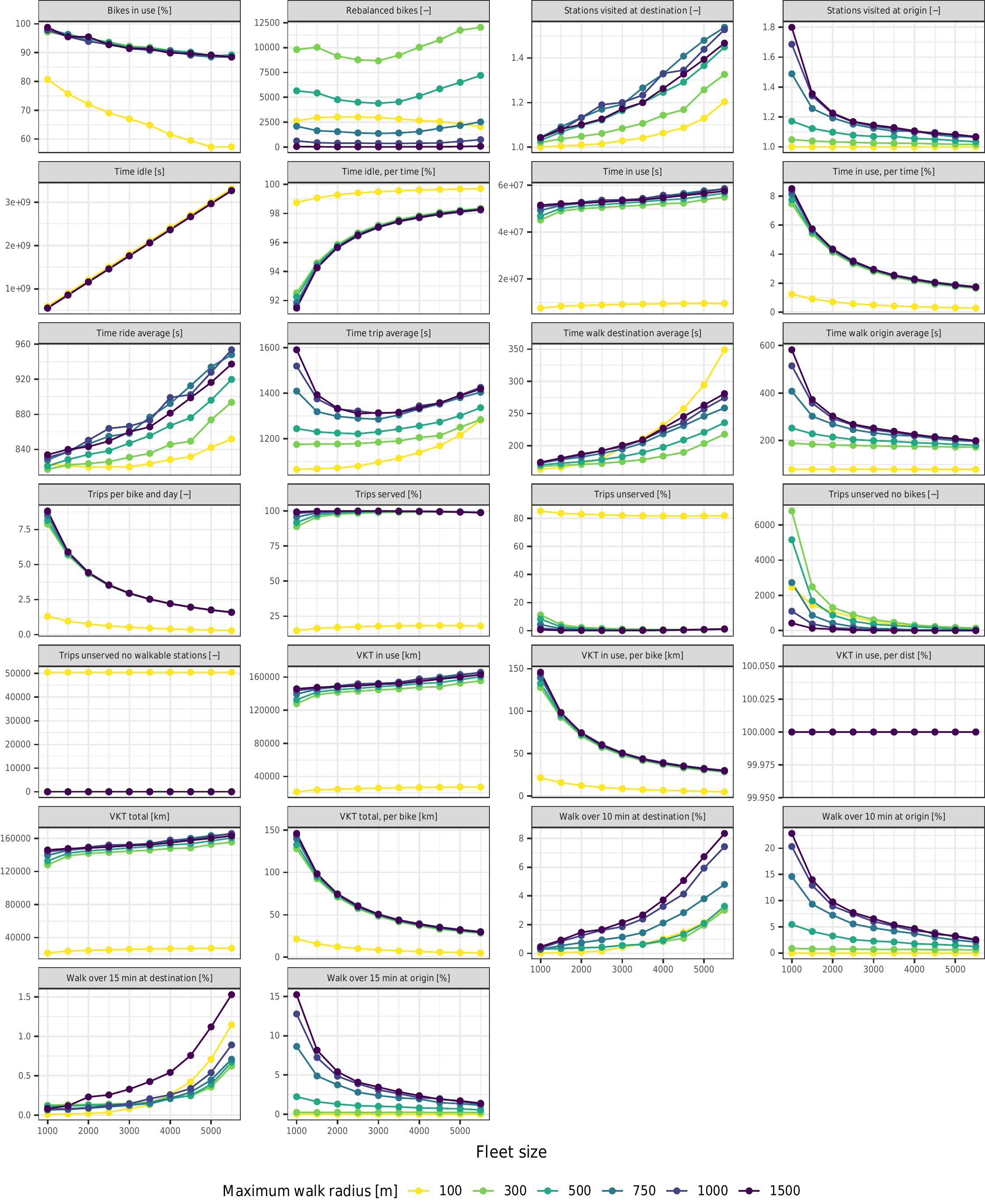}
    \label{fig:influence-SB-walk_radius}
\end{figure}

\begin{figure}[!htb]
    \centering
    \caption{Station-based system: influence of maximum walk radius} 
    \includegraphics[width=\linewidth]{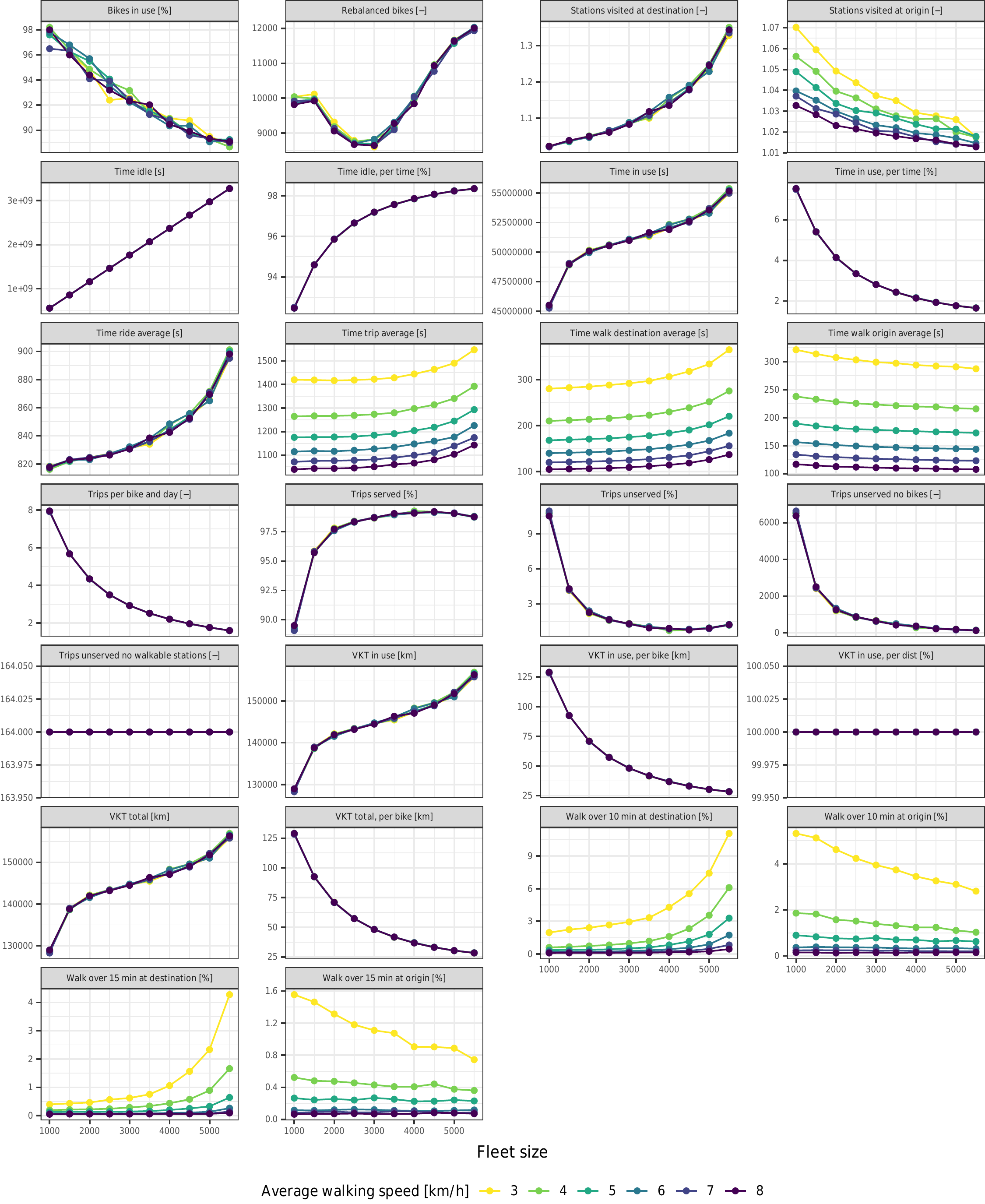}
    \label{fig:influence-SB-walking_speed}
\end{figure}

\begin{figure}[!htb]
    \centering
    \caption{Station-based system: influence of average riding speed} 
    \includegraphics[width=\linewidth]{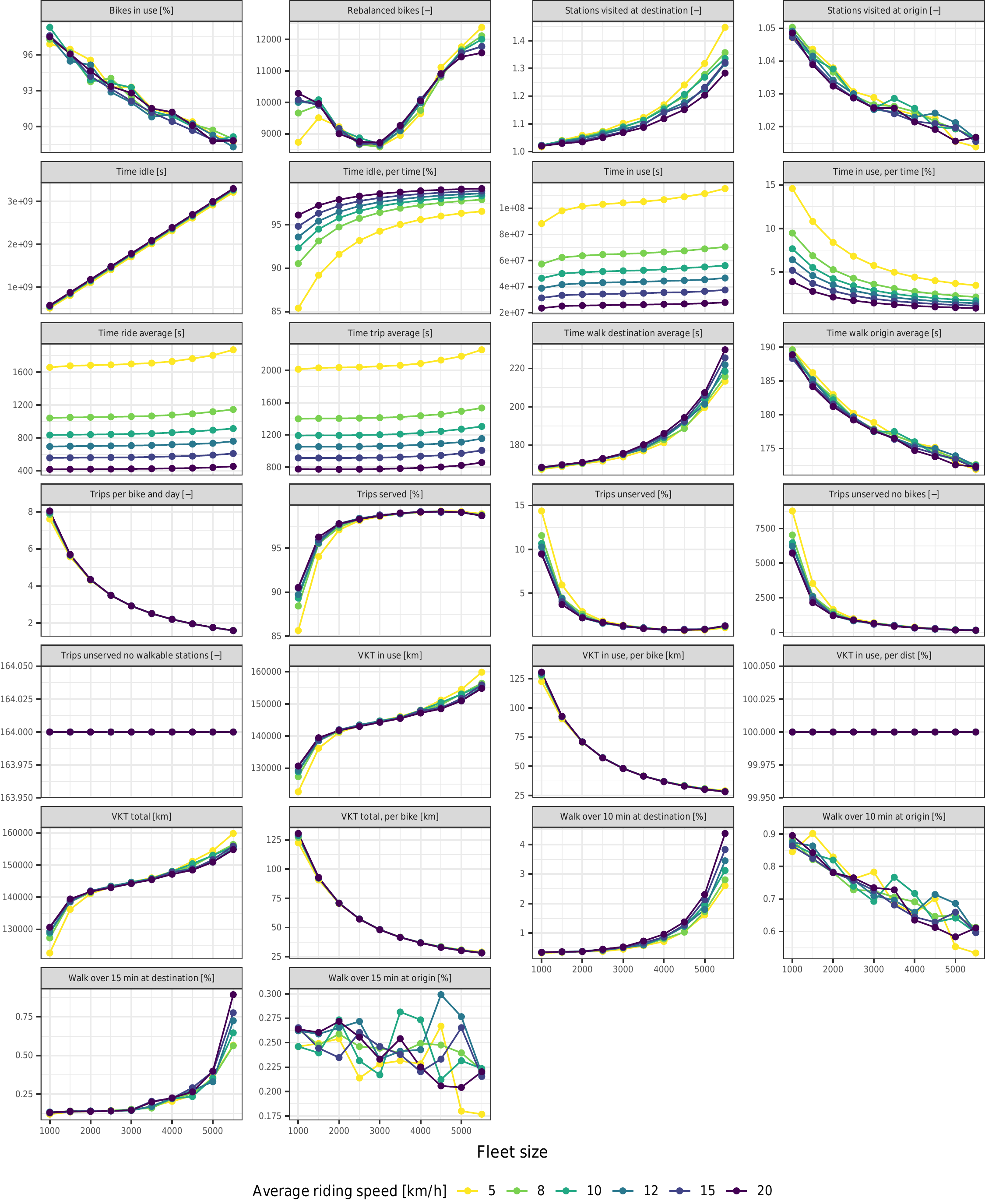}
    \label{fig:influence-SB-riding_speed}
\end{figure}

\begin{figure}[!htb]
    \centering
    \caption{Station-based system: influence of rebalancing parameter} 
    \includegraphics[width=\linewidth]{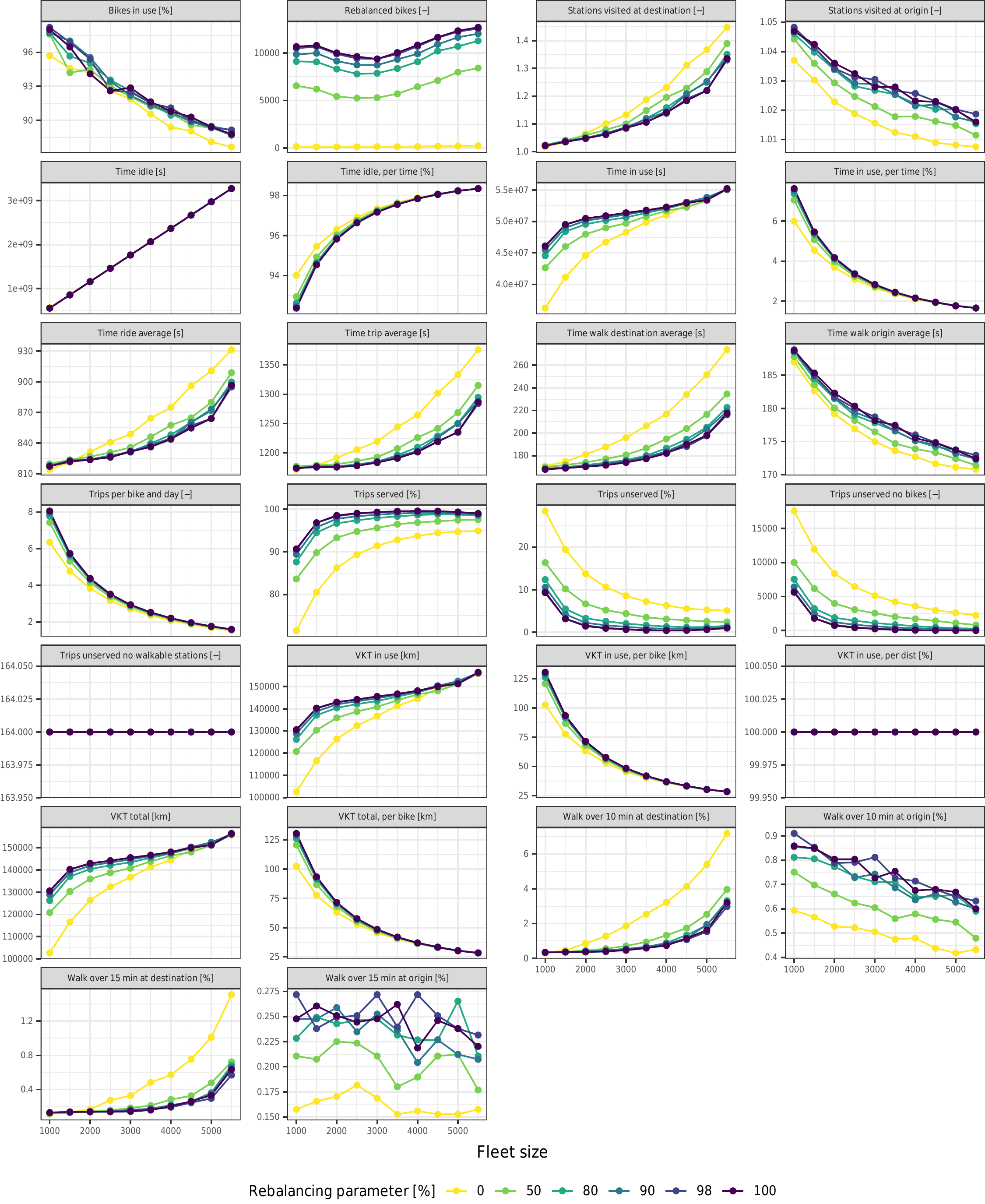}
    \label{fig:influence-SB-magic_beta}
\end{figure}

\begin{figure}[!htb]
    \centering
    \caption{Station-based system: influence of minimum bikes/docks per stations} 
    \includegraphics[width=\linewidth]{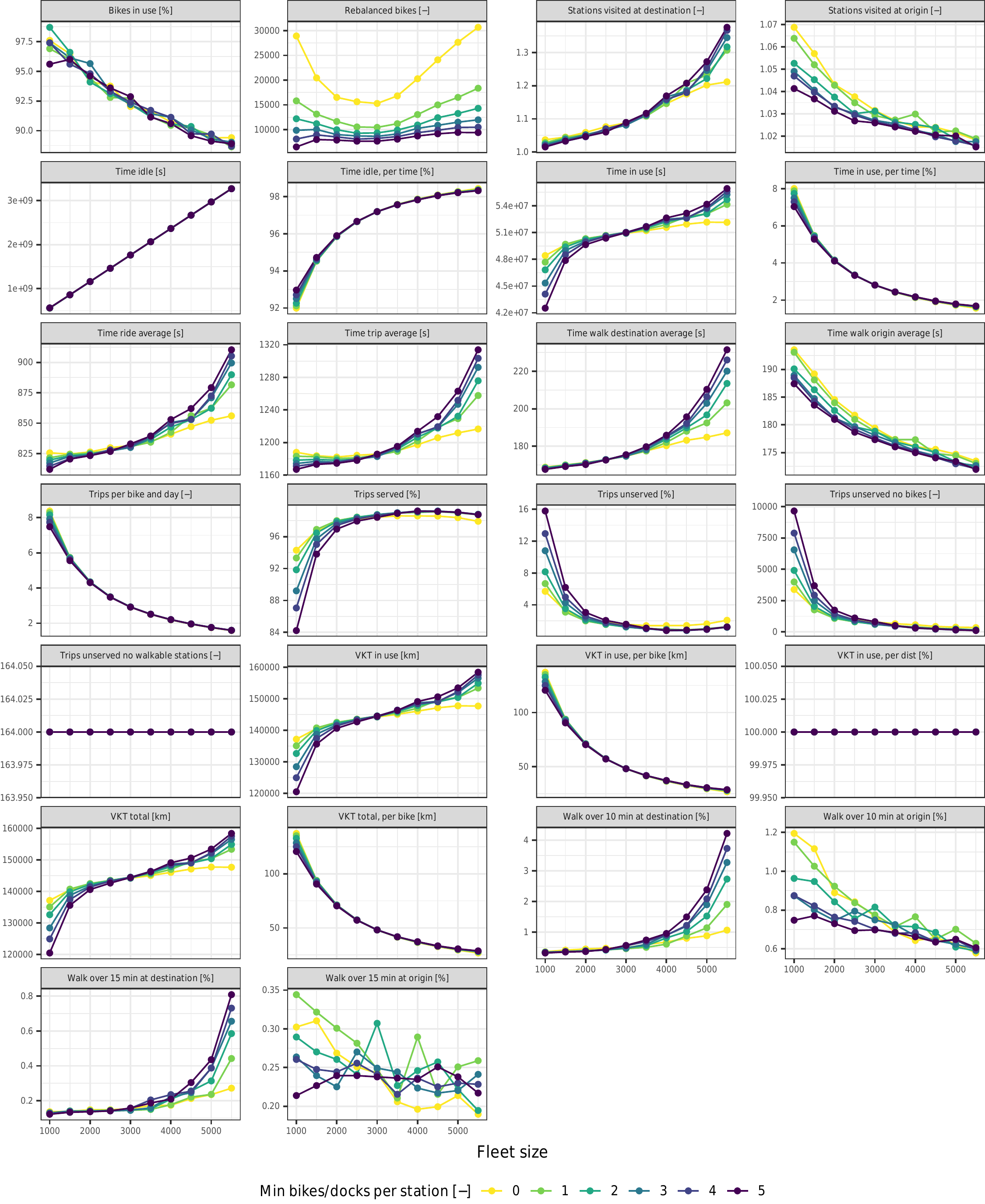}
    \label{fig:influence-SB-magic_min_bikes}
\end{figure}

\subsection{Dockless system}

\begin{figure}[!htb]
    \centering
    \caption{Dockless system: influence of walk radius} 
    \includegraphics[width=\linewidth]{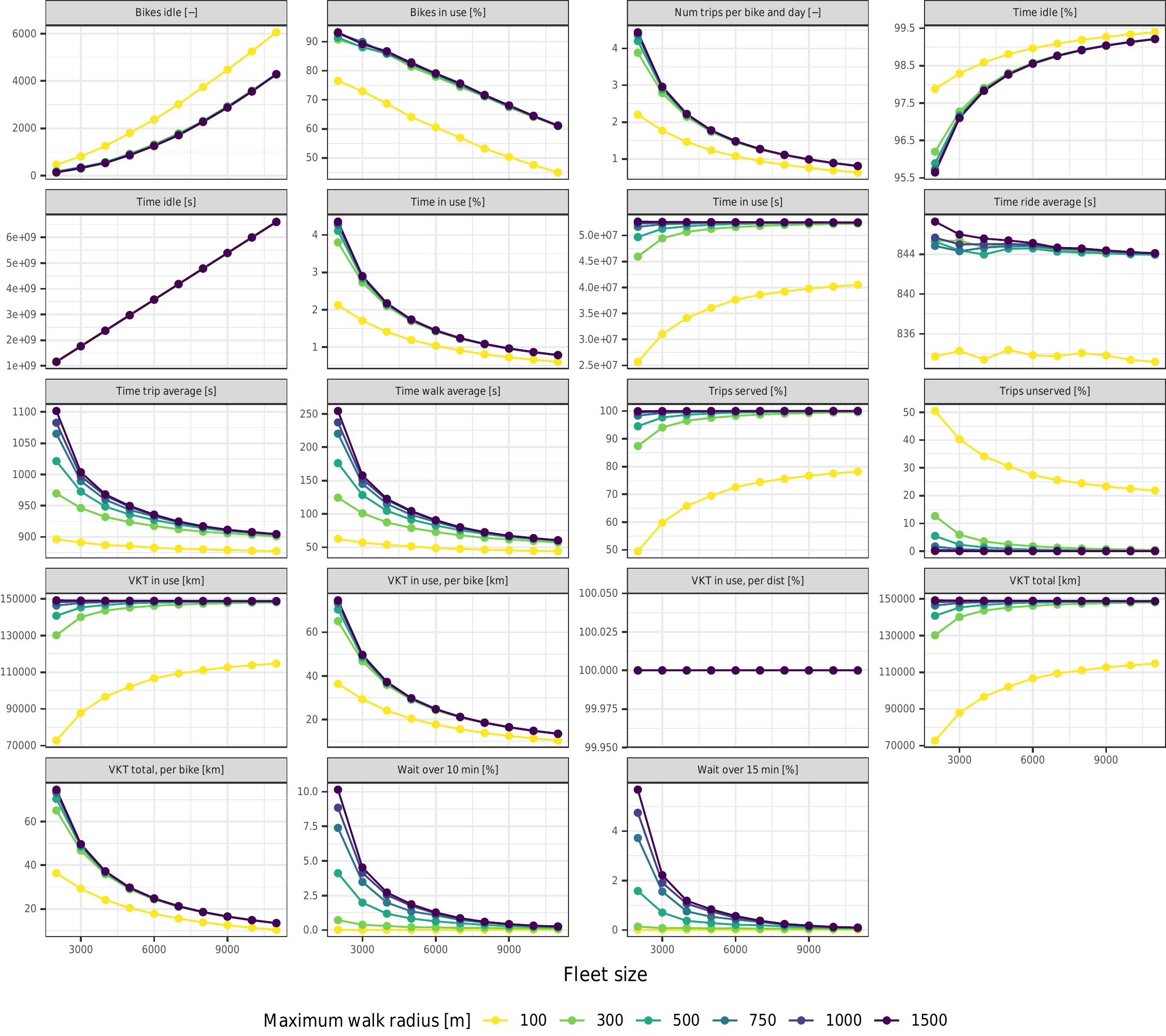}
    \label{fig:influence-DK-walk_radius}
\end{figure}

\begin{figure}[!htb]
    \centering
    \caption{Dockless system: influence of maximum walk radius} 
    \includegraphics[width=\linewidth]{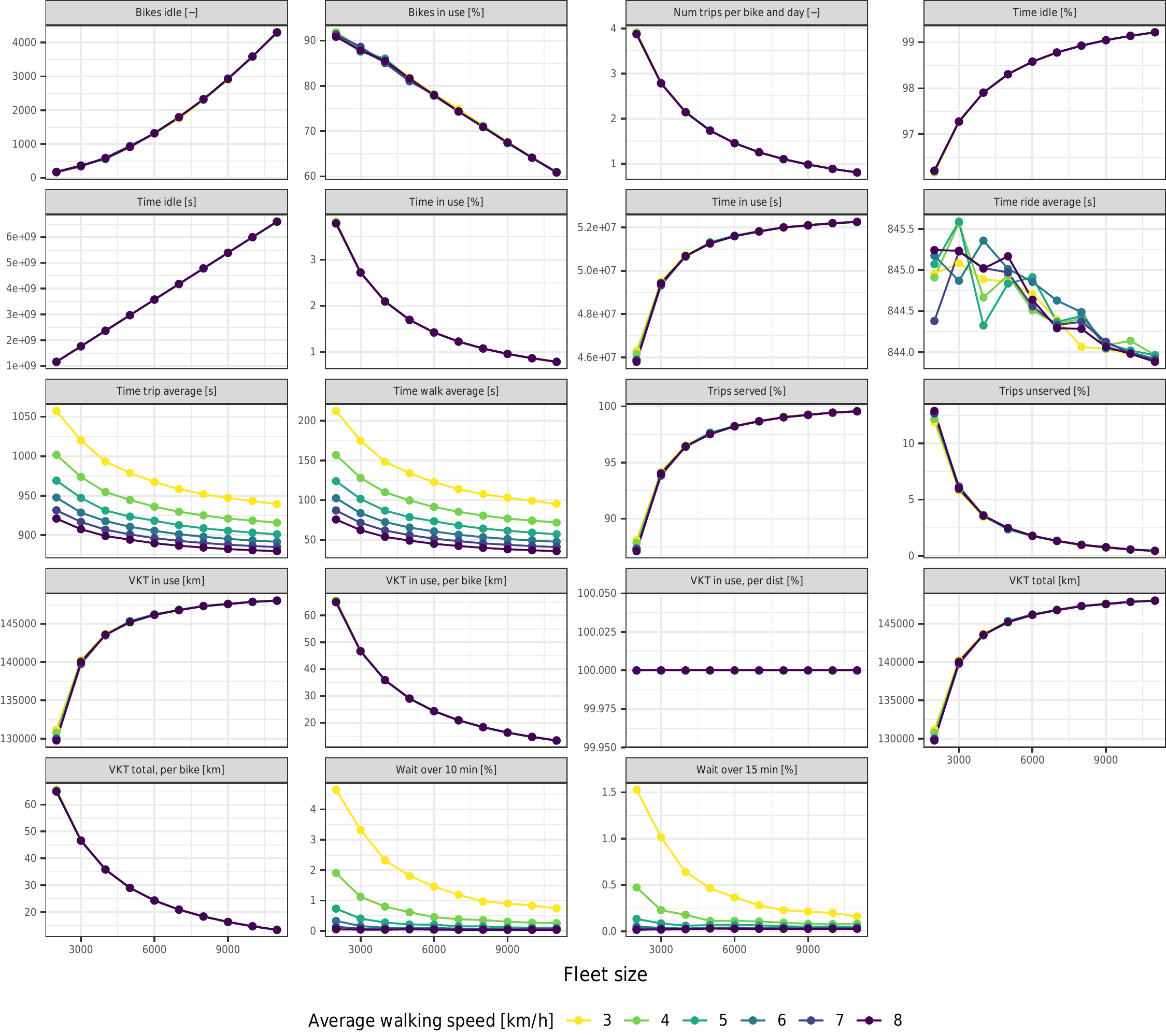}
    \label{fig:influence-DK-walking_speed}
\end{figure}

\begin{figure}[!htb]
    \centering
    \caption{Dockless system: influence of average riding speed} 
    \includegraphics[width=\linewidth]{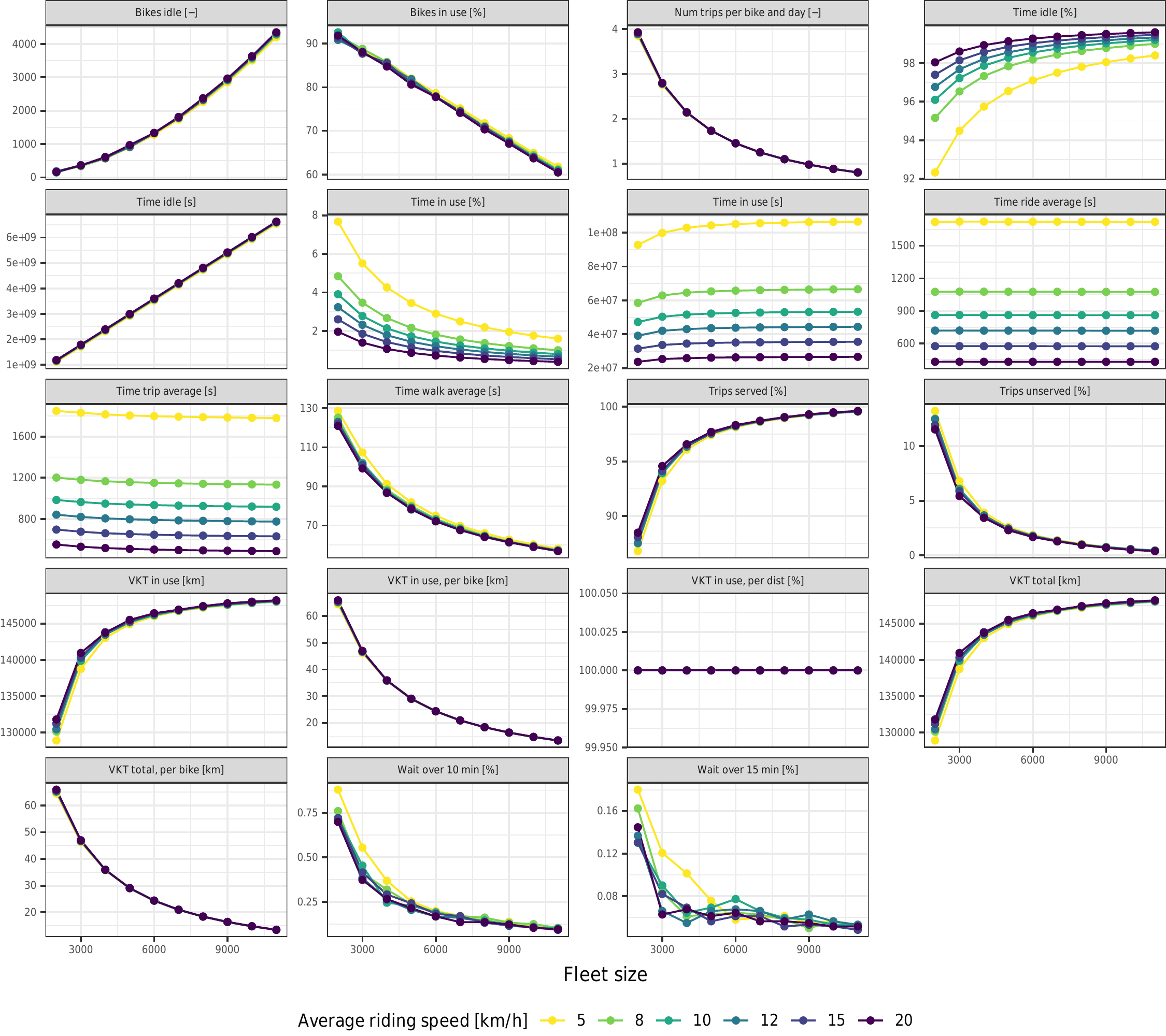}
    \label{fig:influence-DK-riding_speed}
\end{figure}

\subsection{Autonomous system: no rebalancing}

\begin{figure}[!htb]
    \centering
    \caption{Autonomous system without rebalancing: influence of maximum autonomous radius} 
    \includegraphics[width=\linewidth]{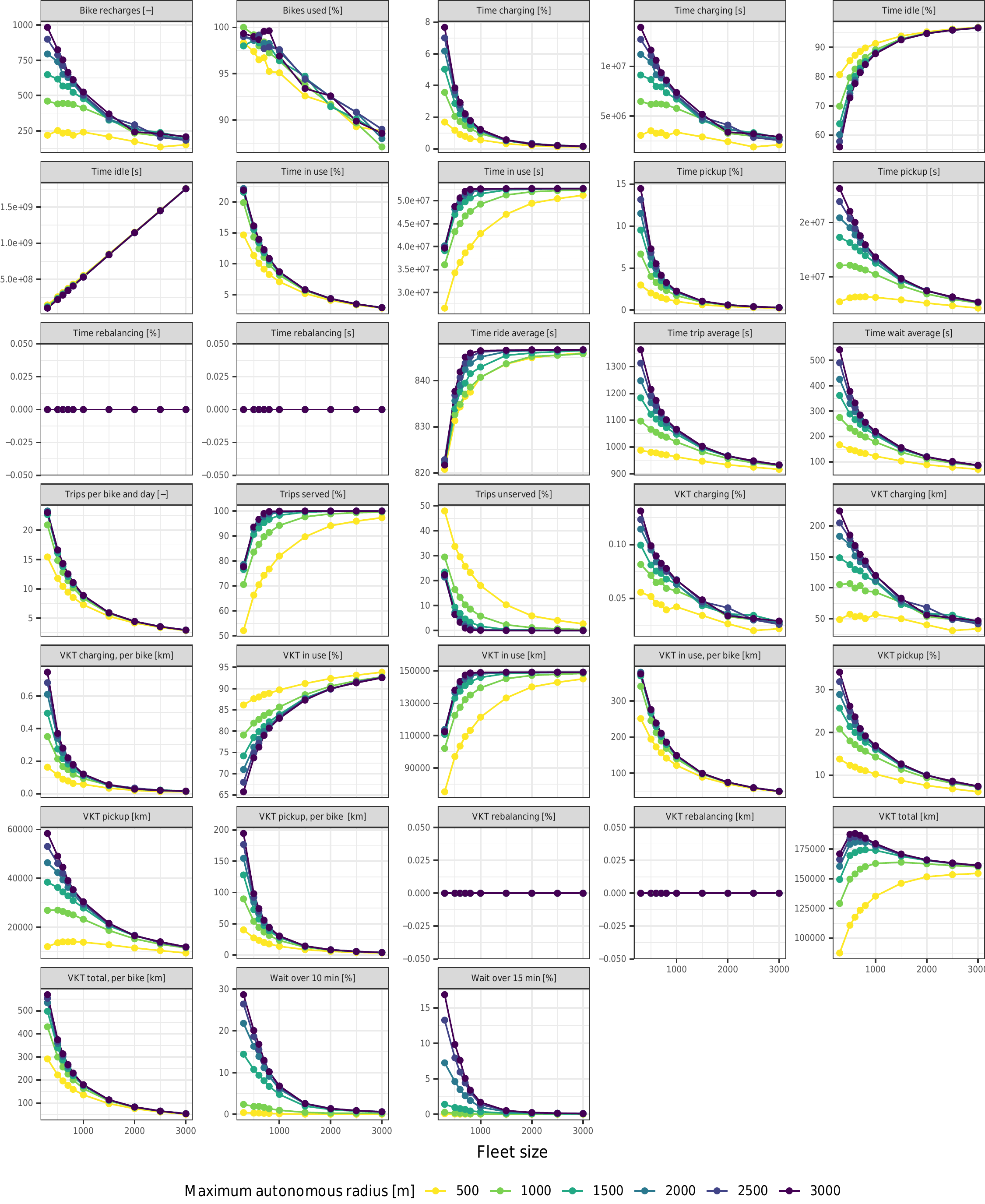}
    \label{fig:influence-AU-NR-autonomous_radius}
\end{figure}

\begin{figure}[!htb]
    \centering
    \caption{Autonomous system without rebalancing: influence of average autonomous driving speed} 
    \includegraphics[width=\linewidth]{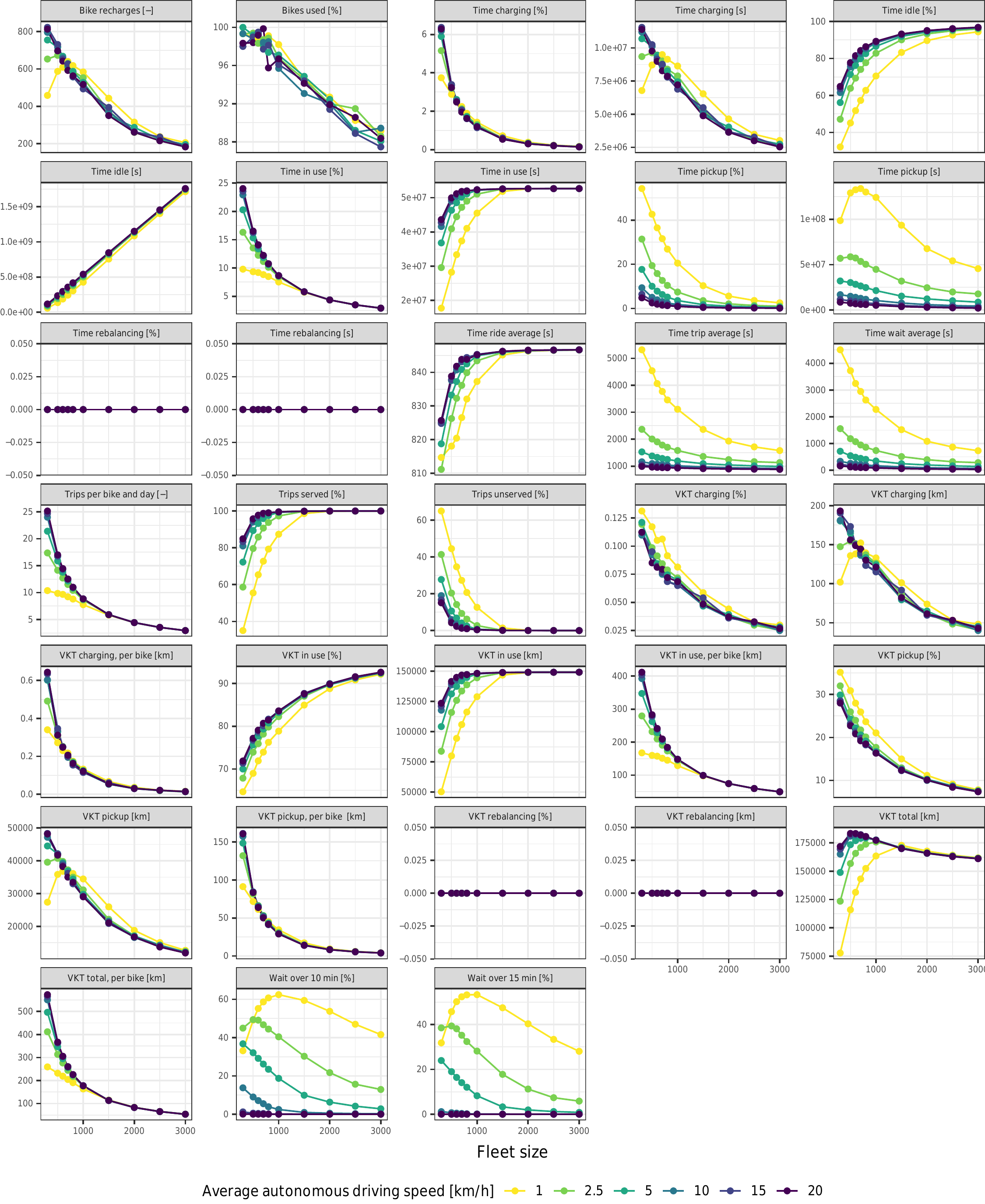}
    \label{fig:influence-AU-NR-autonomous_speed}
\end{figure}

\begin{figure}[!htb]
    \centering
    \caption{Autonomous system without rebalancing: influence of average riding speed} 
    \includegraphics[width=\linewidth]{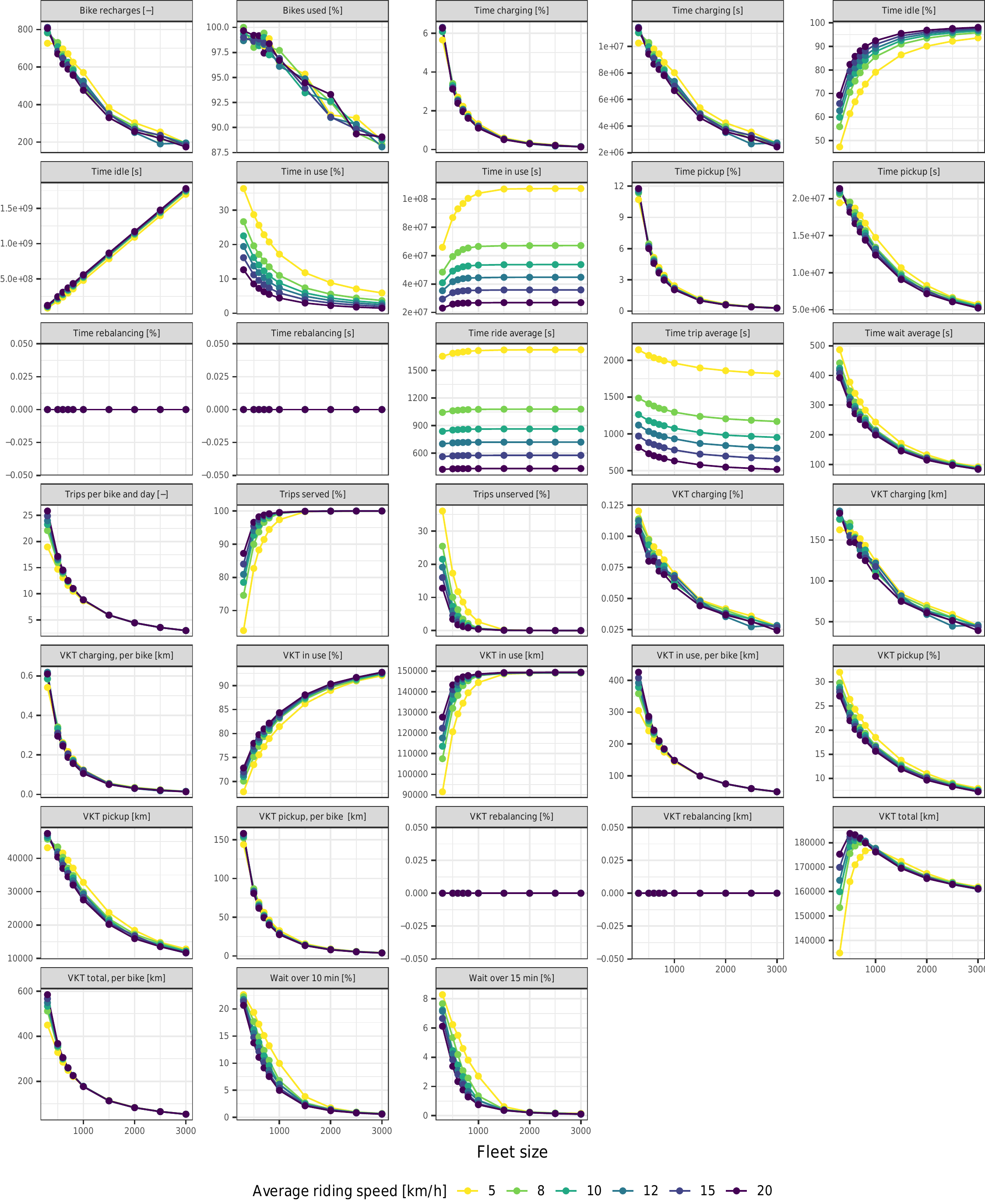}
    \label{fig:influence-AU-NR-riding_speed}
\end{figure}

\begin{figure}[!htb]
    \centering
    \caption{Autonomous system without rebalancing: influence of minimum battery level} 
    \includegraphics[width=\linewidth]{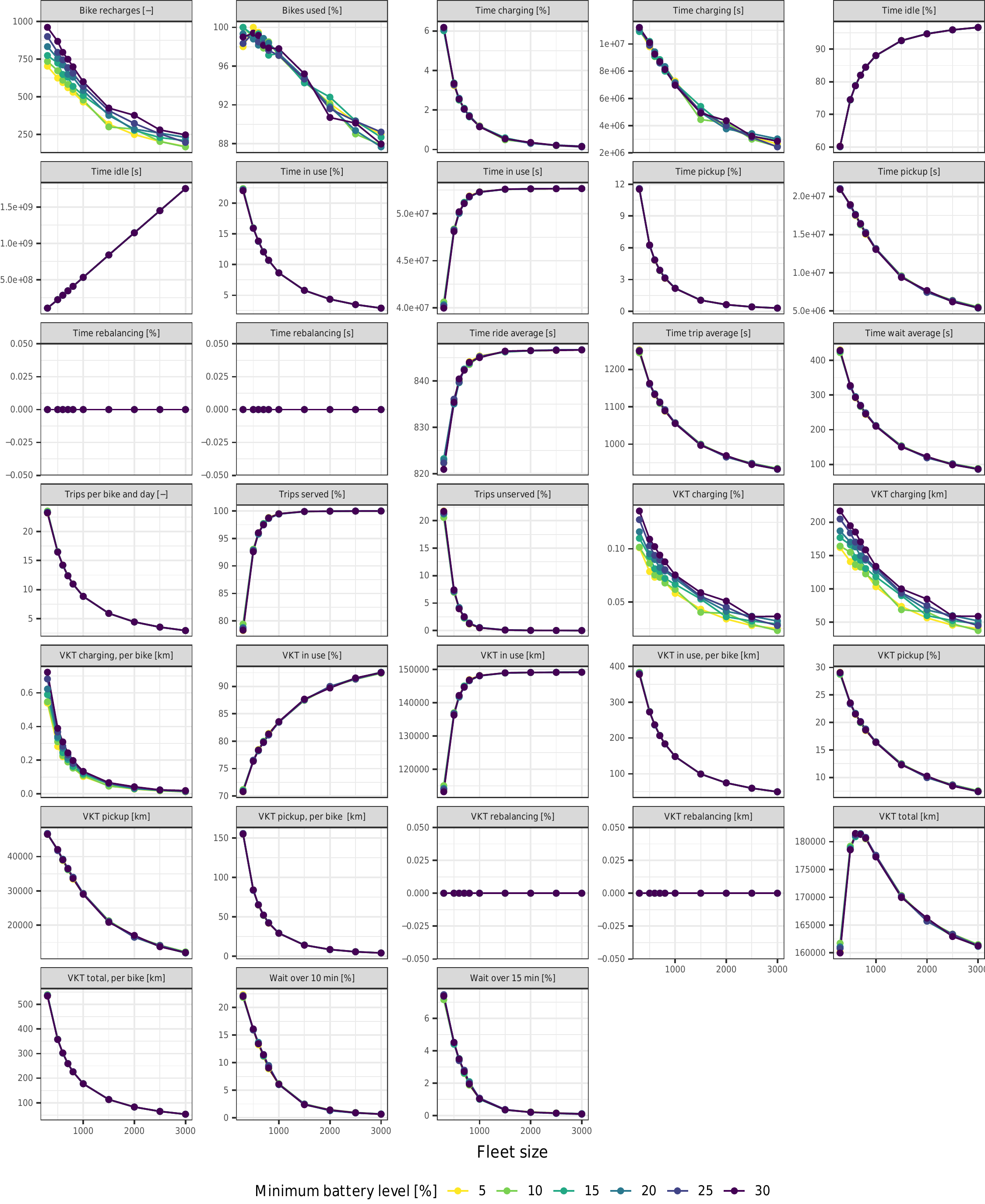}
    \label{fig:influence-AU-NR-battery_min_level}
\end{figure}

\begin{figure}[!htb]
    \centering
    \caption{Autonomous system without rebalancing: influence of battery autonomy} 
    \includegraphics[width=\linewidth]{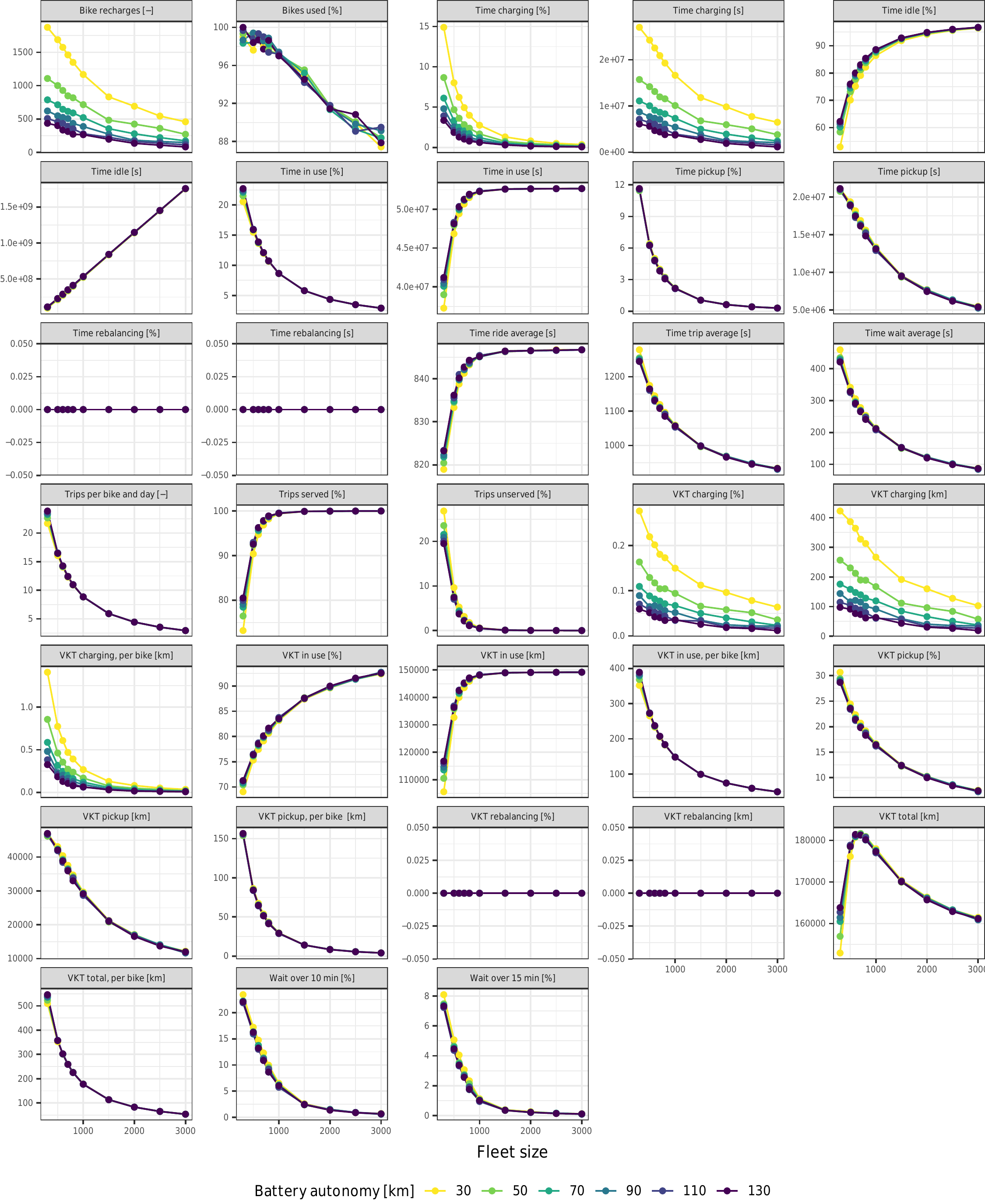}
    \label{fig:influence-AU-NR-battery_autonomy}
\end{figure}

\begin{figure}[!htb]
    \centering
    \caption{Autonomous system without rebalancing: influence of battery recharge time} 
    \includegraphics[width=\linewidth]{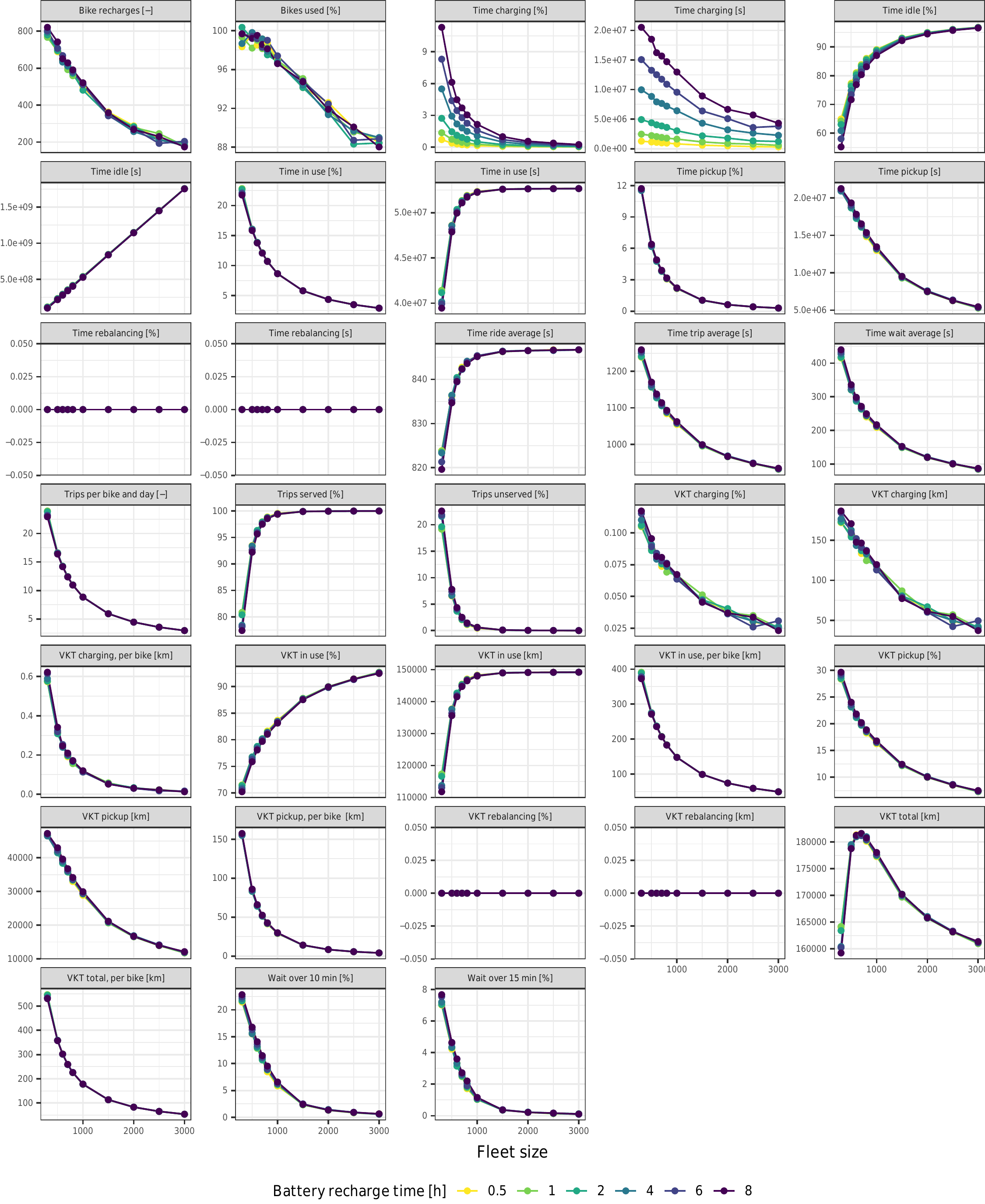}
    \label{fig:influence-AU-NR-battery_charge_time}
\end{figure}

\subsection{Autonomous system: ideal rebalancing}

\begin{figure}[!htb]
    \centering
    \caption{Autonomous system with ideal rebalancing: influence of average autonomous driving speed} 
    \includegraphics[width=\linewidth]{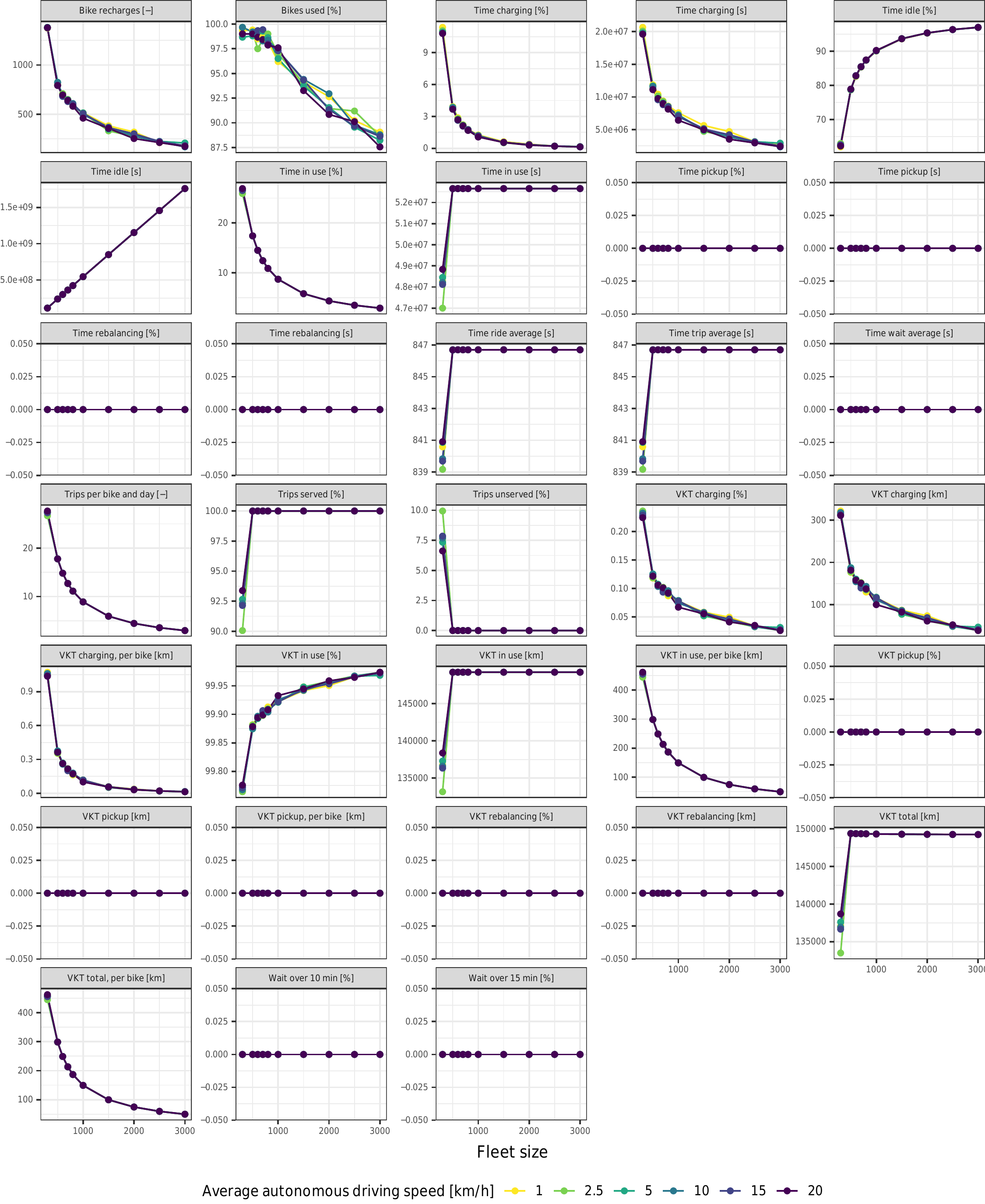}
    \label{fig:influence-AU-IR-autonomous_speed}
\end{figure}

\begin{figure}[!htb]
    \centering
    \caption{Autonomous system with ideal rebalancing: influence of average riding speed} 
    \includegraphics[width=\linewidth]{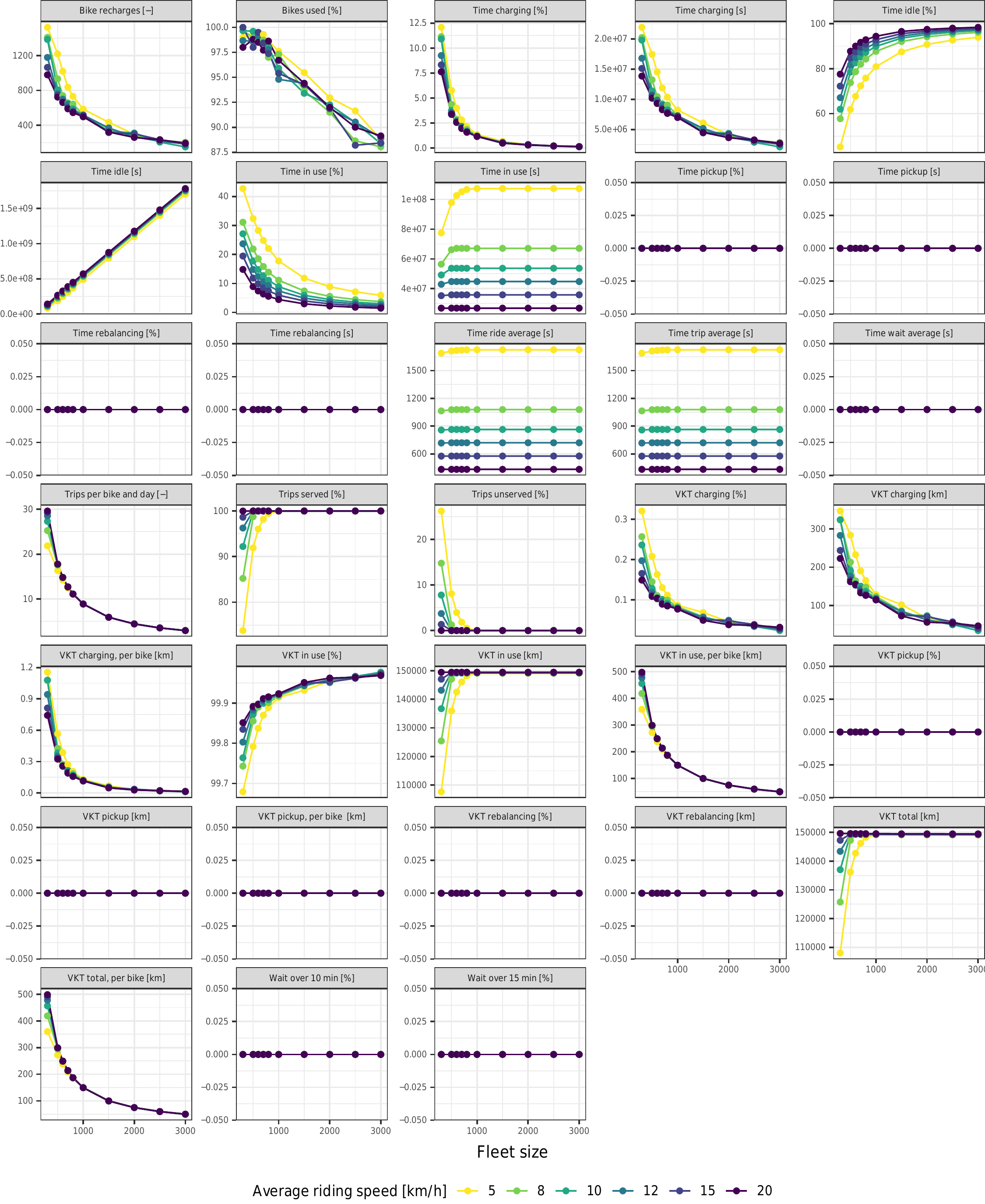}
    \label{fig:influence-AU-IR-riding_speed}
\end{figure}

\begin{figure}[!htb]
    \centering
    \caption{Autonomous system with ideal rebalancing: influence of minimum battery level} 
    \includegraphics[width=\linewidth]{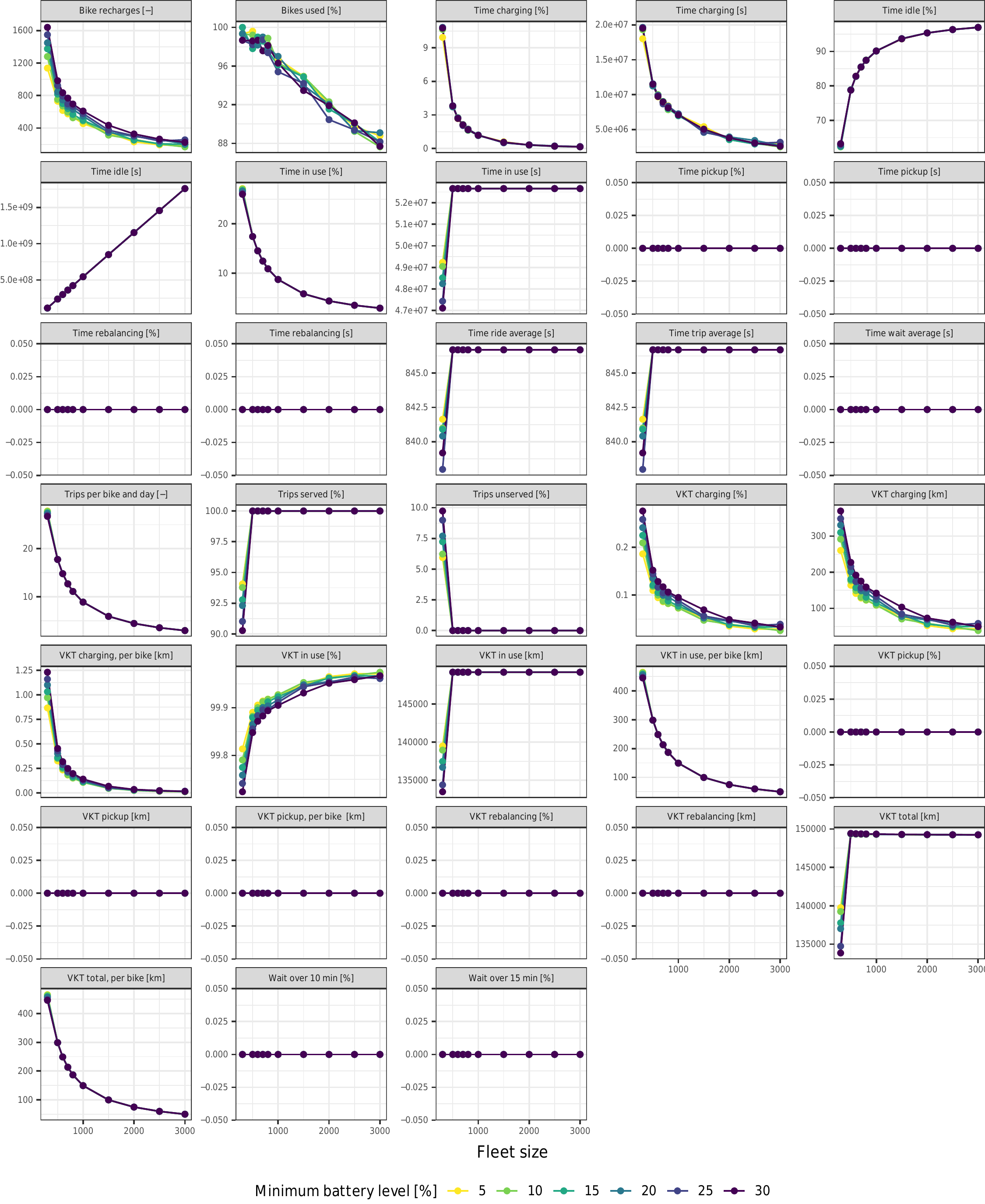}
    \label{fig:influence-AU-IR-battery_min_level}
\end{figure}

\begin{figure}[!htb]
    \centering
    \caption{Autonomous system with ideal rebalancing: influence of battery autonomy} 
    \includegraphics[width=\linewidth]{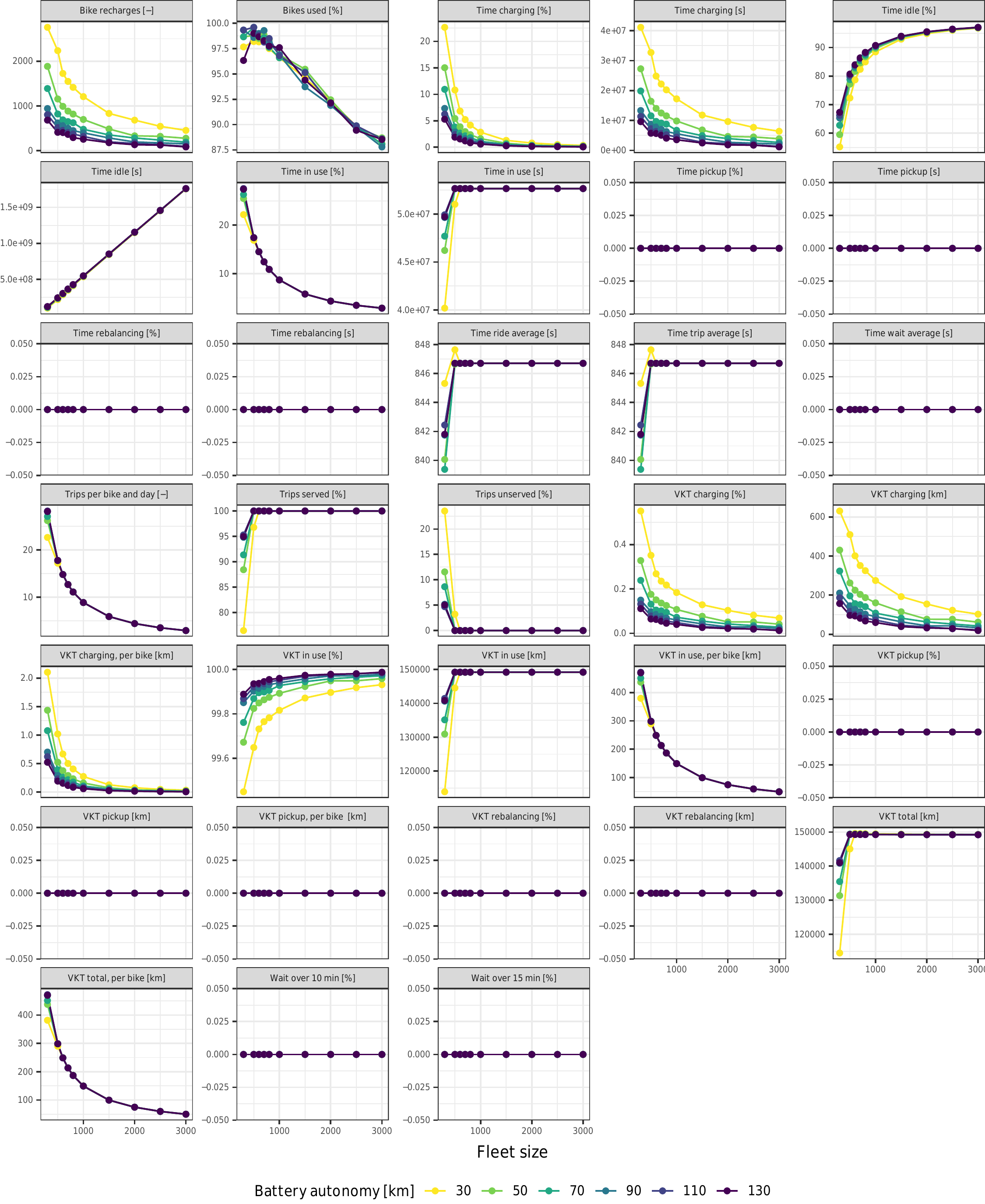}
    \label{fig:influence-AU-IR-battery_autonomy}
\end{figure}

\begin{figure}[!htb]
    \centering
    \caption{Autonomous system with ideal rebalancing: influence of battery recharge time} 
    \includegraphics[width=\linewidth]{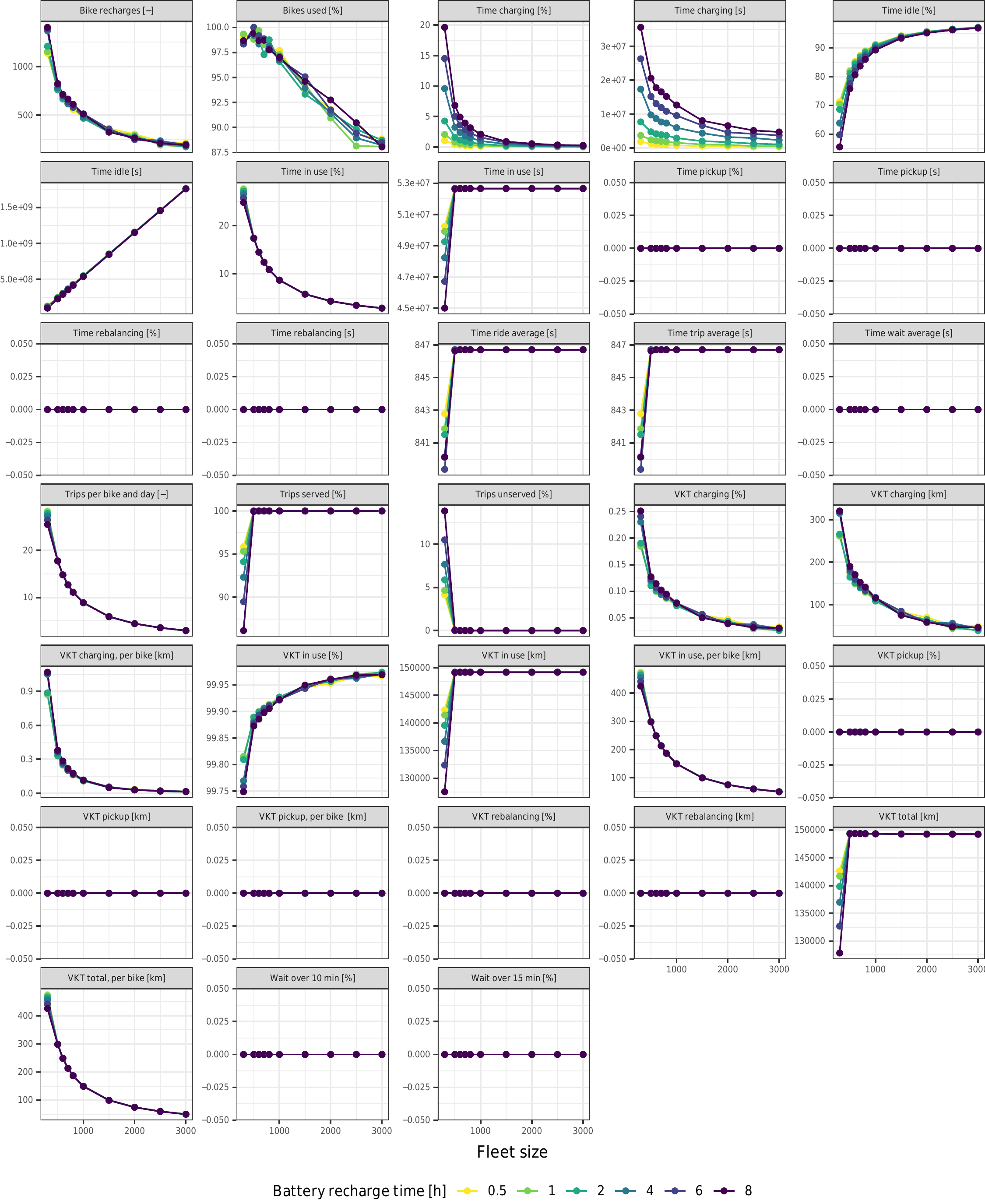}
    \label{fig:influence-AU-IR-battery_charge_time}
\end{figure}

\subsection{Autonomous system: predictive rebalancing based on demand prediction}

\begin{figure}[!htb]
    \centering
    \caption{Autonomous system with predictive rebalancing: influence of average autonomous driving speed} 
    \includegraphics[width=\linewidth]{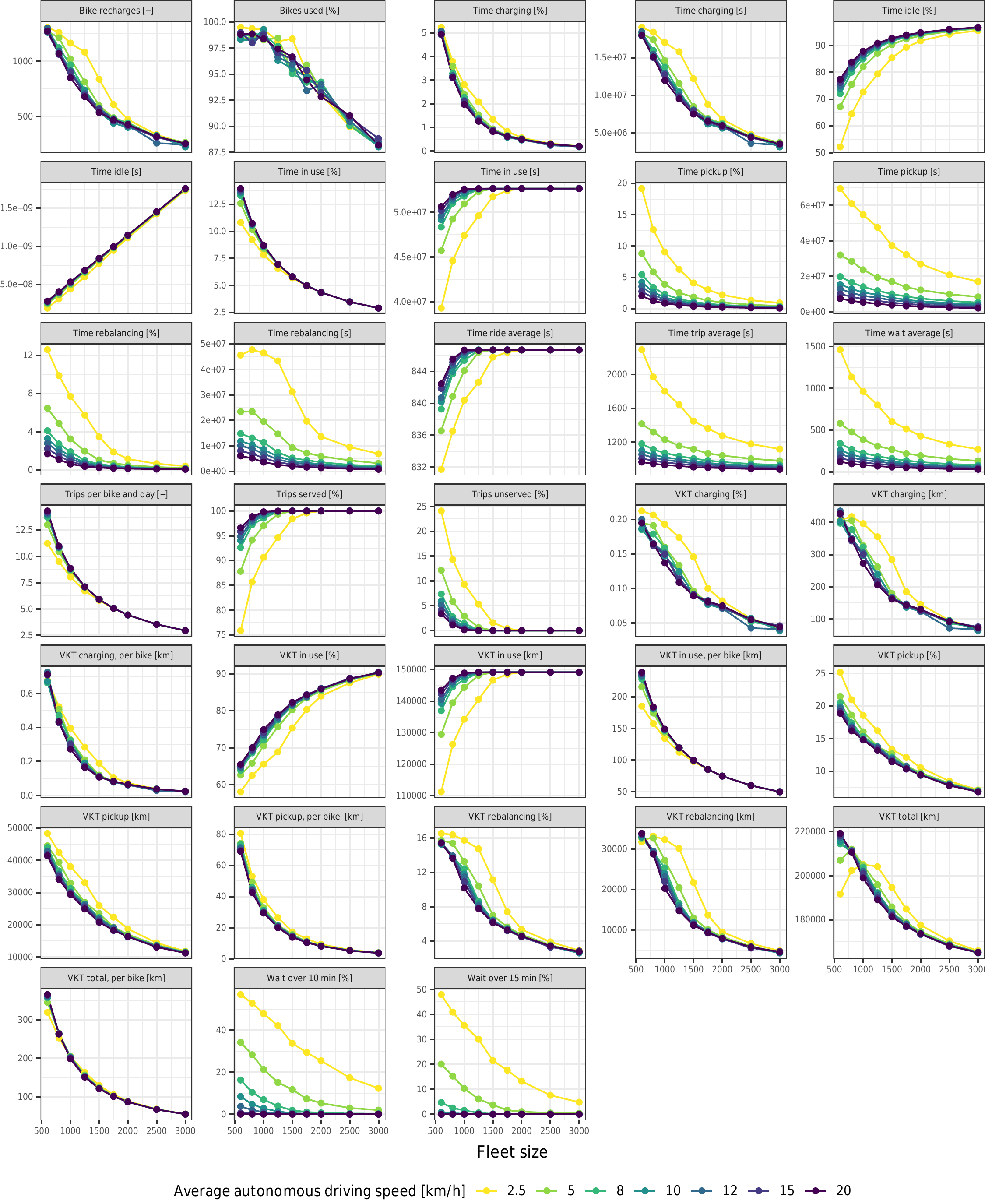}
    \label{fig:influence-AU-PR-autonomous_speed}
\end{figure}

\end{document}